\documentclass[hyper,letterpaper]{JHEP3}
\usepackage{amsmath,amssymb,amsfonts,bm}
\usepackage{graphicx}
\usepackage{multirow}
\usepackage{verbatim}
\usepackage{appendix}
\usepackage{rotating}
\usepackage{multirow}

\numberwithin{equation}{section}

\newcommand{\bea}{\begin{eqnarray}}
\newcommand{\eea}{\end{eqnarray}}
\newcommand{\be}{\begin{equation}}
\newcommand{\ee}{\end{equation}}
\newcommand{\mb}{\mathbf}

\newcommand{\wt}{\widetilde}

\newcommand{\ol}{\overline}

\newcommand{\eg}{\emph{e.g.}}
\newcommand{\ie}{\emph{i.e.}}
\newcommand{\cf}{\emph{cf.}}

\newcommand{\Z}{{\mathbb Z}}
\newcommand{\R}{{\mathbb R}}
\newcommand{\C}{{\mathbb C}}

\renewcommand{\H}{{\mathbb H}}
\newcommand{\Li}{{\rm Li}}

\newcommand{\Tr}{{\rm Tr \,}}
\renewcommand{\Re}{{\rm Re}}
\renewcommand{\Im}{{\rm Im}}
\newcommand{\bs}{\backslash}
\newcommand{\pd}{\partial}

\newcommand{\CA}{\mathcal{A}}
\newcommand{\CB}{\mathcal{B}}
\newcommand{\CC}{\mathcal{C}}

\newcommand{\CH}{\mathcal{H}}

\newcommand{\CL}{\mathcal{L}}
\newcommand{\CM}{\mathcal{M}}
\newcommand{\CN}{\mathcal{N}}
\newcommand{\CO}{\mathcal{O}}
\newcommand{\CP}{\mathcal{P}}

\newcommand{\CT}{\mathcal{T}}

\newcommand{\CW}{\mathcal{W}}
\newcommand{\CX}{\mathcal{X}}

\newcommand{\CZ}{\mathcal{Z}}

\title{Gauge Theories Labelled by Three-Manifolds}

\author{Tudor Dimofte$^{1}$ and Davide Gaiotto$^1$ and Sergei Gukov$^{2,3}$
\\ ~
\\
$^1$ Institute for Advanced Study, Einstein Dr., Princeton, NJ 08540, USA\\
$^2$ California Institute of Technology, Pasadena, CA 91125, USA \\
$^3$ Max-Planck-Institut f\"ur Mathematik, Vivatsgasse 7, D-53111 Bonn, Germany}

\abstract{We propose a dictionary between geometry of triangulated 3-manifolds and
physics of three-dimensional $\CN = 2$ gauge theories.  Under this duality,
standard operations on triangulated 3-manifolds and various invariants thereof (classical as well as quantum)
find a natural interpretation in field theory.
For example, independence of the $SL(2)$ Chern-Simons partition function on the choice of triangulation translates to a statement that $S^3_b$ partition functions
of two mirror 3d $\CN=2$ gauge theories are equal.
Three-dimensional $\CN=2$ field theories associated to 3-manifolds can be thought of
as theories that describe boundary conditions and duality walls in four-dimensional $\CN=2$ SCFTs,
thus making the whole construction functorial with respect to cobordisms and gluing.
\\
\\
\\
\\
\\
\\
{\tt CALT-68-2847}}

\begin{document}

%%%%%%%%%%%%%%%%%%%%%%%%%%%%%%%%%%%%%%%%%%%%%%%%%%%%%%%%%%%%%%%%%%%%%

\section{Introduction}
\label{sec:intro}
One of the predictions of String Theory/M-Theory is the existence of a discrete family
of maximally symmetric six-dimensional conformal field theories, labeled by a simply-laced Lie algebra $\mathfrak{g}$.
These theories lack a Lagrangian definition, but some of their properties are known.
The existence of such six-dimensional SCFT's has a simple, but perhaps surprising, consequence:
it allows a geometric description of many lower-dimensional supersymmetric field theories.
Indeed, one can define large families of $6-d$ dimensional theories $T[M_d,\mathfrak{g}]$
via compactification of the six-dimensional theory labeled by $\mathfrak{g}$ on a $d$-dimensional manifold $M_d$.
If the compactification is accompanied by an appropriate twist, it will lead to theories with $6-d$ dimensional supersymmetry.
In order to fully exploit this type of construction, one should ideally give an alternative explicit definition
of these ``effective'' theories directly in $6-d$ dimensions. If that can be accomplished, the result is a large family
of theories defined in $6-d$ dimensions, whose properties are controlled by the geometry of $d$-dimensional manifolds.

This program was pursued successfully for $d=2$ \cite{Gaiotto-dualities, GMN, AGT}.
The compactification of the six-dimensional theories on a Riemann surface $\CC$
leads to ${\cal N}=2$ supersymmetric gauge theories $T[\CC,\mathfrak{g}]$ in four dimensions.
The geometry of the Riemann surface controls a variety of protected quantities in the four-dimensional gauge theories:
the space of exactly marginal deformations, the space of vacua in flat space and upon compactification on a circle,
the partition function of the $\Omega$-deformed theory, the $S^4$ partition function, the superconformal index, {\it etc.}

It is natural to wonder if there is a similar $d=3$ dictionary.
A twisted compactification of a 6d theory on a three-manifold $M_3$
will give an ${\cal N}=2$ field theory $T[M_3,\mathfrak{g}]$ in three dimensions.
Some properties of these theories follow from the definition.
For example, one of the basic properties of the 6d theories is that they reduce to 5d SYM upon compactification on a circle.
If we consider a 6d SCFT on $S^1 \times M_3$, we find that the moduli space of vacua of $T[M_3,\mathfrak{g}]$
is the same as the space of flat complex $\mathfrak{g}$-connections on $M_3$ \cite{DGH}.

\begin{figure}[htb]
\centering
\includegraphics[width=4.8in]{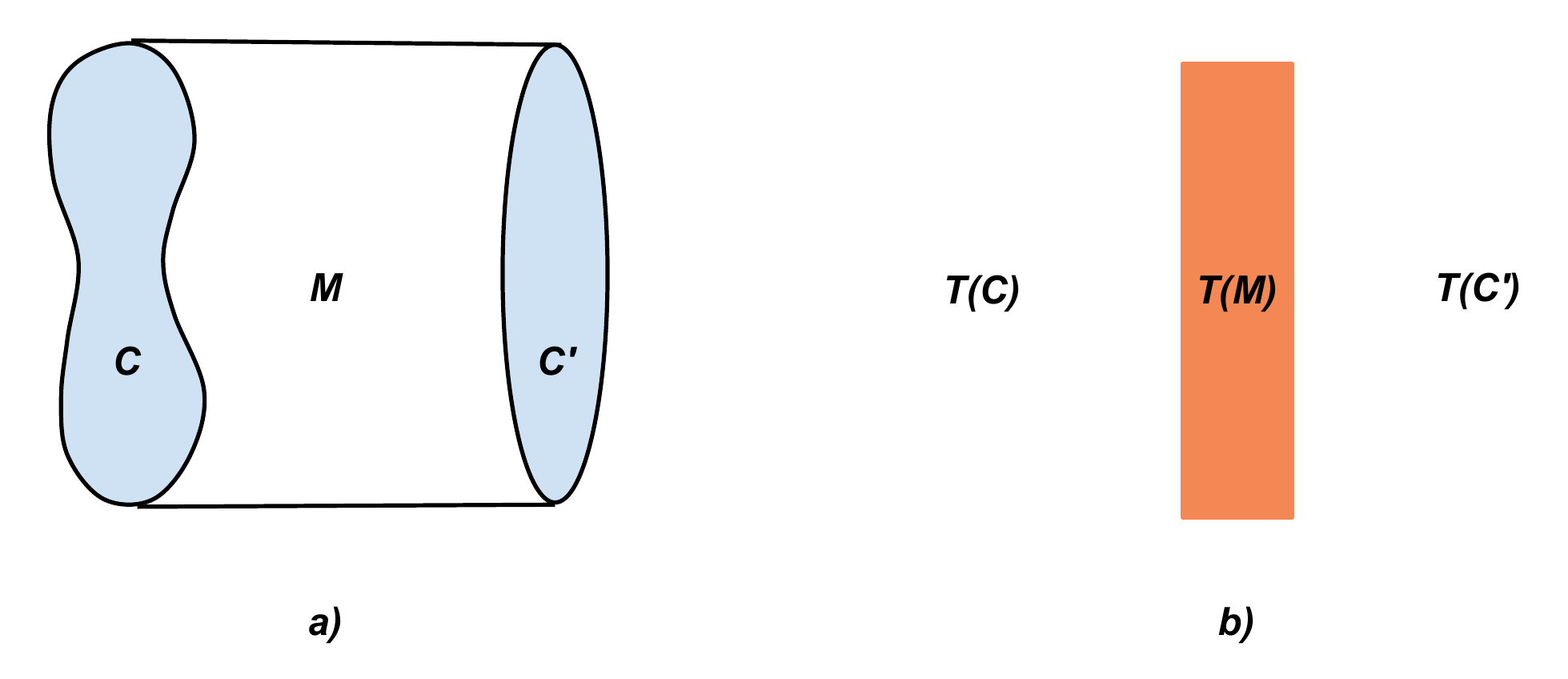}
\caption{$(a)$ A cobordism $M$ between $\CC$ and $\CC'$ gives rise to a domain wall $(b)$ between
4d $\CN=2$ theories $T(\CC)$ and $T(\CC')$.}
\label{fig:cobordism1}
\end{figure}

\begin{figure}[htb]
\centering
\includegraphics[width=4.8in]{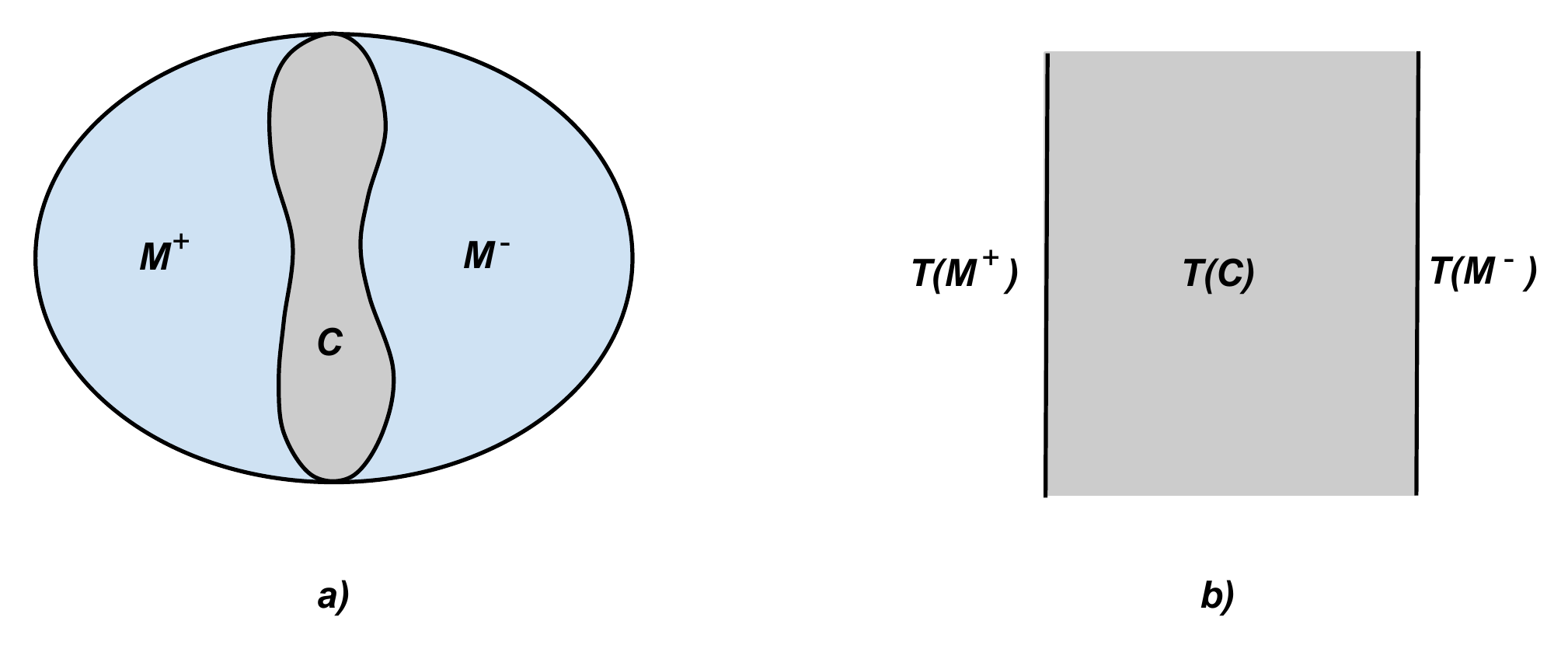}
\caption{$(a)$ A 3-manifold $M$ stretched along a `neck' $\R \times \CC$ becomes a 4d $\CN=2$ superconformal
theory $(b)$ on $\R^3 \times I$ coupled to 3-dimensional theories $T(M^+)$ and $T(M^-)$ at the boundary.
The 4d $\CN=2$ gauge theory in the bulk is determines by the cross-section $\CC$ of the 3-manifold $M$.}
\label{fig:cobordism2}
\end{figure}

One way to find other properties of this $d=3$ correspondence is to draw lessons from its $d=2$ version.
Indeed, consider a three-dimensional cobordism, {\it i.e.} a 3-manifold $M_3$ which interpolates between
two (or, more generally, several) Riemann surfaces, as in Figure~\ref{fig:cobordism1}.
The compactification of the six-dimensional theory on the cobordism should give a domain wall
between the 4d theories associated to the Riemann surfaces.
Note, in particular, that a half-BPS domain wall ({\it cf.} Figure~\ref{fig:cobordism1})
or a boundary condition ({\it cf.} Figure~\ref{fig:cobordism2}) in a 4d ${\cal N}=2$ field theory
preserve the same amount of supersymmetry as a three-dimensional ${\cal N}=2$ field theory.

Therefore, one possible strategy for understanding $T[M_3,\mathfrak{g}]$ is to directly leverage
the $d=2$ correspondence to construct the three-dimensional field theories:
take a closed manifold $M_3$, and stretch it to a configuration of long tubes with a Riemann surface as cross sections,
joined by appropriate plumbing fixtures, {\it i.e.} cobordisms.
One could then reduce the six-dimensional theory on the tubes of section $\CC$
to give known four-dimensional gauge theories $T[\CC,\mathfrak{g}]$ on segments, {\it cf.} Figure~\ref{fig:cobordism2}.
These theories would be coupled through the domain walls associated to the plumbing fixtures,
and the whole setup taken to define a three-dimensional gauge theory in the IR.

This strategy is hampered by the rapid proliferation of possible ``elementary'' plumbing fixtures:
one would need to find a way to construct the corresponding domain walls by hand, and demonstrate a large set of mirror symmetries
which relate different ways to glue together the same manifold. This should be contrasted with a similar approach in $d=2$,
where the tubes are all cylinders with $S^1$ cross-section, and the only plumbing fixture is the pair of pants.

We will follow an alternative, simpler strategy.
Namely, we will abandon the restriction to cut the manifold along tubes only,
and instead propose a candidate ${\cal N}=2$ SCFT $T_{M}$ for the theory $T[M,\mathfrak{su}(2)]$
based on a decomposition (triangulation) of a 3-manifold $M$ into tetrahedra, glued together along the triangular faces.
Note, here and in the rest of the paper we focus (mainly for simplicity) on $\mathfrak{g} = \mathfrak{su}(2)$.
Moreover, since we are interested only in the case $d=3$, so here and in what follows we denote $M_3$ simply by $M$.

We do not derive our construction of the $\CN=2$ theory $T_M$ directly from properties of the six-dimensional theory.
Instead, we wish to associate a simple ``building block'' theory $T_{\Delta}$ to each tetrahedron $\Delta$,
and to define the field theory analogue of the geometric gluing with a simple constraint in mind:
different triangulations of the same manifold must give equivalent definitions of the corresponding theory,
in the sense that they flow to the same SCFT in the IR.
In $d=2$ different decompositions of the same Riemann surface were related by known S-dualities.
In $d=3$ we aim to relate different triangulations of $M$ through known mirror symmetries,
so that every 3-manifold $M$ is associated to a well-defined, triangulation-independent 3d ${\cal N}=2$ SCFT.

We describe the theory $T_M$ as the IR fixed point of an abelian Chern-Simons-matter theory
whose Lagrangian depends on the choice of triangulation of $M$
(plus some extra decoration $\Pi$ that one encounters in $SL(2)$ Chern-Simons theory on $M$).
Intuitively, given a triangulation $M = \bigcup_{i=1}^N \Delta_i$,
we construct a theory for each tetrahedron $\Delta_i$
\be \Delta_i \qquad \leadsto \qquad T_{\Delta_i} \,, \ee
and glue the tetrahedra together to build
\be M \qquad\leadsto \qquad T_M \sim \bigotimes_i T_{\Delta_i} \,. \label{MvsTM} \ee
The gluing of theories $T_{\Delta_i}$ involves a bit more than just taking a tensor product,
and one of the main technical aims of this paper is to develop a proper understanding of the sign `$\sim$' in \eqref{MvsTM}.
Loosely speaking, the gluing involves two steps, which require a careful explanation
and depend on a choice of the extra data $\Pi$ (defined below):
gauging some flavor symmetries, with carefully chosen Chern-Simons couplings,
and adding a superpotential coupling for each internal edge of the triangulation.
The choice of the operators which enter the superpotential couplings is the most subtle part of the construction.
In general, they cannot be simultaneously realized as products of elementary fields,
but are defined as 't Hooft monopole operators.

Regardless of the compactification from six dimensions,
the family of 3d ${\cal N}=2$ SCFTs $T_M$ associated to 3-manifolds $M$ is an interesting object,
and we hope it will lead to interesting connections between three-dimensional SCFTs and three-dimensional geometry and topology.
For example, quantities like the superconformal index of $T_M$ or the partition function on $S^3$
should map to interesting three-manifold invariants,
as summarized in Table \ref{tab:dict}.
In this paper we specialize to a very simple building block theory for the tetrahedron, which is essentially the theory of a single chiral multiplet.
We believe our approach is much more general though, and with an appropriate choice of tetrahedron building block one can produce natural candidates for $T[M,\mathfrak{g}]$.

\begin{table}[htb]
\centering
\renewcommand{\arraystretch}{1.3}
\begin{tabular}{|@{\quad}c@{\quad}|@{\quad}c@{\quad}| }
\hline  {\bf 3-manifold} $M$ & {\bf 3d} $\CN=2$ {\bf theory} $T_M$
\\
\hline
\hline ideal tetrahedron & theory $T_{\Delta}$ \\
\hline change of triangulation & mirror symmetry \\
\hline change of polarization $\Pi$ & $Sp(2N,\Z)$ duality action \\
\hline boundary flip & $F$ transformation \\
\hline gluing along &  superpotential \\[-.2cm]
 an internal edge & coupling \\
\hline Wilson lines & line operators \\
\hline boundary $\CC = \partial M$ & coupling to 4d $\CN=2$ theory \\
\hline flat $SL(2,\C)$ connections & SUSY moduli on $\R^2 \times S^1$ \\
\hline ${\rm Vol} (M) + i {\rm CS} (M)$ & twisted superpotential $\widetilde{\CW}_{{\rm eff}}$ \\
\hline $SL(2)$ Chern-Simons &  \multirow{2}{*}{partition function on $S^3_b$} \\[-.2cm]
 partition function & \\
\hline Seiberg-Witten invariants & superconformal index \\
\hline
\end{tabular}
\caption{The dictionary between geometry and physics.}
\label{tab:dict}
\end{table}

We will be able to motivate our proposal for $T_M = T[M,\mathfrak{su}(2)]$ in a wide variety of ways,
and to check that it has expected properties.
In particular, we take inspiration from two related facts:
\begin{itemize}
\item The moduli space of vacua
of the 3d theory must coincide (with some caveats) with the space of flat $SL(2)$ connections on $M$.
\item The partition function of $T[M,\mathfrak{g}]$ on an ellipsoid $S^3_b$, as in \cite{HHL},
should coincide with the (analytically continued) $\mathfrak{g}$ Chern-Simons partition function on $M$.
\end{itemize}
We engineer $T_M = T[M,\mathfrak{su}(2)]$ in such a way that these two properties are automatically true.

One may wonder why the IR dynamics of the non-abelian six-dimensional theory on
a 3-manifold $M$ should admit a dual 3d description based on abelian gauge fields.
A likely answer is that in a generic vacuum of the 3d theory, the 6d theory is deep in its Coulomb branch on most of $M$.
Far on the Coulomb branch, the 6d theory reduces to an abelian theory of self-dual forms.
It is conceivable that the abelian gauge fields in our description arise from these 6d abelian fields,
and the matter fields arise from excitations localized in the regions of $M$ where the 6d theory is close to the origin of the Coulomb branch.
Similar ideas are useful for $d=2$, but they give rise to IR-free, non-UV complete four-dimensional abelian gauge theories.
On the other hand, a three-dimensional abelian gauge theory is a UV complete description of an IR fixed point.

Finally, we should describe in more detail the class of 3-manifolds $M$ to which our construction applies.
In the $d=2$ case, it is useful to introduce codimension two defects of the six-dimensional $(2,0)$ theory,
which sit at points of the Riemann surface and fill the entire 4d space-time.
These defects do not break any further supersymmetry, and greatly extend the space of
four-dimensional ${\cal N}=2$ theories which are amenable of a geometric construction.
The presence of even a single puncture allows one to use some interesting tools based
on ``ideal'' triangulations of Riemann surfaces, which have vertices at the defects only.
Similarly, in $d=3$ one can add the very same kind of defects, which fill the entire 3d space-time
and are supported on a line (or, better, on a knot/link) inside $M$.
Again, our construction employs an ``ideal'' triangulation: the tetrahedra have vertices at the defects.
In particular, the manifold should have at least one defect.
In $d=2$ a defect can represent a semi-infinite tubular region of the surface, and the same is true in $d=3$.

Our construction does not actually force us to glue all the faces of the tetrahedra pairwise together,
to get a closed manifold with defects. We can also do a partial gluing, and obtain theories associated
to manifolds with boundaries made by faces of the tetrahedra.
The defects and boundaries both have an interpretation in terms of coupling to four-dimensional ${\cal N}=2$ theories.
The difference is that defects represented by semi-infinite tubular region with a cross section $\CC$
correspond to couplings of theories $T_M$ to ${\cal N}=2$ theories in the UV.
In particular, for our theories $T_M = T[M,\mathfrak{su}(2)]$ that come from compactification of
the $(2,0)$ theory of type $\mathfrak{g} = \mathfrak{su}(2)$, the corresponding ${\cal N}=2$ theories
associated to $\CC$ in the UV typically have $SU(2)$ gauge groups.
For example, closed cusps in $M$ represented by semi-infinite tubular region with a 2-torus $\CC = T^2$
as a cross section correspond to coupling to four-dimensional $\CN=4$ super-Yang-Mills with gauge group $SU(2)$.

On the other hand, a big, ``geodesic'' boundary of $M$ (formed from unglued tetrahedron faces) of topology $\CC$ represents coupling
of theory $T_M$ to the IR limit of the $\CN=2$ four-dimensional theory $T[\CC]$ ({\it cf.} Figure \ref{fig:cobordism2}).
In contrast to its UV version, this IR theory is usually abelian.
Therefore, to summarize, each boundary of $M$ corresponds to a possible coupling of the 3d $\CN=2$ theory $T_M$
to either IR or UV limit of the 4d $\CN=2$ gauge theory $T[\CC]$, depending on whether the boundary $\CC$ is big and ``geodesic'' or small and ``defect-like.''
This is very natural because a typical example of a boundary condition for a weakly coupled four-dimensional ${\cal N}=2$ field theory
consists of a three-dimensional ${\cal N}=2$ field theory living at the boundary and coupled to the bulk degrees of freedom.
Looking at the same boundary condition or domain wall in different weakly coupled regions of the bulk parameter space
leads to different descriptions involving different three-dimensional degrees of freedom.\\

The paper is organized as follows.
In section \ref{sec:geom} we will review the geometric properties of triangulated three-manifolds
that will inspire the construction of $T_M$. In fact, we will need to generalize the standard constructions
a little bit in order to describe triangulations of 3-manifolds that support irreducible flat $SL(2,\C)$ connections.
Section \ref{sec:ops} reviews the physical tools needed for the construction of $T_M$,
whereas the definition of the 3d $\CN=2$ theory $T_M$ is presented in section \ref{sec:glue}.
Section \ref{sec:moduli} describes the match between the moduli space of flat connections on $M$
and the moduli space of vacua of the theory $T_M$ on a circle.
Section \ref{sec:S3b} describes a similar match between the $SL(2)$ Chern-Simons partition function of $M$
and the ellipsoid partition function of $T_M$.
Finally, section \ref{sec:lines} extends the dictionary between geometry of $M$ and physics of $T_M$ to line operators.

%%%%%%%%%%%%%%%%%%%%%%%%%%%%%%%%%%%%%%%%%%%%%%%%%%%%%%%%%%%%%%%%%%%%%%%%%%

\section{Geometry of 3-manifolds}
\label{sec:geom}

In this section, we discuss the geometric construction of oriented 3-manifolds $M$ from basic building blocks: ideal tetrahedra. Such ``ideal triangulations'' in three dimensions were initiated by Thurston \cite{thurston-1980}. More precisely, we wish to build 3-manifolds that support irreducible flat $SL(2,\C)$ connections $\CA$. For this purpose, it is often convenient to replace flat $SL(2,\C)$ connections with hyperbolic metrics%
\footnote{The equivalence between flat connections and hyperbolic geometry results from the fact that the isometry group of hyperbolic three-space is $(P)SL(2,\C)$, \cf\ \cite{thurston-1980, Witten-gravCS, gukov-2003, DGLZ}. Almost all flat connections can be realized as (possibly degenerate) hyperbolic metrics; for further remarks on this in the context of ideal triangulations, see \cite{Champ-hypA, Dunfield-Mahler}, and Section 4 of \cite{Dimofte-QRS}.} %
on $M$ --- that is, metrics of constant curvature $-1$. Then the $SL(2,\C)$ structures become geometric, and can be manipulated in a much more intuitive manner.

\FIGURE[l]{
\includegraphics[width=2.8in]{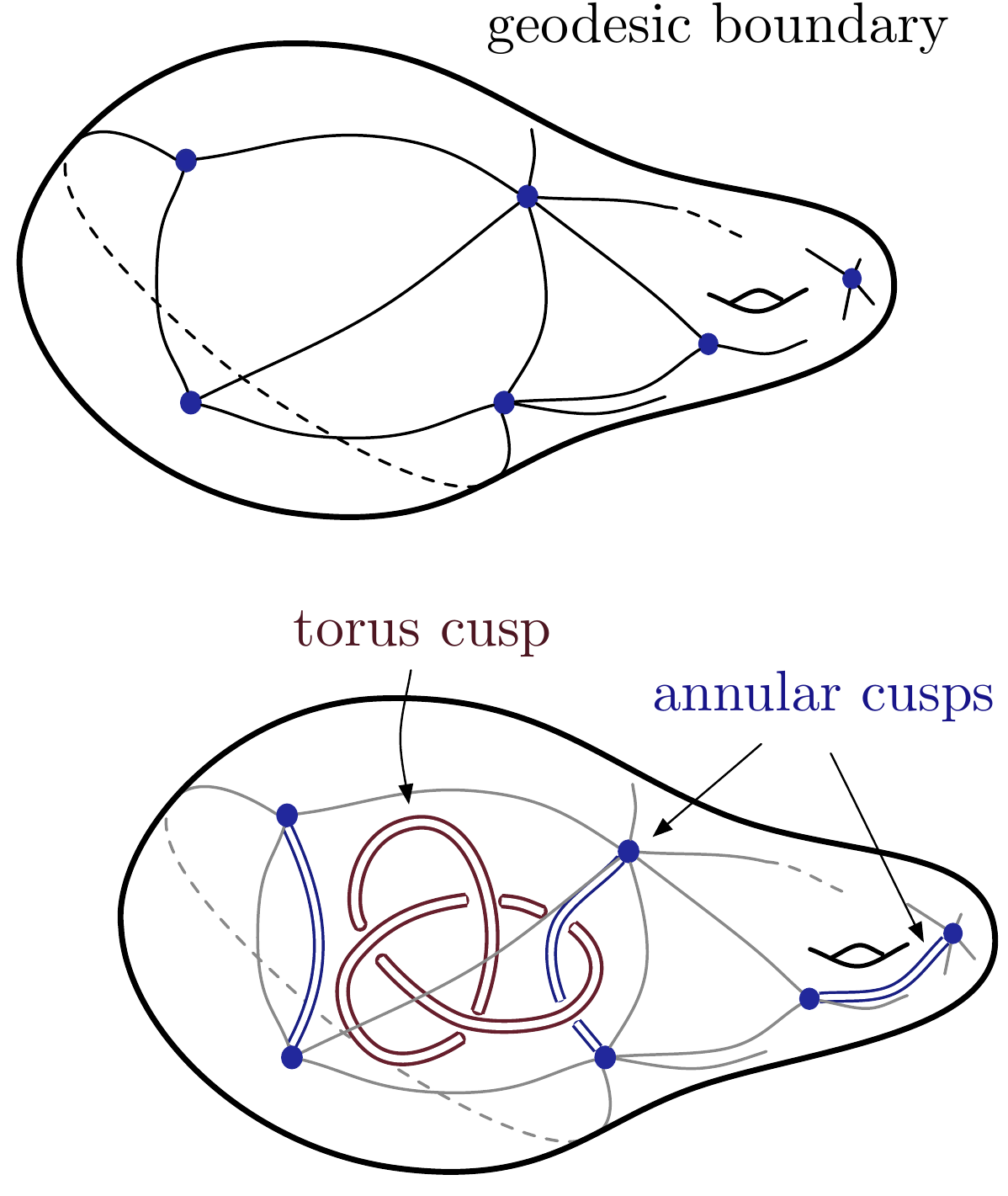}
\caption{Types of boundaries for $M$}
\label{fig:bdies}
}

The 3-manifolds we consider have two different types of boundary, \emph{geodesic boundaries} and \emph{generalized cusps}. Geometrically, the geodesic boundaries are (possibly punctured) geodesic surfaces of any genus, and come with an induced 2-dimensional hyperbolic metric. Any triangulation of $M$ will determine a triangulation of the geodesic boundary, which will be part of the data in eventually defining a 3d gauge theory.

In contrast, ``cusp'' boundaries do not have a triangulation that is relevant in defining 3d gauge theories. Geometrically, cusps are knotted loci where the hyperbolic metric on $M$ develops a cone angle, or the $SL(2,\C)$ connection has a specified monodromy defect.  Such loci can be resolved to boundaries with the topology of either tori $T^2$ or annuli $S^1\times I$. In either case, the induced metric on cusp boundaries is Euclidean. Well-studied examples of 3-manifolds with torus cusps are knot complements in $S^3$. More generally, a cusp might begin and end at punctures on the geodesic boundary of $M$ (Figure \ref{fig:bdies}). Then, the resolved cusp has the topology of an annulus.

The total boundary of $M$, with potential components of both types, determines a boundary moduli space of flat connections,
\be \CP_{\pd M} = \{\text{flat $SL(2,\C)$ connections on $\pd M$}\}\big/(\text{gauge equivalence}). \label{PdM} \ee
This is a symplectic phase space, with a natural holomorphic symplectic form
\be \omega_{\pd M} = \frac{1}{\hbar}\int_{\pd M}\Tr\big(\delta\CA\wedge\delta\CA\big)\,. \label{omegadM} \ee
The semiclassical parameter $\hbar$ here governs the normalization of the symplectic form.
Geometrically, $\omega_{\pd M}$ is an analytic continuation of the Weil-Petersson form in 2-dimensional hyperbolic geometry. In addition to the phase space $\CP_{\pd M}$, we can also define a Lagrangian submanifold \cite{gukov-2003}
\be \CL_M = \{\text{flat $SL(2,\C)$ connections on M}\}\big/(\text{gauge})\quad\subset\quad \CP_{\pd M}\,, \ee
which is the set of flat connections on $\pd M$ that can be extended as flat connections inside the 3-dimensional bulk of $M$. Mathematically, $\CL_M$ is described as the image of the ``character variety'' of $M$ inside the character variety of $\pd M$.

Our goal now is to construct a manifold $M$ together with the pair $(\CP_{\pd M},\CL_M)$ from ideal hyperbolic tetrahedra. This will give us an extremely explicit realization of boundary phase spaces, Lagrangians, and the symplectic structure \eqref{omegadM}, which in turn will enable us in Section \ref{sec:glue} to explicitly build the 3d gauge theory associated to $M$. As previewed in the introduction, this 3d theory will depend on $M$, a triangulation of its geodesic boundary, and a \emph{polarization}  $\Pi$ of its phase space $\CP_{\pd M}$ --- with additional ingredients such as $\CL_M$ playing roles like moduli spaces of vacua.%
\footnote{We note that topologically, one might engineer (resolved) cusp boundaries that look identical to geodesic boundaries. In particular, networks of annular cusps can assume the topology of nontrivial punctured Riemann surfaces \cite{DGGV-hybrid}. Formally, the phase spaces $\CP_{\pd M}$ associated to the two types of boundary would then be equivalent. However, the natural coordinate systems --- and in particular the polarizations --- for phase spaces on cusp and geodesic boundaries are very different. In turn, the 3d gauge theories associated to 3-manifolds with the two different types of boundary will be quite different.} \\

\FIGURE[r]{
\includegraphics[width=1.8in]{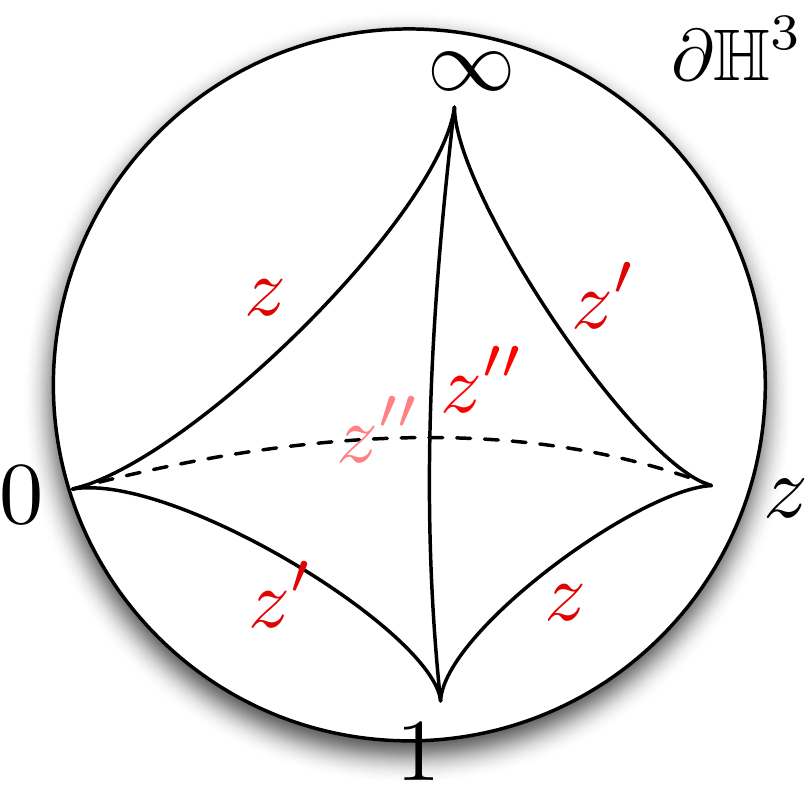}
\caption{An ideal hyperbolic tetrahedron in $\H^3$, with vertices on $\pd\H^3$}
\label{fig:tet}
}

\subsection{Building blocks}

The fundamental building block used in building our 3-manifolds $M$ is an ideal hyperbolic tetrahedron (Figure \ref{fig:tet}). Geometrically, an ideal tetrahedron $\Delta$ has faces that are geodesic surfaces and vertices that lie right on the boundary of hyperbolic 3-space $\H^3$. As shown in Figure \ref{fig:tet}, hyperbolic 3-space can be viewed as the interior of a 3-ball, with the Riemann sphere as its boundary.

The full hyperbolic structure of $\Delta$ is determined by a single complex cross-ratio of the positions of its vertices on $\pd \H^3$. There are three different ways to write this one cross-ratio, encoded in three different \emph{edge parameters} $(z,z',z'')$. Geometrically, the edge parameters are dihedral angles on pairs of opposite edges of the tetrahedron \cite{thurston-1980}. Explicitly,
\be z \,\equiv\, \exp(Z) \qquad\text{with}\qquad Z = \text{(torsion)}+i\,(\text{angle})\,, \ee
and similarly for $z'=\exp(Z')$ and $z''=\exp(Z'')$,
where ``torsion'' measures the twisting of the hyperbolic metric as one moves around an edge. As discussed in \cite{Dimofte-QRS}, the edge parameters satisfy $zz'z''=-1$, which leads to the definition of the boundary phase space
\be \CP_{\pd \Delta} \,=\, \big\{(z,z',z'')\in (\C^*\bs\{1\})^3\;\big|\; zz'z''=-1\big\} \,\simeq\, (\C^*\bs\{1\})^2\,,\ee
or in a lifted, logarithmic form,
\be \CP_{\pd \Delta} \,=\, \big\{(Z,Z',Z'') \in(\C\bs 2\pi i\Z)\;\big|\; Z+Z'+Z''=i\pi \big\}\,. \label{PDelta} \ee
This is an affine linear space, with symplectic form $\omega_{\pd \Delta} = \frac{1}{\hbar}dZ\wedge dZ'$ or, equivalently, a Poisson structure such that
\be \{Z,Z'\} = \{Z',Z''\} = \{Z'',Z\} = \hbar\,. \label{DPoisson} \ee
The edge parameters also obey a second relation $z+z'^{-1}-1=0$, which defines the Lagrangian submanifold
\be \CL_\Delta = \{ z+ z'^{-1}-1=0\} = \{e^Z+e^{-Z'}-1=0\}\quad \subset\;\CP_{\pd \Delta}\,. \label{LDelta} \ee
Any cyclic permutation of the Lagrangian equation (with $z\to z'\to z''\to z$) could also be used.

\FIGURE[l]{
\includegraphics[width=1.5in]{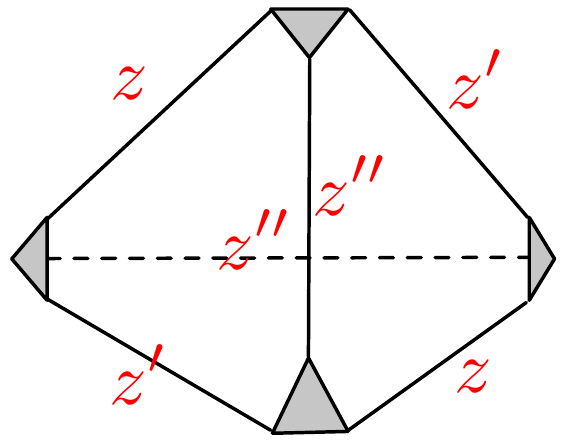}
\caption{A truncated ideal tetrahedron}
\label{fig:trunc}
}

Topologically, it is convenient to truncate or regularize the four vertices of an ideal tetrahedron, as in Figure \ref{fig:trunc}. The tetrahedron then has four large, geodesic boundaries, whose induced metric is hyperbolic; and four small boundaries at the truncated vertices, whose induced metric is Euclidean. In fact, the condition $Z+Z'+Z''=i\pi$ that defines the phase space in \eqref{PDelta} simply says that the sum of angles in the small Euclidean triangles at the vertices is always $\pi$.

While the Lagrangian equation $z+z'^{-1}-1=0$ follows directly from the geometric definition of $(z,z',z'')$ as equivalent cross-ratios, it also has an intrinsic description in terms of $SL(2,\C)$ connections. If we view the boundary $\pd\Delta$ of a tetrahedron as a four-punctured sphere, the phase space $\CP_{\pd \Delta}$ is the set of flat $SL(2,\C)$ connections with \emph{unipotent} monodromy around the four punctures. The Lagrangian $\CL_\Delta$ is then the subspace of flat connections with \emph{trivial} monodromy --- in other words, the flat connections that can be extended from the boundary into the bulk of the tetrahedron. Understanding this description explicitly in coordinates $(z,z',z'')$ requires a bit of further background, which we defer to Section \ref{sec:geobdy}.

\begin{figure}[htb]
\centering
\includegraphics[width=4.8in]{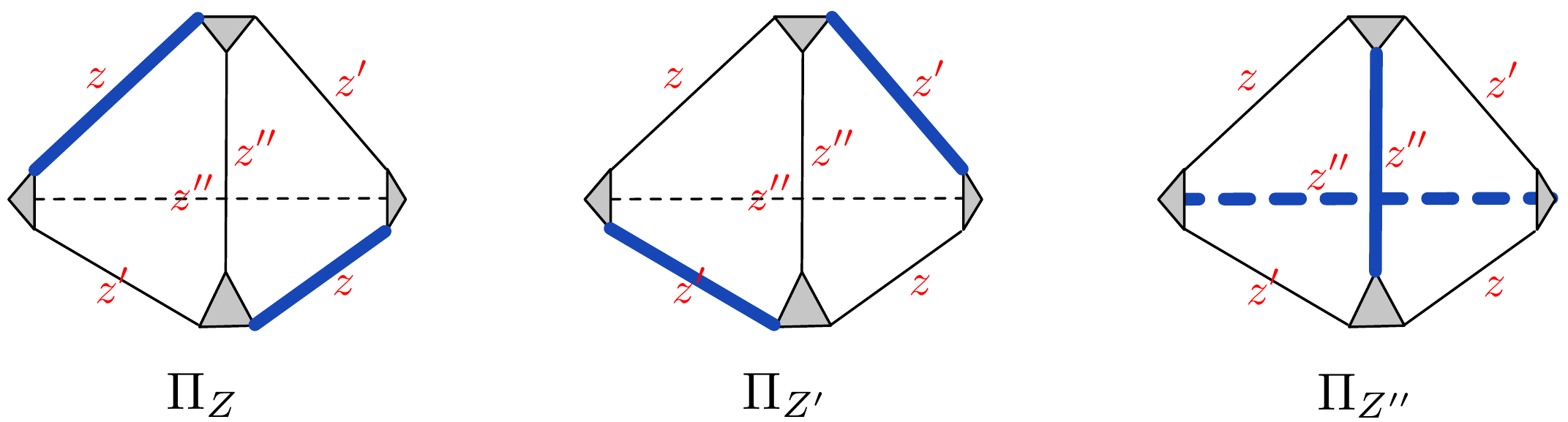}
\caption{Natural polarizations for a tetrahedron, with the thickened pairs of opposite edges corresponding to the ``position'' coordinate.}
\label{fig:tetpol}
\end{figure}

In order to define the gauge theory associated to a tetrahedron, we will need to choose a polarization $\Pi$ for its boundary phase space. This means choosing affine linear coordinates on $\CP_{\pd\Delta}$ that are canonically conjugate to each other with respect to the Poisson structure above, with one coordinate thought of as ``position'' and the other as ``momentum.'' There are three natural possibilities, which we call $\Pi_Z$, $\Pi_{Z'}$, and $\Pi_{Z''}$\,,
\be \begin{array}{lc@{\quad}c@{\qquad}c}  && \text{position $X$} & \text{conjugate momentum $P$} \\[.1cm]
 \Pi_Z &: & Z & 
 Z'' \\[.1cm]
 \Pi_{Z'} &: & Z' & 
 Z \\[.1cm]
 \Pi_{Z''} &: & Z'' & 
 Z'
 \end{array}
\label{tetpols}
\ee
Each of these polarizations can be encoded in a choice of opposite edges on the tetrahedron, such that the edge parameters of the distinguished edges act as ``positions'' (Figure \ref{fig:tetpol}).

We can define a larger class of polarizations by starting with any of those in \eqref{tetpols}, and acting with an affine symplectic transformation $Sp(2,\Z)\ltimes (i\pi\Z)^2$. By this we mean taking the vector $\big(\text{position},\,\text{momentum}\big)$, multiplying by $Sp(2,\Z)\simeq SL(2,\Z)$ matrices, and shifting both position and momentum by integer multiples of $i\pi$. For example, instead of $\Pi_Z$, we could have considered polarization $\Pi^-_Z$ in which $X^-=Z$ is position and $P^-=-Z'$ is momentum; then the transformation from $\Pi_Z$ to $\Pi^-_Z$ is
\be \hspace{-.2in} \Pi_Z\;\to\;\Pi^-_Z\;:\qquad \begin{pmatrix} X^-\\P^- \end{pmatrix}= \begin{pmatrix}1 & 0 \\ 1 & 1 \end{pmatrix}\begin{pmatrix} X\\P \end{pmatrix} + \begin{pmatrix} 0 \\ -i\pi \end{pmatrix}\,. \label{PiZm}
\ee
Similarly, to go from $\Pi_Z$ to $\Pi_{Z'}$, we transform
\be \begin{pmatrix} Z' \\ Z \end{pmatrix} = \begin{pmatrix} -1 & -1 \\ 1 & 0 \end{pmatrix}\begin{pmatrix} Z \\ Z'' \end{pmatrix} + \begin{pmatrix}i\pi \\ 0\end{pmatrix}\,,\ee
where the matrix involved is $ST = \left(\begin{smallmatrix} 0 & -1 \\ 1 & 0\end{smallmatrix}\right)\left(\begin{smallmatrix} 1&0\\1&1 \end{smallmatrix}\right)\in Sp(2,\Z)$. The identity $(ST)^3=I$ corresponds to the fact that three cyclic permutations of shape parameters brings us back where we started.

\subsection{Gluing}
\label{sec:geomglue}

\begin{figure}[htb]
\centering
\includegraphics[width=5in]{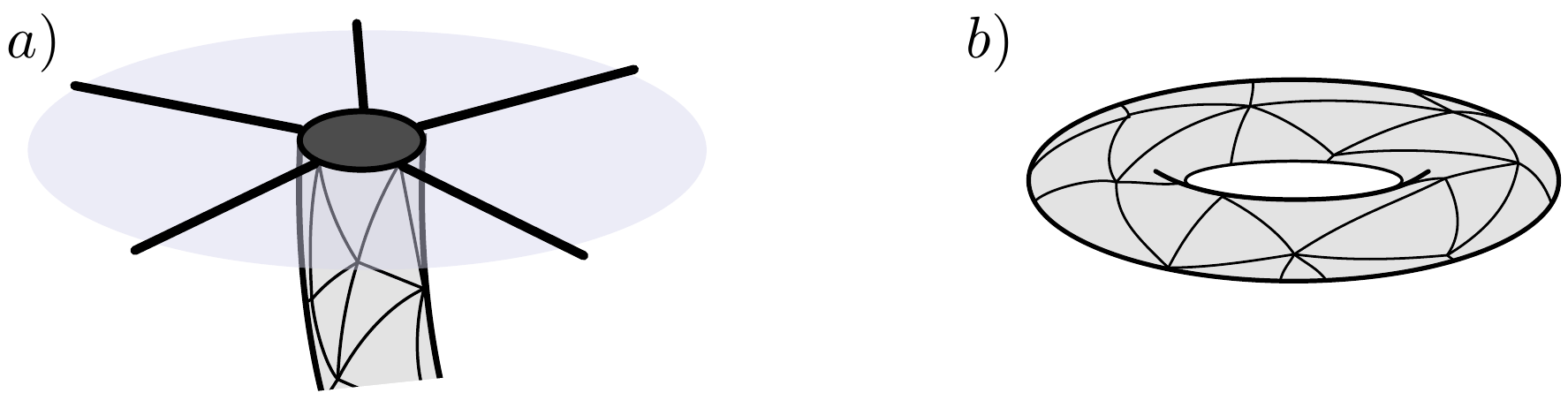}
\caption{Triangulations by Euclidean vertex triangles of (a) an annular cusp attached to a geodesic boundary, and (b) a torus cusp.}
\label{fig:gluingtet}
\end{figure}

Any 3-manifold $M$ with a combination of geodesic and cusp boundaries can be constructed from a collection of ideal tetrahedra $\{\Delta_i\}_{i=1}^N$, by gluing together their faces one pair at a time. Topologically, the geodesic boundary of $M$ comes from faces of tetrahedra that remain unglued. The torus or annular cusps of $M$, however, arise from assembling collections of small truncated-vertex triangles, as in Figure \ref{fig:gluingtet}. Geometrically, it is clear that the geodesic boundary of $M$ will be endowed with a hyperbolic metric, since all the faces of ideal tetrahedra are geodesic, hyperbolic surfaces. Similarly, the cusp boundaries become resolved into Euclidean tori or annuli, triangulated by the Euclidean truncated vertices.

In order for the hyperbolic metric on $M$ resulting from such a gluing to be smooth, one must impose that the total dihedral angle around every internal edge of the triangulation is $2\pi$, and that the hyperbolic torsion vanishes. In other words, for every internal edge $I_j$, the sum of complex edge parameters $Z_i,Z_i',Z_i''$ meeting this edge must equal exactly $2\pi i$. This could be written formally as
\be C_I \equiv \sum_{i=1}^N \Big[n(I,i)Z_i+n'(I,i) Z_i'+n''(I,i)Z_i''\Big] = 2\pi i\qquad (\forall\;\text{internal edges $I$})\,, \label{defCj} \ee
where $n(I,i) \in \{0,1,2\}$ is the number of times the edge $I$ in $M$ coincides with an edge parameter $Z_i$ of tetrahedron $\Delta_i$ in the triangulation $M= \bigcup_{i=1}^N\Delta_i$.

\FIGURE[r]{
\hspace{.1in}\includegraphics[width=1in]{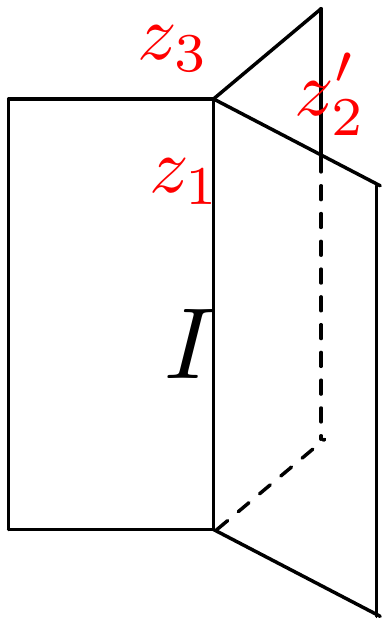}\hspace{.1in}
\caption{Illustration of gluing at an internal edge, with $C_I=Z_1+Z_2'+Z_3$.}
\label{fig:internal}
}

Given individual phase spaces $\CP_{\pd\Delta_i}$ for each tetrahedron $\Delta_i$, one can construct a product phase space $\CP_{\{\pd\Delta_i\}}=\prod_{i=1}^N \CP_{\pd\Delta_i}$ with a product symplectic structure. The edge coordinates in this space obey a Poisson algebra
\be \{Z_i,Z_j'\} = \{Z_i',Z_j''\}= \{Z_i'',Z_j\} = \hbar\,\delta_{ij}\,,\ee
with all other brackets vanishing. It is a wonderful fact that in the product phase space all the ``gluing constraints'' $C_I$ defined in \eqref{defCj} commute with each other \cite{NZ}. It turns out that the remaining linear combinations of edge coordinates in $\CP_{\{\pd\Delta_i\}}$ that commute with (but are independent of) the gluing constraints $C_I$ precisely parametrize the remaining boundary phase space of the glued 3-manifold $M$. This includes both geodesic and cusp-like boundary components, and we will momentarily give explicit examples of both.

Formally, the fact that all gluing constraints $C_I$ commute with each other and with the coordinates of flat connections on $\pd M$ means that $\CP_{\pd M}$ can be obtained as the symplectic quotient of the product phase space $\CP_{\{\pd\Delta_i\}}$ by the flows of the $C_I$ viewed as \emph{moment maps} \cite{Dimofte-QRS},
\be \CP_{\pd M}\,=\, \bigg(\prod_{i=1}^N \CP_{\pd\Delta_i}\bigg)\Big/\!\!\!\Big/\big(C_I=2\pi i\,\big) \,,\ee
where $I$ runs over all internal edges.
The individual Lagrangian submanifolds $\CL_{\Delta_i}$ can also be carried through this symplectic reduction. One forms a product Lagrangian $\CL_{\{\Delta_i\}}=\prod_{i=1}^N\CL_{\Delta_i}\;\subset\;\CP_{\{\pd\Delta_i\}}$ cut out by $N$ polynomial equations $z_i+z_i'{}^{-1}-1=0$; then algebraically eliminates all variables in these equations that do not commute with the $C_I$ (projecting $\CL_{\{\Delta_i\}}$ along the flows of the $C_I$); and sets $C_I=2\pi i$ in the equations that remain (intersecting the projection with the moment map conditions). This leads to a Lagrangian submanifold $\CL_M\,\subset\,\CP_{\pd M}$. Subject to several technical caveats discussed in \cite{Dimofte-QRS}, it is precisely the desired set of flat connections on $M$.

\subsection{Geodesic boundaries}
\label{sec:geobdy}

We proceed to provide some details of the phase spaces $\CP_{\pd M}$ associated to the various types of boundary for $M$, and to give explicit examples of their construction. A more complete, mathematical analysis of boundaries and phase spaces will appear in \cite{DGGV-hybrid}.

It is perhaps simplest to begin with geodesic boundaries. As discussed above, these arise when tetrahedra $\Delta_i$ are impartially glued; then some tetrahedron faces are left over to form one or more disjoint boundaries $\CC \subset \pd M$, each a triangulated, punctured Riemann surface.
The punctures are places where vertices of the tetrahedra $\Delta_i$ are located, and can ultimately be regularized into cusps that end on $\CC$ --- we will say a bit more about this later. The induced 2d triangulation of $\CC$ is ``ideal'' in the sense that all edges begin and end on punctures.

The phase space $\CP_\CC$, a factor in $\CP_{\pd M}$, is the moduli space of flat $SL(2,\C)$ connections on $\CC$, with specified (fixed) holonomy eigenvalues at every puncture. These eigenvalues become central elements in the algebra of functions on $\CP_\CC$. Geometrically, we can also describe $\CP_\CC$ as the complexified Teichm\"uller space of $\CC$, a complexification of the moduli space of 2-dimensional hyperbolic metrics. From this perspective, the puncture eigenvalues reflect the geometric size of holes in $\CC$.

We can construct coordinates on $\CP_\CC$ by associating to every edge $E$ in the triangulation of $\CC$ the total complexified dihedral angle around it.%
\footnote{In fact, these are coordinates on algebraically open patches of $\CP_\CC$ that have the topology of complex tori, \cf\ \cite{FG-Teich, GMNII}.} %
In other words,
\be \text{edge $E$}\;\;\leadsto\;\;
 \text{coordinate $X_E$} = \sum_{i=1}^N\Big[ n(E,i)Z_i+n'(E,i)Z_i'+n''(E,i)Z_i''\Big]\,, \label{defXE}
\ee
where $n(E,i) \in \{0,1,2\}$ is the number of times an edge of tetrahedron $\Delta_i$ with parameter $Z_i$ coincides with the glued edge $E$, and similarly for $n'(E,i)$ and $n''(E,i)$.
This definition is analogous to \eqref{defCj}, except that now $E$ is an external edge of $M$.
\FIGURE[l]{
\hspace{.13in}\includegraphics[width=1.5in]{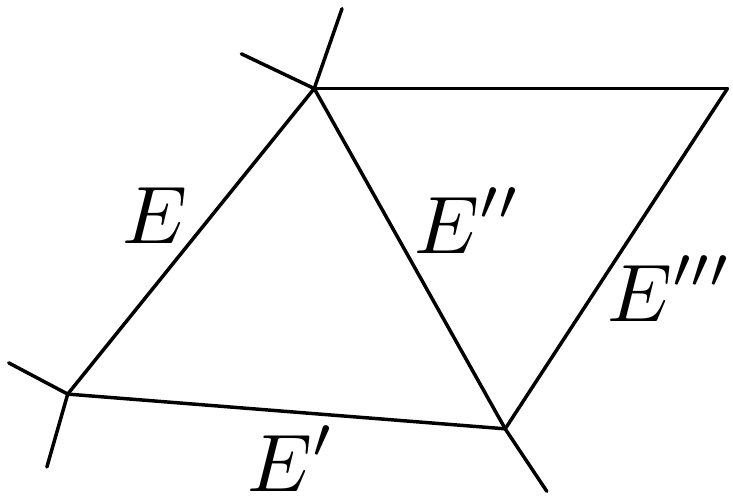}\hspace{.13in}
\caption{Poisson bracket for external edges. Here $\{X_E,X_{E'}\}$ =\,$\{X_{E'},X_{E''}\}$\,=\,$\{X_{E''},X_E\}$ =\,$\{X_{E''},X_{E'''}\}$\,=\,$\hbar$, etc.}
\label{fig:Ebracket}
}
\noindent It turns out that the coordinates $X_E$ are already well known mathematically as complexified ``shear coordinates'' on $\CP_\CC$ \cite{Thurston-shear, Fock-Teich}, defined rigorously in \cite{FG-Teich} in the complex case.%
\footnote{We thank R. Kashaev for first making us aware of this connection.} By following the arguments of \cite{NZ}, one can show that the Poisson structure induced on these edge coordinates is
\be \{X_E,X_{E'}\} = f(E,E')\,, \label{XEbracket} \ee
where $f(E,E')\in\{0,\pm 1,\pm 2\}$ is the number of faces shared by edges $E$ and $E'$, counted with orientation (\cf\ Figure \ref{fig:Ebracket}). Expression \eqref{XEbracket} is precisely the Weil-Petersson Poisson structure on $\CP_\CC$, \cf\ \cite{Fock-Teich}. Moreover, for each puncture $p\in \CC$, one finds that the sum of edge coordinates encircling the puncture is
\be \sum_{\text{$E$ ending on $p$}} \big(i\pi-X_E\big) = 2(\Lambda_p-i\pi)\,, \label{phol}\ee
where $\exp(\pm\Lambda_p)$ are the holonomy eigenvalues at $p$. The elements $\Lambda_p$ form a basis for the center of the Poisson algebra \eqref{XEbracket}.

Shear coordinates on $\CP_\CC$ recently featured prominently in the analysis of BPS states and wall crossing for 4-dimensional $\CN=2$ theories associated to punctures Riemann surfaces $\CC$ \cite{GMNII}. In particular, we note that \cite{GMNII} considered edge coordinates $\CX_E = \exp(i\pi-X_E)$, which could be identified as the exponentiated central charges for a generating set of BPS states in 4d gauge theory. The electric-magnetic pairing of BPS charges was given by \eqref{XEbracket}.

\FIGURE[r]{
\includegraphics[width=1.5in]{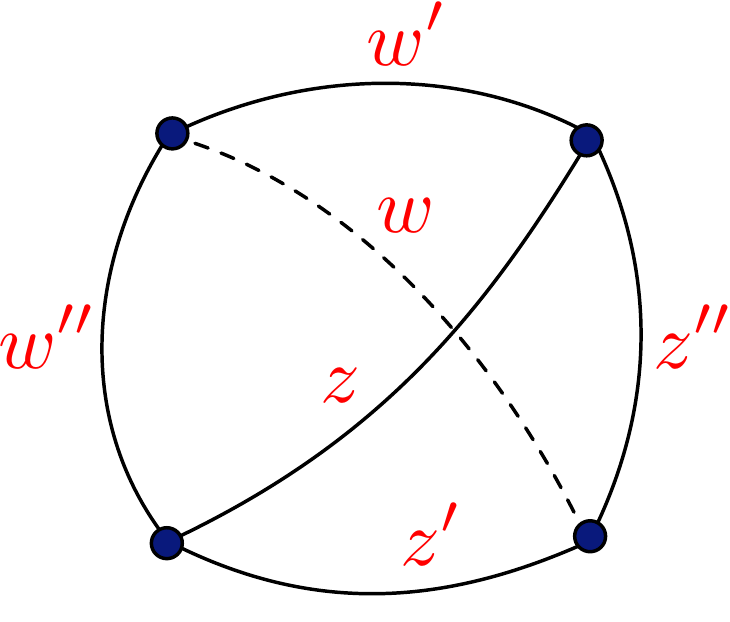}
\caption{$\pd\Delta$ as a four-punctured sphere.}
\label{fig:tet6}
}

The simplest example of shear/edge coordinates already appeared above, when we described the phase space $\CP_{\pd\Delta}$ \eqref{PDelta} of an ideal tetrahedron. If we view the boundary $\pd\Delta$ as a triangulated four-punctured sphere, we should start with six (logarithmic) edge coordinates $(Z,Z',Z'',W,W',W'')$ that obey a Poisson algebra
\begin{align} &\{Z,Z'\}=\{Z',Z''\}=\{Z'',Z\}=\{Z,W'\} \notag \\ &\quad =\{Z',W''\}=\{Z'',W\}=\{W,Z'\}=\{W',Z''\} \\ &\quad =\{W'',Z\}= \{W,W'\}=\{W',W''\}=\{W'',W\} = \hbar\,, \notag
\end{align}
according to the faces shared by these edges, with all other brackets vanishing. Then we impose conditions \eqref{phol} that the holonomy eigenvalue around each vertex $p$ is $\Lambda_p=2\pi i$ --- in other words, we require that the holonomy be unipotent:
\be W+W'+W''=Z+Z'+W''=Z+W'+Z''=W+Z'+Z''=i\pi \ee
This forces opposite edges to have equal parameters, $W=Z,\,W'=Z',\,W''=Z''$, and cuts down the phase space to $\CP_{\pd \Delta} = \{(Z,Z',Z'')\,|\,Z+Z'+Z''=i\pi\}$, with Poisson structure \eqref{DPoisson}

\begin{figure}[htb]
\centering
\includegraphics[width=4.8in]{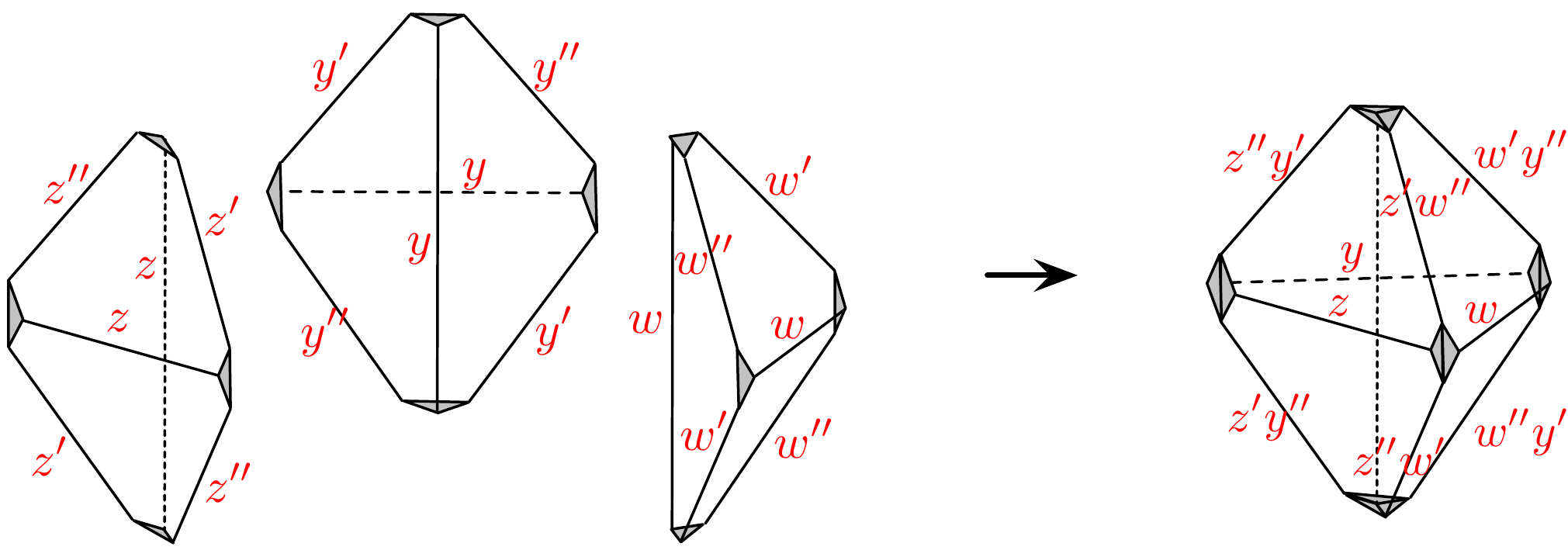}
\caption{Forming a bipyramid from three tetrahedra.}
\label{fig:bip3}
\end{figure}

As an example involving a nontrivial gluing, we can consider the ``bipyramid'' $M$ of Figure \ref{fig:bip3}. Its boundary is a 5-punctures sphere $\CC$. Here, we form the bipyramid from three ideal tetrahedra, with respective shape parameters $Z{}^{(}{}'{}^{)(}{}''{}^{)},\,W{}^{(}{}'{}^{)(}{}''{}^{)},\,Y{}^{(}{}'{}^{)(}{}''{}^{)}$.\,%
\footnote{Any solid 3-ball whose boundary is an $n$--punctured sphere ($n\geq 4$), with unipotent holonomy at each puncture, can be obtained via a similar gluing.} %
This leads to a 6-dimensional product phase space $\CP_{\{\pd\Delta_i\}} \approx \{(Z,Z',Z'',W,W',W'',Y,Y',Y'')\}$ with relations $Z+Z'+Z''=W+W'+W''=Y+Y'+Y''=i\pi$. Inside $\CP_{\{\pd\Delta_i\}}$ there is a single gluing constraint
\be C \equiv Z+W+Y \,\to\, 2\pi i\, \label{bipC} \ee
corresponding to the internal, vertical edge of the bipyramid;
it should be used as a symplectic moment map to reduce $\CP_{\{\pd\Delta_i\}}$ to the 4-dimensional phase space $\CP_{\pd M}=\CP_\CC$.

Explicitly, coordinates on $\CP_\CC$ are given by the dihedral angles of the nine external edges of the bipyramid:
\begin{subequations} \label{bipedges}
\be Z\,,\quad W,\,\quad Y  \ee
for the three equatorial edges, and
\be Z'+W'',\,\quad Z''+W',\,\quad W'+Y''\,,\quad W''+Y'\,,\quad Y'+Z''\,,\quad Y''+Z' \ee
\end{subequations}
for the six longitudinal edges. It is easy to check that, as functions on the product phase space $\CP_{\{\pd\Delta_i\}}$, the nine external shear/edge coordinates \eqref{bipedges} all commute with $C$. Moreover, modulo the gluing constraint \eqref{bipC}, one can check using formula \eqref{phol} that the total logarithmic holonomy eigenvalue around each of the five punctures $p$ of $\CC$ is $\Lambda_p=2\pi i$. The resulting five relations among the nine external edge coordinates cut the dimension of $\CP_\CC$ down to four.

As in the case of a single tetrahedron, the punctures on the boundary of the bipyramid carry unipotent holonomy (with logarithmic eigenvalue $2\pi i$). This is related to the fact that, upon truncating tetrahedron vertices as in Figures \ref{fig:trunc}, \ref{fig:bip3}, the small vertex triangles come together to form Euclidean 2d discs. These discs effectively cap off the punctures and force unipotent holonomy. In general one can build 3-manifolds that have annular cusps, rather than discs, ending at the punctures of a geodesic boundary. The annular cusps will then allow any holonomy eigenvalues to be realized. Constructions of this type are extremely interesting in the context of 3d and 4d gauge theory, but will mainly be deferred to future work \cite{DGGV-hybrid}.

For simple manifolds such as the tetrahedron and the bipyramid, whose boundaries carry unipotent punctures and whose interiors have the topology of 3-balls, the Lagrangian submanifolds $\CL_M\,\subset\,\CP_{\pd M}$ are also very simple. They are always cut out by the condition that the puncture holonomies are actually trivial (not just unipotent) --- so that a flat connection on the boundary can be extended to the bulk of $M$.

\FIGURE[l]{
\includegraphics[width=1.7in]{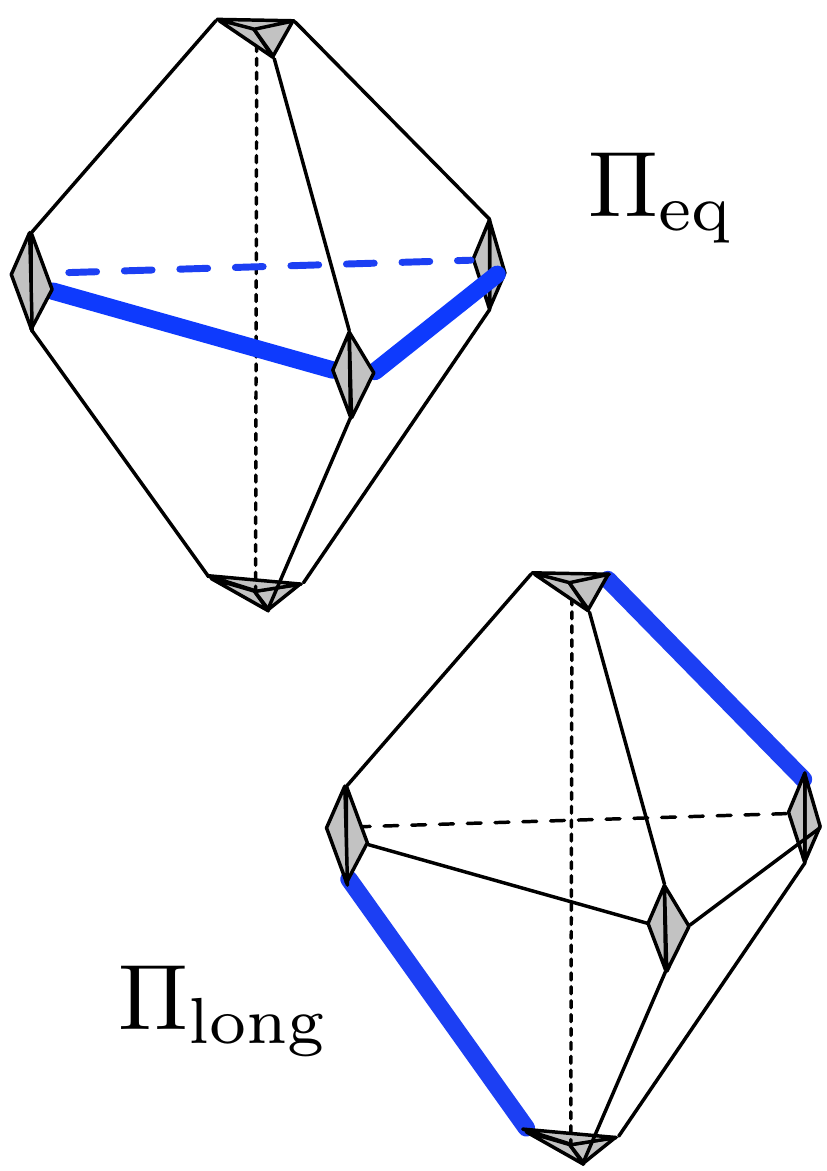}
\caption{Two polarizations for the bipyramid.}
\label{fig:bippol}
}

To conclude the discussion of geodesic boundaries, we observe that several natural polarizations $\Pi$ for a phase spaces $\CP_\CC$ can be specified by choosing maximal subsets of commuting edges on $\CC$. In other words, we choose a maximal set of independent edges that share no common faces. The corresponding coordinates $X_E$ then correspond to ``positions'' in $\CP_\CC$. Their conjugate momenta can be constructed (not quite uniquely) as combinations of the remaining edges.

For example, in the case of the bipyramid, two such polarizations are shown in Figure \ref{fig:bippol}, one using ``positions'' on equatorial edges and the other on longitudinal edges. (Note that the three equatorial edges all commute, but obey a constraint $Z+W+Y=C=2\pi i$, so only two of them, say $Z$ and $W$, are independent.) The respective positions $X_{1,2}$ and momenta $P_{1,2}$ in these polarizations are summarized as
\be \label{bipcoord3}
 \begin{array}{lc@{\;\;}c@{\qquad}c}
  && \text{positions} & \text{momenta} \\[.1cm]
  \Pi_{\rm eq} &:\;& X_1 = Z\,,\quad X_2=W & P_1 = Z''+Y'\,,\quad P_2 = W''+Y'\,\\[.1cm]
  \Pi_{\rm long} &:\;& X_1'=W'+Y''\,,\quad X_2' =Z'+Y'' & P_1' =Z''+Y'\,,\quad P_2' = W''+Y'\,
\end{array}
\ee
In equatorial coordinates $(X_i,P_i)$, the Lagrangian $\CL_M$ (\ie\ the set of connections with trivial holonomy) can be shown to have the simple description
\be \CL_M\;:\qquad p_1+\frac{p_2}{x_1}-1=0\,,\qquad p_2+\frac{p_1}{x_2}-1=0\,, \label{Lbipeq} \ee
while in longitudinal coordinates we have
\be \CL_M\;:\qquad p_1'+x_1'^{-1}-1=0\,,\qquad p_2'+x_2'^{-1}-1=0\,,\ee
with $x_i=\exp(X_i)$, $p_i=\exp(P_i)$, etc.

Different polarizations for a geodesic boundary phase space $\CP_\CC$ are related to one another by affine $Sp(\dim_\C \CP_\CC,\Z)$ transformations. From the above discussion, it should be easy to see that the complex dimension of $\CP_\CC$ must be
\be \dim_\C\,\CP_\CC = (\text{\# external edges on $\CC$})-(\text{\# punctures on $\CC$})\,,
\ee
which by an Euler character argument agrees with the standard formula $\dim_\C\,\CP_\CC = 6g-6+2n$, where $g$ is the genus and $n$ is the number of punctures of $\CC$. The affinely extended group $Sp(6g-6+n,\Z)$ is a subgroup of the full affine group $Sp(2N,\Z)$ of transformations on the product phase space $\CP_{\{\pd\Delta_i\}}=\prod_{i=1}^N\CP_{\pd\Delta_i}$. Therefore, we can always choose a polarization of $\CP_{\{\pd\Delta_i\}}$ that is compatible with the final desired polarization of the quotient space $\CP_\CC$.

\subsection{Torus cusps}
\label{sec:cuspbdy}

The cusp boundaries of a 3-manifold $M$ arise from the resolution of line defects, and have the topology of annuli or tori, depending on whether the defects are open or closed. For simplicity, we will only consider the closed, toroidal case in the present paper, though we note that annular cusps share many of the the same properties, and can be analyzed in a similar way.

\FIGURE[r]{
\includegraphics[width=2.1in]{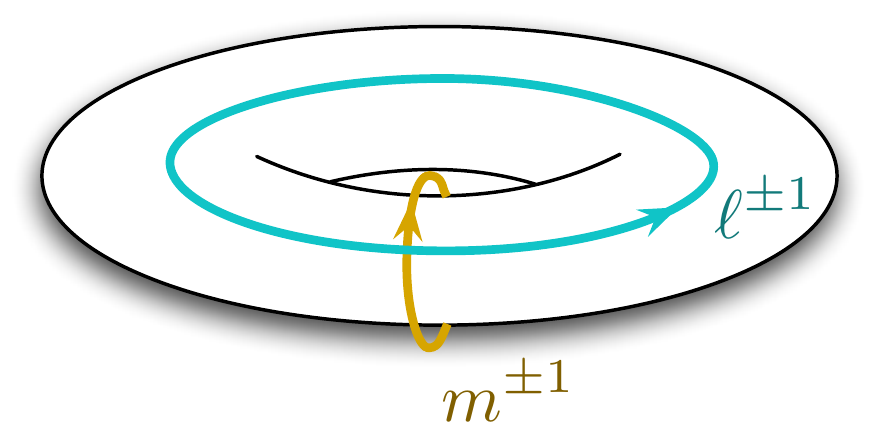}
\caption{Holonomy eigenvalues on a torus boundary.}
\label{fig:torusLM}
}

Suppose, then, that $\pd M$ contains a toroidal cusp boundary $T^2$. For example, $M$ could be the complement of a knot in $S^3$. To describe the associated phase space $\CP_{T^2}$, we can choose a basis of ``A and B cycles'' on the torus --- typically called meridian and longitude cycles in the case of knot complements.%
\footnote{For a knot complement in $S^3$, $M=S^3\bs K$, there is actually a canonical choice of cycles. The meridian is an infinitesimally small loop linking the knot $K$ once, while the longitude intersects the meridian once and is nullhomologous in $M$ (in particular, it has zero linking number with the knot). Presently, however, we will allow ourselves the freedom of choosing any basis of cycles whatsoever.} %
Since the fundamental group $\pi_1(T^2)$ is abelian, the $SL(2,\C)$ holonomies along these cycles are simultaneously diagonalizable, and $\CP_{T^2}$ is simply parametrized by their eigenvalues, \cf\ \cite{cooper-1994}:
\be \CP_{T^2} = \big\{(m,\ell)\,\in\,\C^*\times \C^*\big\}\big/\Z_2\,,\ee
where the Weyl group $\Z_2$ acts by inversion $(m,\ell)\mapsto (m^{-1},\ell^{-1})$. As above, it is also convenient to take logarithms%
\footnote{As discussed in \cite{Dimofte-QRS}, the shift by $i\pi$ in $v+i\pi=\log\ell$ characterizes the correct lift from $PSL(2,\C)$ structures (most naturally computed by triangulation data) to $SL(2,\C)$.} %
$\boxed{u\equiv \log m}$ and $\boxed{v+i\pi\equiv\log \ell}$ and to lift the phase space to
\be \CP_{T^2} = \big\{(u,v)\,\in\,\C\times\C\big\}/\Z_2\,. \ee
Then the symplectic structure of $\CP_{T^2}$ becomes $\omega_{T^2}=\frac{2}\hbar dv\wedge du$ \cite{gukov-2003}, or
\be \{v,u\}=\hbar/2\,. \ee

\begin{figure}[htb]
\centering
\includegraphics[width=6.3in]{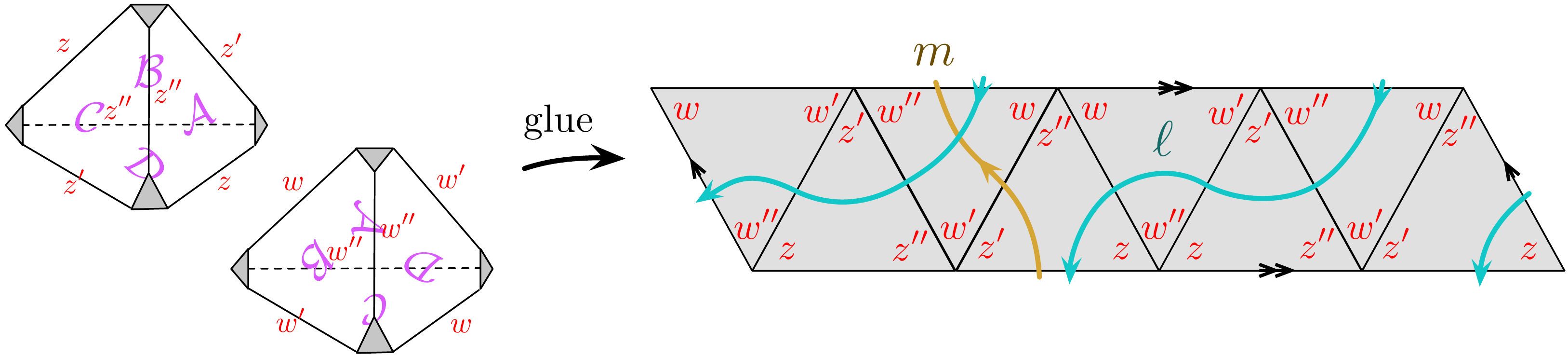}
\caption{Gluing two tetrahedra, as indicated by calligraphic letters on the faces, to form the figure-eight knot complement. On the right is a map of the resulting torus cusp boundary, triangulated by Euclidean vertex triangles.}
\label{fig:fig8cusp}
\end{figure}

The logarithmic eigenvalues $u$ and $v$ can both be computed as linear combinations of edge parameters $Z_i,\,Z_i',\,Z_i''$ of tetrahedra in a triangulation of $M$. To see this, recall that a cusp boundary $T^2$ is composed of small truncated-vertex triangles of the tetrahedra $\Delta_i$. Thus, it comes with a (Euclidean) 2d triangulation, as illustrated in Figure \ref{fig:fig8cusp}. The dihedral angles of tetrahedra $\Delta_i$ become actual (complexified) angles in the 2d triangles. Logarithmic holonomies can be computed by adding and subtracting the angles subtended by a given path, then dividing by two \cite{NZ, Neumann-combinatorics}. For example, in Figure \ref{fig:fig8cusp} we have drawn the meridian and longitude of the figure-eight knot complement on a boundary $T^2$. The corresponding holonomies are
\begin{subequations} \label{Uv41}
\begin{align} U\equiv 2u &= Z'-W \\
 2v &= 2(Z-Z')
\end{align}
\end{subequations}
As functions on the product phase space $\CP_{\{\pd\Delta_i\}}\simeq \{(Z,Z'Z'',W,W'Z'')\,|\,Z+Z'+Z''=W+W'+W''=i\pi\}$, these satisfy the expected commutation relation $\{v,U\}=\hbar$.

Continuing with the example of the figure-eight knot complement, we find that the triangulation of Figure \ref{fig:fig8cusp} has two internal edges, with corresponding gluing constraints
\be \label{C41}
C_1 = 2Z+Z''+2W+W'' \to 2\pi i\,,\qquad C_2 = 2Z'+Z''+2W'+W'' \to 2\pi i\,. \ee
(It is easy to read these off from the map of the cusp, since every internal edge begins and ends at a ``vertex'' on the cusp triangulation. One just adds the angles surrounding the vertex.) Note that $C_1$ and $C_2$ both commute with $U$ and $v$. Moreover, \emph{prior to} enforcing the condition $C_1=C_2=2\pi i$, there is an automatic relation $C_1+C_2=4\pi i$, so that one of the two gluing constraints is redundant. In general, for every closed torus cusp in a 3-manifold $M$, there will be one such redundant gluing constraint. In the end, for our figure-eight example, we see that $\CP_{\pd M}=\CP_{T^2} = \CP_{\{\pd\Delta_i\}}\big/\!\!\big/(C_1=2\pi i) = \CP_{\{\pd\Delta_i\}}\big/\!\!\big/(C_2=2\pi i)$.

The Lagrangian submanifold for the figure-eight knot complement is obtained by the symplectic reduction procedure described at the end of Section \ref{sec:geomglue} above. One starts with the product Lagrangian
\be \CL_{\{\Delta_i\}} = \{z+z'^{-1}-1=0,\,w+w'^{-1}-1=0\}\;\subset\; \CP_{\{\pd\Delta_i\}}\,, \ee
where $z=e^Z,z=e^{Z'},w=e^{W},$ and $w'=e^{W'}$; rewrites the equations in terms of $m^2=e^U$, $\ell=-e^v$, and one of the gluing monomials $c_j=e^{C_I}$; eliminates all remaining variables that do not commute with $c_j$; and sets $c_j=1$. The end result is
\be \label{A41}
 \CL_{M} = \{\ell-(m^4-m^2-2-m^{-2}+m^{-4})+\ell^{-1}=0\}\;\subset\;\CP_M\,,\ee
and this equation is the well known ``A-polynomial'' of the figure-eight knot \cite{cooper-1994, gukov-2003}.

\subsection{Changing the triangulation}
\label{sec:geom23}

\begin{figure}[htb]
\centering
\includegraphics[width=5in]{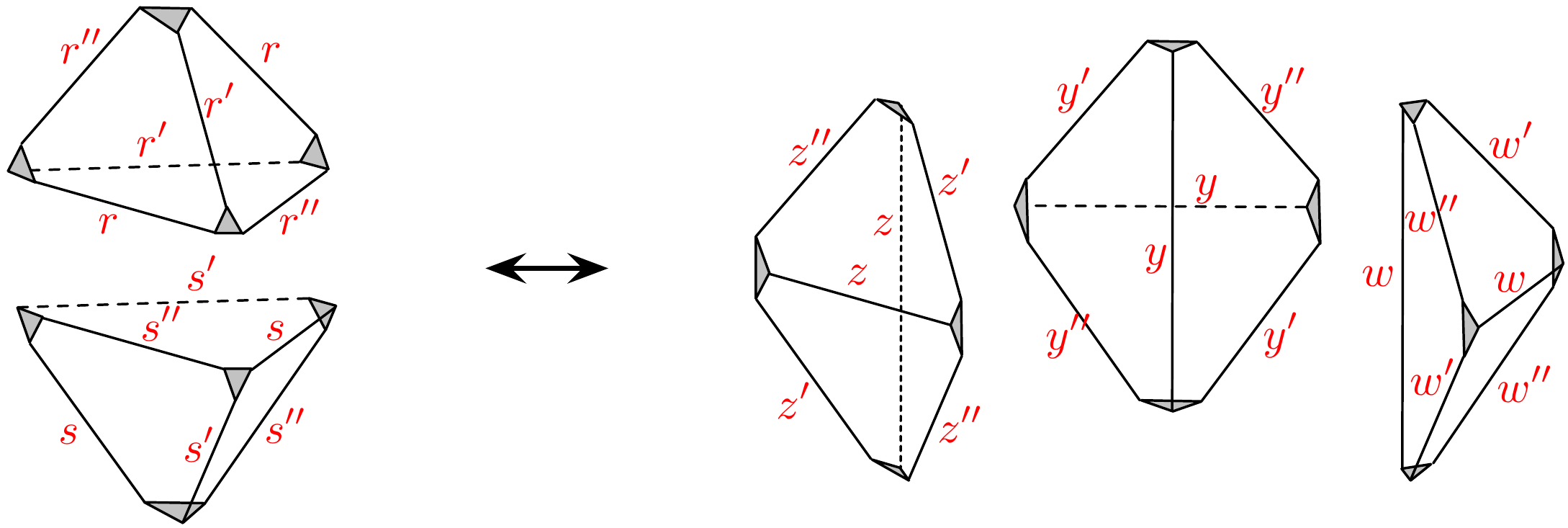}
\caption{The 2--3 Pachner move}
\label{fig:Pach}
\end{figure}

We have explained, in principle, how to construct 3-manifolds $M$, phase spaces $\CP_{\pd M}$, and Lagrangians $\CL_M$ by gluing together ideal tetrahedra $\Delta_i$. It would be useful to verify that such constructions do not depend on a precise choice of triangulation $\{\Delta_i\}$. Geometrically, once we fix the triangulation of geodesic boundaries, any two triangulations of $M$ are related by a sequence of ``2--3 Pachner moves,'' \cf\ \cite{Matveev-book}. These replace two tetrahedra glued along a common face with three tetrahedra glued along three faces and a common edge, and vice versa, as shown in Figure \ref{fig:Pach}.

Invariance of phase spaces and Lagrangians under the 2--3 move was verified%
\footnote{Again we note that the invariance of Lagrangians comes with a few subtle caveats, as discussed in \cite{Champ-hypA, Dunfield-Mahler} and reviewed in Sections 4--5 of \cite{Dimofte-QRS}. For sufficiently generic triangulations, these caveats can be safely ignored.} %
in detail in (\eg) \cite{Dimofte-QRS}, guaranteeing the internal consistency of our present gluing constructions. For example, for phase spaces, the essence of the argument is that the product phase spaces corresponding to the bipyramid on the left of Figure \ref{fig:Pach} is the symplectic reduction of the product phase space on the right,
\be \hspace{-.5in}\CP_{\pd(\text{bipyramid})}\;\; =\qquad \CP_{\Delta_R}\times \CP_{\Delta_S} \;=\; \big(\CP_{\Delta_Z}\times \CP_{\Delta_W}\times \CP_{\Delta_Y}\big)\big/\!\!\big/(C=2\pi i)\,, \label{bip23} \ee
where $C$ is the gluing constraint coming from the internal edge. In fact, we already described the right-hand side of \eqref{bip23} in Section \ref{sec:geobdy}. The left-hand side is even easier to analyze. In the same two polarizations $\Pi_{\rm eq}$ and $\Pi_{\rm long}$ of Figure \ref{fig:bippol}, we now find coordinates for $\CP_{\Delta_R}\times\CP_{\Delta_S}$:
\be \label{bipcoord2}
\begin{array}{lc@{\;\;}c@{\qquad}c}
  && \text{positions} & \text{momenta} \\[.1cm]
  \Pi_{\rm eq} &:\;& X_1 = R+S''\,,\quad X_2=R''+S & P_1 = R''\,,\quad P_2 = S''\\[.1cm]
  \Pi_{\rm long} &:\;& X_1'=R\,,\quad X_2' =S & P_1' =R''\,,\quad P_2' = S''
\end{array}
\ee
The two equivalent descriptions \eqref{bipcoord3}--\eqref{bipcoord2} of $\CP_{\pd(\text{bipyramid})}$ are related by combining or splitting the coordinates associated to the external dihedral angles, for example splitting $Z\leftrightarrow R''+S''$.

The 2--3 Pachner moves always preserve the triangulations of geodesic boundaries of $M$. In contrast, they do not preserve the ``small'' triangulations of cusp boundaries; but the triangulations of cusp boundaries are never important for defining phase spaces here, or 3d gauge theories later on.

\FIGURE[l]{
\includegraphics[width=2.9in]{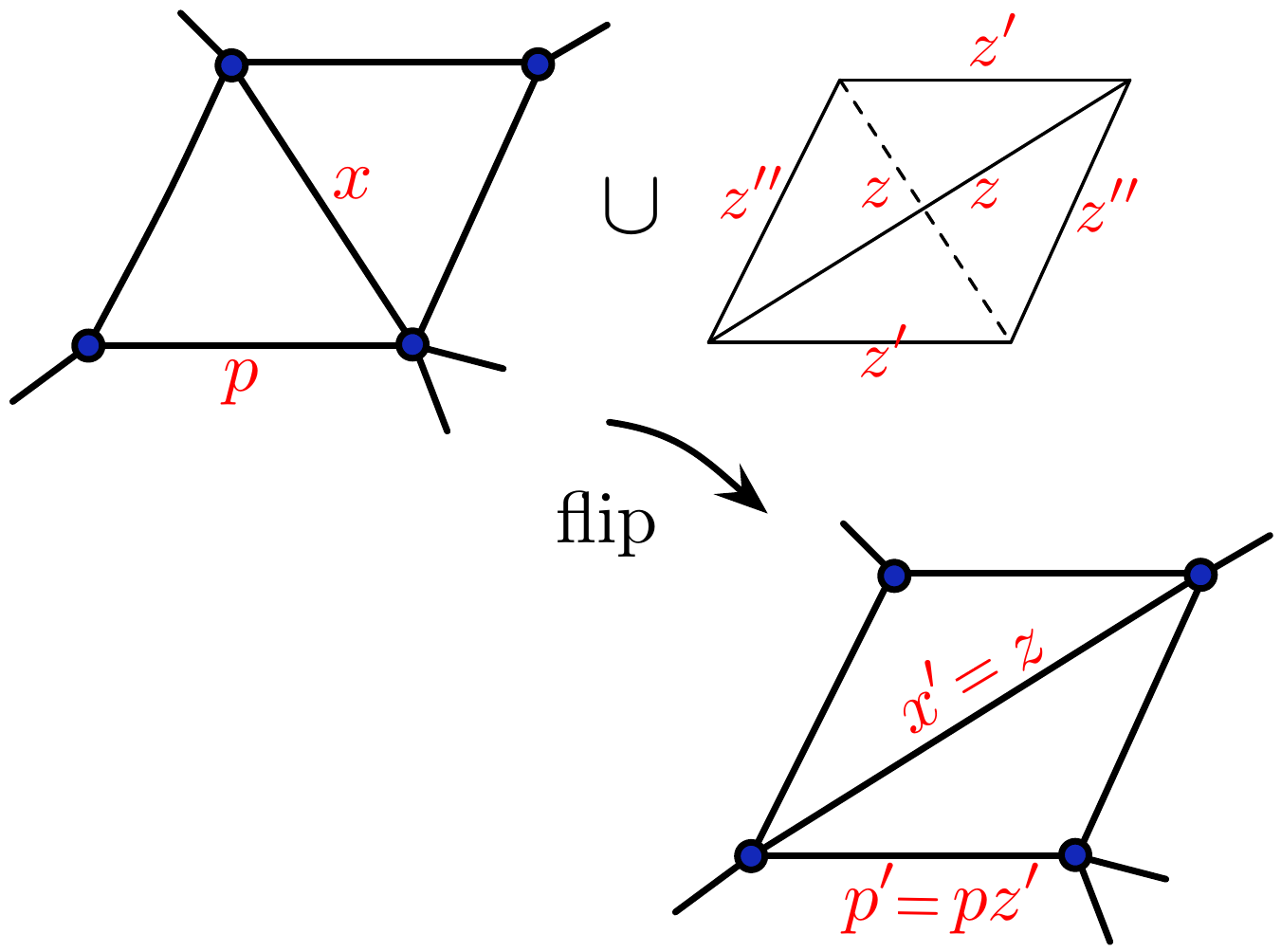}
\caption{Flipping an external edge by attaching a tetrahedron.}
\label{fig:flip}
}

If we want to change the triangulation of a geodesic boundary $\CC\subset\pd M$, we must consider another type of fundamental move: a flip. The flip acts by gluing an additional tetrahedron $\Delta_F$ onto a quadrilateral in $\CC$, as in Figure \ref{fig:flip}, and effectively ``flipping'' the diagonal of this quadrilateral. In the process of attaching $\Delta_F$, a new internal edge $I_F$ is created, which imposes a new gluing constraint $C_{I_F}$. The flipped phase space $\CP_{\CC'}$ is therefore related to $\CP_\CC$ by a symplectic reduction
\be \CP_{\CC'} = \big(\CP_\CC\times \CP_{\pd\Delta_F}\big)\big/\!\!\big/(C_{I_F}=2\pi i)\,. \ee
Obviously $\CP_{\CC'}$ and $\CP_{\CC}$ must be isomorphic, but the two have different ``natural'' polarizations.

To illustrate this explicitly, if we start with a polarized phase space $\CP_\CC$ in which one of the canonical position--momentum pairs $(X,P)$ corresponds to dihedral angles as in Figure \ref{fig:flip}, then gluing on the tetrahedron $\Delta_F$ yields an internal edge constraint
\be C_{I_F}  = X + Z \;\to\, 2\pi i\,. \label{Cflip} \ee
Now, let us attach a new position coordinate $X'$ to the newly flipped diagonal, and its conjugate momentum $P'$ to the same edge as $P$. After the symplectic reduction (in particular, imposing \eqref{Cflip}), we find that
\be X' = 2\pi i - X\,,\qquad P' = -(P+Z')\,. \ee
If we also keep track of Lagrangians, we would find that the flipped $\CL_{M'}$ is related to $\CL_M$ by substituting $x\to x'^{-1},\; p\to p'^{-1}(1-x')$ in the defining equations for $\CL_M$.%
\footnote{We suggest the verification of this statement as an exercise for the reader.}

The flip transformation, described here from a 3-dimensional viewpoint, is very familiar in 2-dimensional Teichm\"uller (and quantum Teichm\"uller) theory, \cf\ \cite{Penner-decorated, Fock-Teich, FockChekhov, Kash-Teich, FG-Teich}. This should not be surprising, given the above observation that shear coordinates of Teichm\"uller theory should be identified with 3d dihedral angles.

%%%%%%%%%%%%%%%%%%%%%%%%%%%%%%%%%%%%%%%%%%%%%%%%%%%%%%%%%%%%%%%%%%%%%%%%%%
%%%%%%%%%%%%%%%%%%%%%%%%%%%%%%%%%%%%%%%%%%%%%%%%%%%%%%%%%%%%%%%%%%%%%%%%%%
\section{Operations on 3d abelian theories}
\label{sec:ops}

Our next goal is introduce the basic ingredients and building blocks necessary to understand the field theory side of the correspondence $(M,\Pi)\leftrightarrow T_{M,\Pi}$. We will see a clear parallel with the construction of 3-manifolds in Section \ref{sec:geom}, which will lead
us to the definition of the theory $T_{M,\Pi}$ in Section \ref{sec:glue}.

\subsection{An $Sp(2N,\Z)$ action on 3d CFTs with $U(1)^N$ flavor symmetry}
\label{sec:dualities}

\subsubsection{Generalities}

There is a beautiful $SL(2,\Z)$ action on the space of 3-dimensional conformal field theories with $U(1)$ flavor symmetry. This action was first described in \cite{Witten-SL2} as a way to understand
the meaning of different choices of boundary conditions for an abelian gauge field in $AdS_4$ in the context of $AdS_4/CFT_3$.

To be precise, $SL(2,\Z)$ acts on the space of 3d theories equipped with a specific way to couple a $U(1)$ flavor symmetry to a background $U(1)$ gauge field.
The $SL(2,\Z)$ action can be defined by specifying the action of its generators $S$ and $T$, which obey the relations
\be S^4 = (ST)^3 = id.\ee
The generator $T$ does not change the underlying 3d CFT. It only modifies the prescription
of how to couple the theory to the background gauge field $A$, by adding to the conserved current for the background flavor symmetry the Hodge dual field strength $*F=*dA$. In terms of a Lagrangian, this is simply accomplished by adding a background Chern-Simons interaction at level $k=1$,
\be T\,:\quad \CL\to \CL + \frac{1}{4\pi} A\wedge dA\,. \ee

In contrast, the $S$ generator changes the structure of the 3d theory by making the background gauge field $A$ dynamical.%
\footnote{One can add a Yang-Mills kinetic term at intermediate stages in the calculation. But for $S$ to have the correct properties, one must flow to the IR at the end, and then $g_{{\rm YM}}\to \infty$ and this term is removed.} %
The new 3d theory is then prescribed a coupling to a new background $U(1)$ gauge field $A_{\rm new}$: the new flavor current is the Hodge dual field strength $*F$ of the old, now dynamical, gauge field. Equivalently, one prescribes a Lagrangian coupling
\be S\,:\quad \CL\to \CL+ \frac{1}{2 \pi} A_{\mathrm{new}} \wedge dA \qquad\text{($A$ dynamical)}\,.\ee
It is the monopole operators for $A$ that are charged under the new $U(1)$ flavor symmetry; thus this $U(1)$ is sometimes called ``topological.'' From the definitions of $S$ and $T$, one can prove that the relations $S^2=C$ and $(ST)^3=id.$ hold, where the transformation $C$ (charge conjugation) just inverts the sign of the background gauge field. We will generally denote the action of an $SL(2,\Z)$ group element $g$ on a theory $\CT$ as $g \circ \CT$.

There is a useful alternative interpretation of this $SL(2,\Z)$ action: it is the action of electric-magnetic duality
on the space of conformally invariant boundary conditions for a free abelian four-dimensional gauge theory.
Indeed, given a three-dimensional CFT with a prescribed coupling to a background gauge field,
we can build a boundary condition by coupling the CFT to the value of the 4d gauge field at the boundary.
This gives a generalization of Neumann boundary conditions: the normal component of the 4d field strength
at the boundary becomes proportional to the conserved current of the 3d CFT. If we denote the 3d theory as $\CT$,
we can denote the resulting boundary condition as $\CB[\CT]$.

\begin{figure}[htb]
\centering
\includegraphics[width=4.5in]{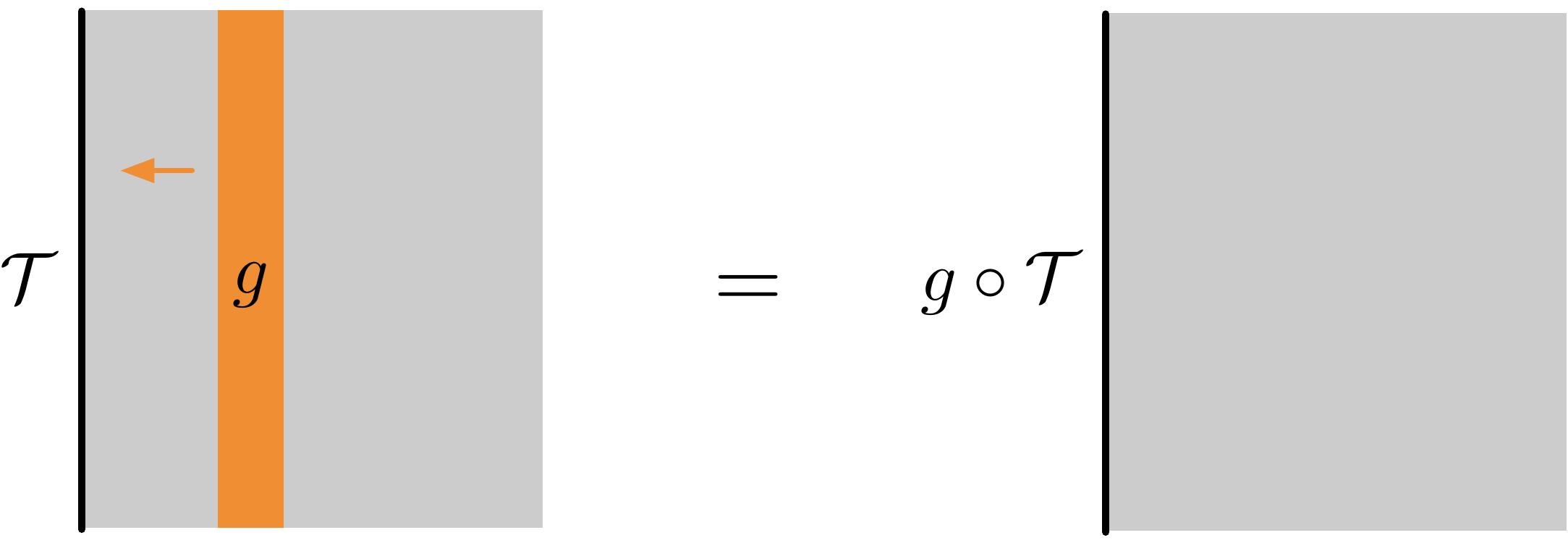}
\caption{The action of duality domain walls on boundary conditions. A duality transformation $g \in Sp(2N,\Z)$ maps
a generalized Neumann boundary condition defined by coupling to a 3d theory $\CT$ into a boundary condition
associated with a boundary CFT $g \circ \CT$.}
\label{fig:boundarywall}
\end{figure}

Next, we can do an electric-magnetic duality transformation $g \in SL(2,\Z)$ in the four-dimensional bulk,
and ask how the boundary condition $\CB[\CT]$ looks in the new duality frame.
This ``new'' boundary condition $g \circ \CB[\CT]$ turns out to coincide with $\CB[g \circ \CT]$.
This fact  can be shown readily with the help of ``duality domain walls'' \cite{GW-Sduality}:
the action of bulk dualities on boundary conditions can be interpreted as the collision (or OPE)
of these domain walls with the boundary, as illustrated in Figure \ref{fig:boundarywall}.
For an abelian gauge theory the duality walls are very easy to construct from the definition of electric-magnetic duality.
The collision with the boundary then reproduces the $SL(2,\Z)$ action defined above.

The $SL(2,\Z)$ action on boundary conditions is a little bit more general than
the $SL(2,\Z)$ action on 3d theories with a coupling to a background gauge field.
For example, there exists an extra $SL(2,\Z)$ orbit of boundary conditions which includes the pure Dirichlet boundary condition on the 4d gauge field.
This boundary condition is invariant under $T$, and it is sent to the pure Neumann boundary condition by $S$.

Now, it is rather obvious how to generalize this $SL(2,\Z)$ action to an $Sp(2N,\Z)$ action on boundary conditions for a general four-dimensional $U(1)^N$ abelian gauge theory,
or to an action on 3d CFTs with $U(1)^N$ flavor symmetry: it is the action of the electric-magnetic duality
group of the $U(1)^N$ four-dimensional gauge theory. Notice that in this case, there are several orbits
of boundary conditions which involve at some point Dirichlet boundary conditions
for some of the bulk gauge fields. These orbits will look a bit singular from the point of view of an action on 3d CFTs.
Concretely, they signal situations where the flavor symmetry is spontaneously broken to a subgroup in the IR \cite{GW-Sduality}.

To make this a little more explicit, suppose we are given a Lagrangian description of
a 3d CFT with $U(1)^N$ global symmetry, whose current is coupled to $N$ background
gauge fields $\vec A=(A_1,...,A_N)$.
The generators of $Sp(2N,\Z)$ fall into three basic categories: ``$T$-type,'' ``$S$-type,''
and ``$GL$-type'' (\cf\ \cite{HuaReiner}). Representing them as matrices in $N\times N$ blocks,
we find Lagrangian transformations
\be \text{``$T$-type''}\quad g=\begin{pmatrix} I & 0 \\ B & I \end{pmatrix},\; \text{$B$ symmetric}\quad:\qquad \CL[\vec{A}]\to\CL[\vec{A}_{\rm new}]+\frac{1}{4\pi} \vec A_{\rm new} \cdot B\,d\vec{A}_{\rm new}\,; \label{SpT} \ee
\be \text{``$S$-type''}\quad g= \begin{pmatrix}I-J & -J \\ J & I-J \end{pmatrix}\quad:\qquad \CL[\vec A]\to\CL[\vec A]+\frac{1}{2\pi} \vec{A}_{\rm new}\cdot J\,d\vec A\, \label{SpS} \ee
(where $J={\rm diag}(j_1,...,j_N)$ with $j_i\in \{0,1\}$, and we have gauged every $A_i$ for which $j_i=1$, replacing its $U(1)$ with a new topological flavor symmetry); and
\be \text{``$GL$-type''}\quad g = \begin{pmatrix} U & 0 \\ 0 & U^{-1\,t}\end{pmatrix},\;U\in GL(N,\Z)\quad:\qquad \CL[\vec A]\to \CL[U^{-1}\vec A_{\rm new}]\,. \label{SpGL}
\ee
The latter $GL$-type action simply redefines the flavor currents by an invertible, integral transformation.

\subsubsection{Adding supersymmetry}
\label{sec:susySp}

The $Sp(2N,\Z)$ action can be supersymmetrized to give an $Sp(2N,\Z)$ action on supersymmetric 3d theories
equipped with a supersymmetric coupling to a background abelian gauge supermultiplet.
This can be done for any amount of supersymmetry, but it is important to make a specific choice, as different choices give different group actions.

In the reference \cite{GW-Sduality} this was applied to theories with $\CN=4$ supersymmetry.
As a useful example of the $S$ action in the context of $\CN=4$ theories, we can consider a single hypermultiplet
of unit flavor charge canonically coupled to an $\CN=4$ background gauge field.
If we make the $\CN=4$ background gauge field dynamical --- performing an $S$ operation --- we have a familiar 3d theory: $\CN=4$ SQED with one flavor. This is the canonical setup for 3d mirror symmetry \cite{IS, dBHOY}, which provides an alternative description of the theory in terms of a free twisted hypermultiplet
that arises as a monopole operator in the original description.
In particular, it carries unit flavor charge under the new $\CN=4$ background gauge field. So the transformation
$S$ acts rather trivially on this simple 3d theory: it sends it back to itself \cite{KS-mirror}. On the other hand, $T$ acts non-trivially.

Any $\CN=4$ statement can be reinterpreted as an $\CN=2$ statement, but a little care is needed: the $\CN=4$ 3d gauge multiplet consists of an $\CN=2$ 3d gauge multiplet plus a chiral multiplet.  The $\CN=4$ $Sp(2N,\Z)$ action is the combination of an $\CN=2$ $Sp(2N,\Z)$ action plus additional operations involving 3d chiral multiplets and superpotential couplings. This anticipates a central theme of this paper: the interplay between the ``gauge'' $Sp(2N,\Z)$ action and a ``matter'' action which involves adding new chiral multiplets with appropriate superpotential couplings.
Indeed, the 3d $\CN=2$ theories $T_M$ associated to 3-manifolds $M$ with boundary will be coupled both to background gauge fields and background chiral multiplets.

As a first step towards understanding this statement, let us describe the ``gauge'' $Sp(2N,\Z)$ action for $\CN=2$ theories. Suppose we have a theory with $U(1)^N$ flavor symmetry, coupled to $N$ background vector multiplets $V_i$. Each $V_i$, containing a real scalar field $\sigma_i$ and two Majorana fermions $\lambda^\alpha_i$ in addition to the gauge field $A_i$, can also be dualized to a linear multiplet \cite{AHISS, NishinoGates}
\be V_i\quad\leftrightarrow\quad \Sigma_i = \ol{D}^\alpha D_\alpha V\,, \ee
where the lowest component of $\Sigma_i$ is $\sigma_i$. Now, in order to supersymmetrize the $Sp(2N,\Z)$ action \eqref{SpT}--\eqref{SpGL}, one simply has to substitute $AdA'\to V\Sigma'$ for all relevant Chern-Simons or FI terms:
\begin{align} \text{``$T$-type''}\quad g=\begin{pmatrix} I & 0 \\ B & I \end{pmatrix} \;\;&:\qquad \CL[\vec{V}]\to\CL[\vec{V}_{\rm new}]+\frac{1}{4\pi}\int d^4\theta\,\vec\Sigma_{\rm new} \cdot B\,\vec V_{\rm new}\,; \label{Sp2T} \\
\text{``$S$-type''}\quad g= \begin{pmatrix}I-J & -J \\ J & I-J \end{pmatrix}\;\;&:\qquad \CL[\vec V]\to\CL[\vec V]+\frac{1}{2\pi}\int d^4\theta\, \vec{\Sigma}_{\rm new}\cdot J\,\vec V\, \label{Sp2S} \\
\text{``$GL$-type''}\quad g = \begin{pmatrix} U & 0 \\ 0 & U^{-1\,t}\end{pmatrix}\;\;&:\qquad \CL[\vec V]\to \CL[U^{-1}\vec V_{\rm new}]\,. \label{Sp2GL}
\end{align}
Note that a $GL(N,\Z)$ linear transformation $U^{-1}$ can be applied both to a collection of vector multiplets $\vec V$ and linear multiplets $\vec \Sigma$, wherever they occur in the Lagrangian.

\subsection{A $\Z_2$ action on 3d $\CN=2$ theories with a chiral operator}
\label{sec:F}

The basic ``matter'' action on 3d $\CN=2$ theories begins with a theory that has a coupling to a background 3d chiral multiplet $\phi$. In practice, what we mean is a choice of chiral operator $\CO$
that can be inserted in a superpotential
\begin{equation}
\CW = \phi\,\CO \,.
\end{equation}
Here and elsewhere, we will not keep track of the normalization of superpotential terms. In particular, we will view a rescaling of $\CO$ as a trivial operation.

We can define an operation $F$ that makes $\phi$ dynamical (thus, setting $\CO$ effectively to zero). The new theory can be coupled to a new background chiral field $\phi'$ by coupling to the new chiral operator $\CO'=\phi$, namely by the superpotential
\begin{equation}
\CW = \phi' \phi \,.
\end{equation}
It is easy to see that $F^2=1$. We can simply look at the combined superpotential
\begin{equation}
\CW = \phi'' \phi'+\phi' \phi +\phi\,\CO
\end{equation}
and integrate out $\phi'$.

Much like the ``gauge'' $Sp(2N,\Z)$ action of the previous subsection, the operation $F$ can be given an interesting four-dimensional interpretation. One can consider possible boundary conditions on a four-dimensional hypermultiplet. If we split the four real scalar fields in the hypermultiplet into two complex scalar fields, which we can denote as $X$ and $Y$, then the two basic boundary conditions are either Neumann for $X$ and Dirichlet for $Y$, or vice versa. A way to understand this is that $Y$ sits in a multiplet of the unbroken supersymmetry which contains the normal derivative of $X$.

If we introduce extra degrees of freedom at the boundary, say a 3d theory with a preferred chiral operator $\CO$,
then we can consider a deformed Dirichlet boundary condition $Y=\CO$. This will be accompanied by a
corresponding deformation of the Neumann boundary conditions for $X$, involving the corresponding piece of the supermultiplet $\CO$.
This defines a certain class of boundary conditions which we denote $\CB_Y$,
so that the boundary condition associated to a 3d theory $\CT$ is denoted by $\CB_Y[\CT]$.
An alternative way to describe this boundary condition is to start with the undeformed boundary condition
and add the boundary superpotential coupling
\begin{equation}
\CW = X \CO \,.
\end{equation}

Naively, one can construct a completely different class of boundary conditions $\CB_X$ as Dirichlet boundary conditions
with $X=\CO'$, where $\CO'$ is a chiral operator in a 3d boundary theory $\CT'$.
It turns out that these two classes actually coincide, as every member $\CB_Y[\CT]$ of one class has a mirror $\CB_X[\CT']$ in the other class.
One simply takes $\CT'$ to be the image of $\CT$ under $F$, with $\CO' = \phi$; then we claim that
\be \CB_Y[\CT] = \CB_X[F\circ\CT]\,. \ee
To see this, we simply follow the definition of $F\circ\CT$ to obtain an overall superpotential coupling
\begin{equation}
\CW = Y \phi + \phi\,\CO \,.
\end{equation}
Integrating out $\phi$ sets $Y = - \CO$. Furthermore, the boundary condition $X=\phi$ means that we can simply ``absorb'' $\phi$ into $X$,
thus relaxing the Dirichlet boundary conditions. It takes a bit more work to make sure that $X$ acquires Neumann boundary conditions, but it
follows from the fact that the normal derivative $\partial_n X$ plays the role of auxiliary field in the $Y$ supermultiplet.
In summary, if we begin with Dirichlet boundary conditions for $X$ and perform an $F$ transformation --- adding a single boundary chiral multiplet $\phi$ and a superpotential $\CW = Y\phi$ --- we will flow in the IR to Dirichlet boundary conditions for $Y$.

\subsection{Useful $\CN=2$ mirror symmetries}
\label{sec:MS}

From the above, it should be clear that the $\CN=4$ $S$ operation consists of a combination of $\CN=2$ $S$ and $F$ operations.
Indeed, the $\CN=4$ conserved current supermultiplet contains a complex moment map operator $\mu$, which is a chiral operator for
an $\CN=2$ subalgebra. The $\CN=4$ gauge multiplet contains an $\CN=2$ chiral multiplet $\phi$, which is coupled
to the complex moment map operator $\mu$ by the superpotential coupling
\begin{equation}
\CW = \phi \mu \,.
\end{equation}
Thus, for example, the basic $\CN=4$ mirror symmetry statement of Section \ref{sec:susySp} can be recast as a statement about
a 3d $\CN=2$ theory $\CT_2$ of two chiral multiplets $u$ and $\tilde u$ with opposite flavor charges,
and an operator $\CO = \mu = u \tilde u$.
This theory is invariant under the combined $\CN=2$ $S$ and $F$ operations, {\it i.e.} it satisfies $S F \circ \CT_2 = \CT_2$.

A basic consequence, pointed out in \cite{AHISS, dBHO},
is that the two theories that are obtained from a $S$ operation or from a $F$ operation on the theory
of two chiral multiplets are actually the same in the IR, {\it i.e.} they are $\CN=2$ mirror duals.\footnote{Notice that the coupling of the background gauge field to $\CT_2$ is unaffected by charge conjugation $C$.}
The theory $S \circ \CT_2$ is just $\CN=2$ SQED with $N_f=1$. The theory $F \circ \CT_2$, or rather $CF \circ \CT_2$,  is the so-called XYZ model,
a theory of three chiral fields $\phi$, $u$, $\tilde u$ with a superpotential
\begin{equation}
\CW = \phi u \tilde u \,.
\label{Wuuu}
\end{equation}
These two theories are mirror to each other:
\be
\begin{array}{ll}
\text{\bf SQED}~(S\circ \CT_2): & \quad \text{gauged $U(1)$ with two chirals of charge} \; +1 \; \text{and} \; -1 \;  \\[.1cm]
\text{\bf XYZ}~(CF\circ \CT_2): & \quad \text{three chirals with superpotential} \; \CW = \phi u \tilde u
\end{array} \label{XYZmir}
\ee

There is actually a bit more structure to this problem. In $\CN=2$ language, each of the two chiral multiplets in $\CT_2$
can be rotated independently, and the theory really has $U(1)^2$ flavor symmetry.
Similarly, the XYZ model has a $U(1)^2$ flavor symmetry that rotates the phase of $\phi$, $u$, $\tilde u$ and leaves
the superpotential $\CW$ invariant. The two $U(1)$'s map via the mirror symmetry \eqref{XYZmir} to an axial $U(1)$ and a topological $U(1)$ in $N_f=1$ SQED. In terms of SQED, the topological $U(1)$ symmetry is carried by two chiral monopole operators $v_\pm$ with charges $\pm 1$. It is slightly nontrivial (\cf\ \cite{AHISS}) to see that the monopole operators also transform with charge $-1$ under the axial $U(1)$. We summarize these various flavor symmetries in Table \ref{tab:charges}.

\begin{table}[htb]
\be
\begin{array}{l@{\;}|@{\;}cc@{\;}|@{\;}ccc}
\multicolumn{6}{c}{\text{$N_f=1$ SQED}} \\[.1cm]
& u & \tilde u & \mu & v_+ & v_- \\\hline
U(1)_{\rm gauge} & 1 & -1 & 0 & 0 & 0 \\
U(1)_{\rm axial} & 1 & 1 & 2 & -1 & -1 \\
U(1)_{\rm top} & 0 & 0 & 0 & 1 & -1
\end{array} \qquad\qquad
\begin{array}{l@{\;}|@{\;}ccc}
\multicolumn{4}{c}{\text{XYZ}} \\[.4cm]
& \phi & u & \tilde{u} \\\hline
U(1)_{\rm axial} & 2 & -1 & -1\\
U(1)_{\rm top} & 0 & 1 & -1
\end{array}
\notag \ee
\caption{Correspondence of symmetries in $N_f=1$ SQED and the XYZ model. The designations ``axial'' and ``topological'' in the XYZ model are only introduced for comparison to SQED.}
\label{tab:charges}
\end{table}

 Eventually, we will investigate the properties of these theories under $Sp(4,\Z)$ transformations.
For now, we would like to derive yet another useful $\CN=2$ mirror pair by a mass deformation of this theory.

We aim to understand the properties of a simple theory, consisting of a single chiral multiplet $\phi$ of charge $1$.
In order to define a coupling to a background gauge field,
we need to face a subtlety: a single chiral multiplet canonically coupled to a background gauge field has an anomaly,
which can be cancelled by adding a half-integral Chern-Simons coupling
for the background gauge field. This fact is closely related to another important fact.
If we integrate out a massive chiral multiplet coupled with charge $q$ to a background gauge field,
we generate an effective (supersymmetric) Chern-Simons interaction at level $k = \tfrac{1}{2}\,q^2\,{\rm sign} (m)$.

Thus we define an $\CN=2$ theory $\CT_1$ as a chiral field of charge $1$, coupled to a background field with an extra Chern-Simons interaction at level $-\tfrac{1}{2}$. We want to show that $ST \circ \CT_1$ coincides with $\CT_1$. This is certainly compatible with $(ST)^3 = 1$.
In particular, we want to show that a $U(1)$ CS theory at level $k=\tfrac{1}{2}$ coupled to a single chiral multiplet of charge $+1$
is mirror to a free chiral multiplet of charge $+1$.
To demonstrate this statement, we will go back to the XYZ model.

\begin{figure}[htb]
\centering
\includegraphics[width=3.4in]{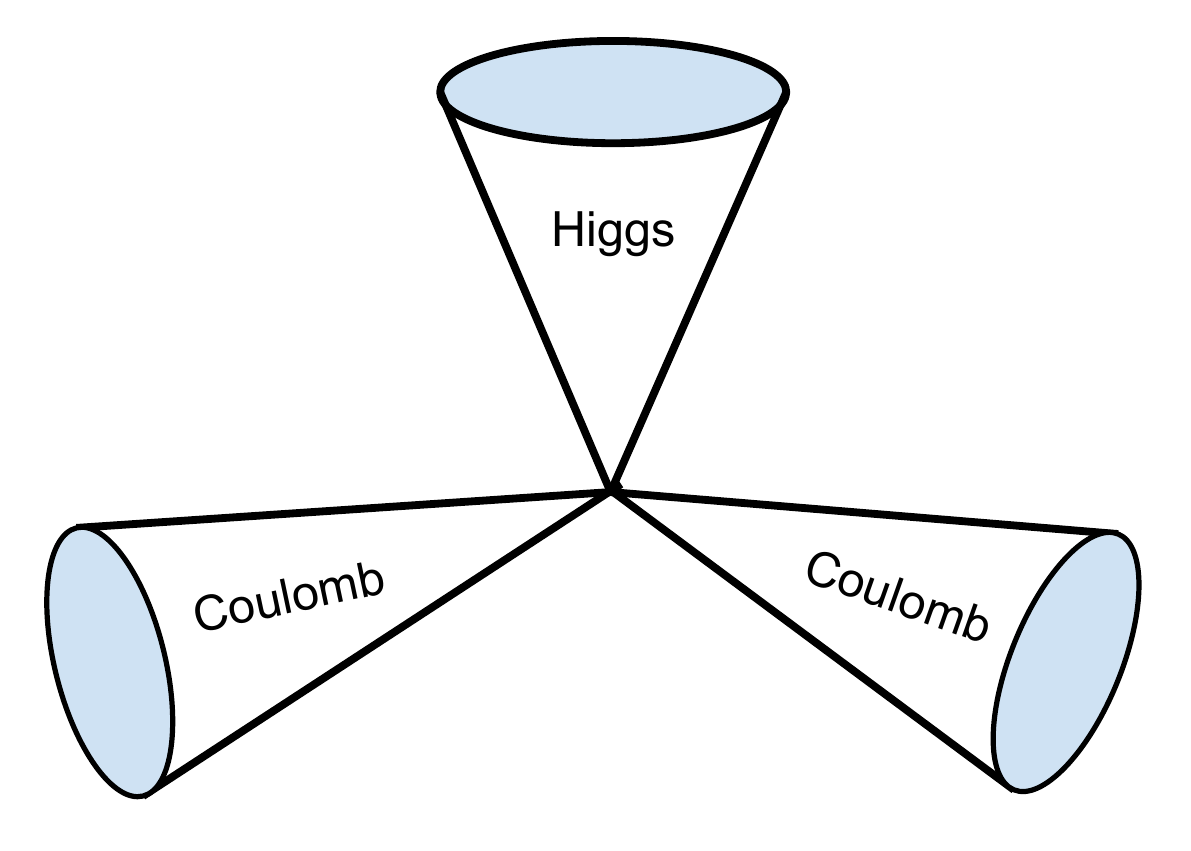}
\caption{The quantum moduli space of $\CN=2$ SQED is identical to the moduli space of vacua in the $XYZ$ model.
It has three branches, permuted by the quantum $\Z_3$ symmetry.}
\label{fig:moduli3}
\end{figure}

The XYZ model, or $N_f=1$ SQED, has a triality property. In the XYZ model this is just permutation of the three chiral fields.
The theory has three 1-complex-dimensional branches of SUSY vacua.
Indeed, the superpotential \eqref{Wuuu} leads to the scalar potential
\be
V \; = \; \Big| \frac{\partial \CW}{\partial \phi} \Big|^2 + \Big| \frac{\partial \CW}{\partial u} \Big|^2
+ \Big| \frac{\partial \CW}{\partial \tilde u} \Big|^2
\; = \; |\phi u|^2 + |\phi \tilde u|^2 + |u \tilde u|^2
\ee
which is minimized on field configurations where one of the chiral fields has a vev, while the other two vanish.
The resulting three branches parametrized by the vevs of $\phi$, $u$, or $\tilde u$ meet at the origin.
In the $N_f=1$ SQED, on the other hand, the classical moduli space is controlled by the term
$\sigma^2 (|u|^2 + |\tilde u|^2)$ in the scalar potential
\be
V \; = \; \frac{e^2}{2} \Big( |u|^2 - |\tilde u|^2 - \zeta \Big)^2 + \sigma^2 |u|^2 + \sigma^2 |\tilde u|^2
\ee
that forces either $\sigma = 0$ or $u = \tilde u = 0$. The quantum corrected moduli space of
the $N_f=1$ SQED is the same as that of the XYZ model, as shown in Figure \ref{fig:moduli3}.
One of the branches in the moduli space of SUSY vacua is the Higgs branch, parameterized by the vev of the meson $\mu =u \tilde u$.
The other two are halves of the Coulomb branch, where $\sigma$ is real and positive, or real and negative.
The two halves of the Coulomb branch are parameterized by the vevs of the corresponding vortex-creation (monopole) operators.

Now, if one turns on opposite twisted mass%
\footnote{By ``twisted mass'' in three dimensions, we mean a mass term arising as a background value for the real scalar field in a vector multiplet. Sometimes this is also called a ``real mass.''} %
for two of the chiral fields in the XYZ model, it kills two branches and makes the third smooth:
\be
\CM_{{\rm SUSY}} ~=~~
{\raisebox{-0.7cm}{\includegraphics[width=3.0cm]{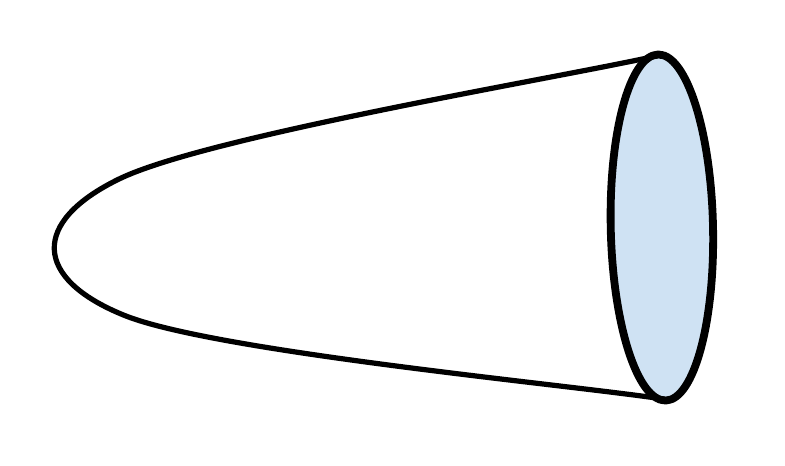}}\,}
\ee
In the $N_f=1$ SQED description, this statement can take three equivalent forms.
The first form is simple: an FI parameter is the same as a twisted mass for the monopole operators.
It kills the Coulomb branch and smoothens the Higgs branch.
The other two forms of the statement --- which are really what we need --- are more subtle.
We must to turn on twisted masses for the other two choices of flavor symmetry in the XYZ description.
They rotate only one of the monopole operators, and the meson.

Let us start with the XYZ model. By consulting Table \ref{tab:charges}, we see that if we turn on a large and (say) positive twisted mass $m_{\rm axial}$ for the axial $U(1)$, and an equal mass $m_{\rm top}$ for the ``topological'' $U(1)$,
\be m_{\rm axial} \approx m_{\rm top} \gg 0\,, \ee
we can integrate out the chirals $\phi$ and $\tilde u$. We are left with a single free chiral $u$, which still transforms under the difference of $U(1)_{\rm top}$ and $U(1)_{\rm axial}$. Explicitly, defining a new background gauge multiplet $V_{\rm top}'\equiv V_{\rm top}-V_{\rm axial}$, which can still have a small twisted mass parameter $m_{\rm top}'=m_{\rm top}-m_{\rm axial}$, we find that $u$ is coupled to $V_{\rm top}'$ with charge $1$. Integrating out the multiplet $\tilde{u}$ generates a background Chern-Simons term $\frac{k}{4\pi}\int d^4\theta\,\Sigma_{\rm top}'\,V_{\rm top}'$ at level $k=-1/2$. Thus we obtain our theory $\CT_1$. (Alternatively, we could have chosen $m_{\rm axial}=-m_{\rm top}\gg 0$, which would allow us to integrate out $u$ and keep $\tilde u$. This leads to an equivalent description of $\CT_1$.)

In terms of $N_f=1$ SQED, the topological mass $m_{\rm top}\approx-m_{\rm axial}$ becomes an FI parameter
\be \frac{2}{4\pi}\int d^4\theta\, \Sigma_{\rm top}\,V_{\rm gauge}\,=\, \frac{1}{2\pi}\int d^4\theta\,m_{\rm top}\,V_{\rm gauge}\,. \label{SQEDFI} \ee
It is this large FI term which ultimately allows us to keep the monopole $v_+$ light. This may look a bit mysterious, but it is easily motivated by looking at what happens to the fundamental matter fields of $N_f=1$ SQED in the presence of a large axial mass. Both $u$ and $\tilde u$ become very heavy, unless we tune $\sigma$ to $\pm m_{\rm axial}$, so that either $\tilde u$ is light and $u$ is heavy, or vice versa. Let us choose $\sigma=-m_{\rm axial}$; or, more appropriately, let us redefine the dynamical gauge multiplet as $V_{\rm gauge}\to V_{\rm gauge}-V_{\rm axial}=V_{\rm gauge}-\theta\bar\theta\, m_{\rm axial}$. Then we can integrate out $\tilde u$, and in the process generate a Chern-Simons term of level $1/2$ for the combination $-V_{\rm gauge}+2V_{\rm axial}$, under which $\tilde u$ is charged. Hidden in the cross-term of the supersymmetric Chern-Simons interaction is an FI term $-\frac{1}{2\pi}\int d^4\theta\,m_{\rm axial}\,V_{\rm gauge}$, which cancels \eqref{SQEDFI}, leaving behind a small difference $\frac{1}{2\pi}\int d^4\theta\,m_{\rm top}'\,V_{\rm gauge}$\,! Thus we end up with a fundamental chiral $u$, coupled with charge $1$ to a $U(1)$ gauge multiplet $V_{\rm gauge}$, which has a level $\frac12$ Chern-Simons interaction $\frac{1}{8\pi}\int d^4\theta\,\Sigma_{\rm gauge}\,V_{\rm gauge}$. The theory has a single light monopole operator $v_+$, transforming with charge $1$ under the new topological $U(1)_{\rm top}'$. This is precisely the description of $ST\circ \CT_1$. We have therefore derived
\be \boxed{ST\circ \CT_1 \,\simeq\, \CT_1} \label{STmir} \ee
as a consequence of the basic $\CN=2$ mirror symmetry \eqref{XYZmir}.

In section \ref{sec:dualities}, we discussed the interpretation of the $Sp(2N,\Z)$ action on a 3d theory $\CT$
as the action of electric-magnetic duality in 4d abelian gauge theory on a corresponding boundary condition $\CB[\CT]$.
In this interpretation, our simple 3d theory $\CT_1$ can define a boundary condition for a 4d theory with gauge group $U(1)$ --- by identifying the 4d gauge symmetry with the 3d flavor symmetry. If we start with a 4d duality frame in which the chiral multiplet of $\CT_1$ carries 4d electric charge,
then by acting with $SL(2,\Z)$ duality we obtain all other variants of the theory $\CT_1^{(p,q)}$,
where a distinguished chiral operator transforms as a dyon of electric charge $p$ and magnetic charge $q$.
In particular, the $ST$ element of $SL(2,\Z)$ acts as
\be
\CT_1^{(1,0)} \; \xrightarrow{ST} \;
\CT_1^{(0,1)} \; \xrightarrow{ST} \;
\CT_1^{(-1,1)} \; \xrightarrow{ST} \;
\CT_1^{(1,0)}\,.
\label{STonT1}
\ee
The mirror symmetry \eqref{STmir} actually guarantees that, just like $\CT_1^{(1,0)}$, the theories $\CT_1^{(0,1)}$ and $\CT_1^{(-1,1)}$ are equivalent to theories of free chirals coupled to the appropriate (magnetic or dyonic) 4d $U(1)$ gauge field with Chern-Simons level $-\frac12$. The chain of equivalences \eqref{STonT1} should remind us of \eqref{tetpols}. \\

This concludes our quick tour of the basic operations and mirror symmetries in 3d $\CN=2$ gauge theories.
Of particular importance in the rest of the paper is the basic relation $ST \circ \CT_1 = \CT_1$
and the mirror symmetry between the XYZ model and $N_f=1$ SQED. These basic duality relations admit
many generalizations in various directions (to theories that include larger gauge groups and / or larger spectrum of matter fields),
which have an elegant interpretation in terms of triangulations of 3-manifolds.

One simple generalization, which we mention only briefly, is that the XYZ model and $N_f=1$ SQED appear as the first mirror pair in the infinite family of mirror abelian gauge theories:
\begin{align} 
%\begin{array}{ll}
\text{\bf Theory A}~: & \quad U(1)^r \; \text{with} \; k \; \text{neutral chirals and} \; N \; \text{charged hypermultiplets} \label{mirrorR} \\
\text{\bf Theory B}~: & \quad \widehat{U(1)}{}^{N-r} \; \text{with} \; N-k \; \text{neutral chirals and} \; N \; \text{charged hypermultiplets} \notag 
%\end{array}
\end{align}
where the charges of the hypermultiplets in the two theories, $R_i^a$ and $\widehat R_i^a$,
obey the ``orthogonality'' constraints
\be
\sum_{i=1}^N R_i^a \widehat R_i^b = 0 \qquad \forall \; a, b \,.
\label{rrmirror}
\ee
In addition, both mirror theories A and B have gauge invariant cubic superpotential of the form
\be
\CW \; = \; \sum_{\alpha=1}^k \sum_{i=1}^N \; y_{\alpha i} \, \phi_{\alpha} \tilde Q_i Q_i
\ee
with Yukawa couplings $y_{\alpha i}$ (resp. $\widehat{y}_{\beta i}$) which obey a relation similar to \eqref{rrmirror}:
\be
\sum_{i=1}^N y_{\alpha i} \widehat{y}_{\beta i} = 0 \qquad \forall \; \alpha, \beta \,.
\ee
All four matrices $R$, $\widehat{R}$, $y$, and $\widehat{y}$ are assumed to be of maximal rank.
It is easy to see that if we take $N=r=1$ and $k=0$, then Theory~A is $\CN=2$ SQED with $N_f=1$, whereas Theory~B is the XYZ model.
The next simplest case, $N=r=k=1$, gives another prominent pair of mirror 3d theories that we also mentioned earlier:
a free hypermultiplet and $\CN=4$ SQED.
More generally, in this class of examples Theory A contains a total of $2N+k$ chiral multiplets (with charges $-1$, $0$, and $+1$),
whereas Theory B contains a total of $3N-k$ chiral multiplets. For this reason, the mirror symmetry of such a mirror pair could be
referred to as a ``$(2N+k)-(3N-k)$ move.''

%%%%%%%%%%%%%%%%%%%%%%%%%%%%%%%%%%%%%%%%%%%%%%%%%%%%%%%%%%%%%%%%%%%%%%%%%%

\section{Construction of $T_M$}
\label{sec:glue}

In this section, we will now combine the ingredients of Sections \ref{sec:geom} and \ref{sec:ops} to provide the map from a pair $(M,\Pi)$, where $M$ is a 3-manifold and $\Pi$ a polarization of its boundary phase space $\CP_{\pd M}$, to a 3d SCFT $T_{M,\Pi}$, with specified couplings to background gauge fields and chiral multiplets. We will do so in two steps. First, we attach a 3d theory to any triangulation $\{\Delta_i\}_{i=1}^N$ of the three-manifold $M$,
and then we show that different triangulations of the same three-manifold give mirror descriptions of the 3d SCFT.

\subsection{Definition}
\label{sec:defTM}

In order to implement the first step, we begin by defining a theory $T_{\Delta, \Pi_Z}$ that we associate to
a single tetrahedron $\Delta$ in polarization $\Pi_Z$ (as in \eqref{tetpols}):
\be
\boxed{\phantom{\int}
T_{\Delta, \Pi_Z} = \CT_1\,.
\phantom{\int}}
\label{TDelta}
\ee
Recall from Section \ref{sec:MS} that $\CT_1$ is a theory of a single chiral multiplet coupled to a background $U(1)$ gauge field, with a level $-\frac12$ Chern-Simons term turned on. We will say from now on that the free chiral is associated to the edges of the tetrahedron $\Delta$ labelled by $Z$, and denote it as $\phi_Z$ or $\CO_Z$. It is also useful to think of the twisted mass of $\CO_Z$ as $\Re(Z)$, and its R-charge as $\Im(Z)/\pi$, where $Z$ is the classical edge/shape parameter of $\Delta$. We use this interpretation here as an intuitive aid to motivate our gluing construction; it will be made much more precise in Sections \ref{sec:moduli} and \ref{sec:S3b}.

We can extend the definition \eqref{TDelta} to any other polarization $\Pi$ obtained by an (affine) $SL(2,\Z)$ transformation $g$ on $\Pi_Z$:
\be
T_{\Delta, g \circ \Pi_Z} = g \circ \CT_1 \label{TDeltaPi}
\ee
For example, in a polarization $\Pi_Z^-$ as in \eqref{PiZm}, we would find $T_{\Delta,\Pi_Z^-}=T\circ \CT_1$ to be the theory of a free chiral coupled to a background $U(1)$ with Chern-Simons level $k=+\frac12$.
This definition is consistent with the $\Z_3$ symmetry of the tetrahedron: the triality symmetry
permutes three equivalent polarizations $\Pi_{Z}$, $\Pi_{Z'}$, $\Pi_{Z''}$
in \eqref{tetpols} which, on the $\CN=2$ gauge theory side,
correspond to the three duality frames \eqref{STonT1} of the theory $\CT_1$ permuted
by the $ST$ element of $SL(2,\Z)$.

The second step is the definition of $T_{\{ \Delta_i\}, \tilde\Pi}$, the theory associated to
the union of $N$ tetrahedra $\Delta_i$, in a generic polarization $\tilde\Pi$.
We can always write $\tilde\Pi = g \circ \{ \Pi_i \}$ for some $g\in Sp(2N,\Z)$, where  $\{ \Pi_i \}$ is a polarization defined as a product of independent polarizations
$\Pi_i$ of the individual tetrahedra. We choose each $\Pi_i$ to be either $\Pi_{Z_i}$, $\Pi_{Z'_i}$ or $\Pi_{Z''_i}$.
Then we define
\be
\boxed{\phantom{\int}
M = \bigcup_{i=1}^N \Delta_i^{(\Pi)} \qquad \leadsto \qquad T_{ \{\Delta_i\} , \tilde\Pi} \; = \; g \circ \bigotimes_{i=1}^N T_{\Delta_i,\Pi_i}
\phantom{\int}}
\label{TMoutofTD}
\ee
where we regard the product of $N$ copies of $\CT_1$ theories as a theory with a canonical coupling to a $U(1)^N$ background gauge field. We should think of each $U(1)$ as corresponding to an \emph{independent} position coordinate in the polarization $\tilde \Pi$.
This definition is independent of the choice of $\Pi_i\in \{\Pi_{Z_i},\Pi_{Z'_i},\Pi_{Z''_i}\}$ due to the the symmetry $ST \circ \CT_1 = \CT_1$.

In order to define the actual SCFT $T_{M,\Pi}$ associated to the 3-manifold $M$, we need to implement a field-theory version of the gluing constraints $C_I\to 2\pi i$ for each internal edge $I$ in the triangulation. The basic idea is to choose a polarization $\tilde\Pi = g\circ\{\Pi_i\}$ for the collection of tetrahedra such that
\begin{itemize}
\item[1)] it is compatible with the final desired polarization $\Pi$ of the boundary $\CP_{\pd M}$; and
\item[2)] all the internal edge coordinates $C_I$ are ``positions'' in $\tilde\Pi$.
\end{itemize}
If we are careful, we can then construct operators $\CO_I$ in the theory $T_{\{\Delta_i\},\tilde\Pi}$, one for each internal edge. These operators will be charged under a subset of $U(1)$ flavor symmetries, also associated to the edges $C_I$ --- or rather to independent linear combinations of them. We can then define $T_{M,\Pi}$ by adding a superpotential to $T_{\{\Delta_i\},\tilde\Pi}$ of the form
\be
\boxed{\phantom{\int}
\CW \; = \; \sum_{{I \in \, \text{internal} \atop \text{edges of~} M}} \CO_I\,.
\phantom{\int}}
\label{WforTM}
\ee
This superpotential breaks all the $U(1)$ symmetries under which the $\CO_I$ are charged. It also sets the R-charge of each $\CO_I$ equal to $2$. We will see later that this is precisely equivalent to setting $C_I=2\pi i$.

In addition to the internal edge operators $\CO_I$, the theories $\CT_{\{\Delta_i\},\tilde\Pi}$ and $\CT_{M,\Pi}$ also have a set of operators $\CO_E$ associated to the external edges%
that are ``positions'' in $\Pi \subset \tilde\Pi$. These operators are charged precisely under the $U(1)$ gauge symmetries that persist as symmetries of $T_{M,\Pi}$ --- one for each independent position in $\Pi$. Indeed, it is easy to see that the flavor group of $T_{M,\Pi}$ will contain exactly $\frac12\dim \CP_{\pd M}$ $U(1)$'s. In summary, we have built a correspondence:
\be
\begin{array}{c@{\quad}|@{\quad}c}
\text{geometry} & \text{gauge theory} \\\hline
\Delta,\,\Pi_Z & T_{\Delta,\Pi_Z}= \CT_1 \\[.1cm]
\{\Delta_i\},\, \{\Pi_i\} & T_{\{\Delta_i\},\{\Pi_i\}}=\otimes_i T_{\Delta_i,\Pi_i} \\[.1cm]
\text{positions, \eg\ $Z_i$} & \text{operators $\CO_{Z_i}$ with $U(1)$ symmetries} \\[.1cm]
\{\Pi_i\} \to \tilde\Pi=g\circ \{\Pi_i\} & T_{\{\Delta_i\},\{\Pi_i\}}\to \CT_{\{\Delta_i\},\tilde\Pi}=  g\circ T_{\{\Delta_i\},\{\Pi_i\}} \\[.1cm]
\text{internal edges}\;C_I & \text{operators}\; \CO_I \\[.1cm]
\text{external positions, \eg\ $X_E$} & \text{operators $\CO_E$} \\[.1cm]
C_I \to 2\pi i \;\text{(symp$^\text{c}$ reduction)} & \CW = \sum_I \CO_I \\[.1cm]
M = \cup_i\Delta_i,\;\;\Pi  & T_{M,\Pi} = \CT_{\{\Delta_i\},\tilde\Pi}\;\text{+ superpotential $\CW$}
\end{array}
\ee
\bigskip

The construction of operators $\CO_I$ (and also $\CO_E$) in the product theory $\CT_{\{\Delta_i\},\tilde\Pi}$ is a little tricky. In order to describe it, we must distinguish two classes of edges. We call an edge ``easy'' if its classical coordinate $C_I$ (or $X_E$) is a sum containing at most one of the edge parameters $Z_i,\,Z_i',\,Z_i''$ for any tetrahedron $\Delta_i$; otherwise the edge is ``hard.'' Thus, $C_I = Z_1+Z_2$ or $C_I=2Z_1''+Z_3+Z_4'$ would be examples of easy edges, while the internal edges \eqref{C41} in the standard triangulation of the figure-eight knot complement are hard.

Suppose that a triangulation $M=\{\Delta_i\}_{i=1}^N$ only contains easy edges, and let us focus on the internal ones $C_I$. For every edge $I$, we can define a polarization $\{\Pi_i^I\}$ so that the tetrahedron parameters appearing in $C_I$ are all position coordinates. Due to the definition of easy edges, we can always choose $\Pi_i\in \{\Pi_Z,\Pi_{Z'},\Pi_{Z''}\}$ so that the product polarization has this property. Then, in the theory $T_{\{\Delta_i\},\{\Pi_i^I\}}$ there will automatically exist an operator $\CO_I$ for the edge $C_I$, constructed as a product of elementary chiral fields. For example, if our easy edge is $C_I = 2Z_1''+Z_3+Z_4'$, we choose a product polarization $\{\Pi_i^I\}$ that includes $\Pi_1^I = \Pi_{Z_1''},\,\Pi_3^I=\Pi_{Z_3},$ and $\Pi_4^I=\Pi_{Z_4'}$. Then $T_{\{\Delta_i\},\{\Pi_i^I\}}$ will have operators $\CO_{Z_1''}$, $\CO_{Z_3}$ and $\CO_{Z_4'}$, all elementary chiral fields, from which we define $\CO_I = (\CO_{Z_1''})^2\CO_{Z_3}\CO_{Z_4'}$.

Now, we are really interested in the theory $T_{\{\Delta_i\},\tilde\Pi}$, associated to the polarization $\tilde\Pi$ in which every internal edge is a position coordinate. For each individual $C_I$, there exists an (affine) $Sp(2N,\Z)$ transformation $g_I$ such that
\be \tilde\Pi = g_I\circ \{\Pi_i^I\}\,,\qquad T_{\{\Delta_i\},\tilde\Pi} = g_I\circ T_{\{\Delta_i\},\{\Pi_i^I\}}\,. \label{gI} \ee
This is not quite an arbitrary transformation. In particular, since $C_I$ is a position coordinate in both $\{\Pi_i^I\}$ and $\tilde \Pi$, the action of $g_I$ cannot gauge any of the $U(1)$ flavor symmetries under which the operator $\CO_I$ transforms. Therefore, we can easily pull $\CO_I$ through the transformation on the right of \eqref{gI} to define the corresponding internal edge operator in $T_{\{\Delta_i\},\tilde\Pi}$.

If a triangulation only contains easy edges, we can repeatedly use this construction to define all the operators appearing in the superpotential \eqref{WforTM}. Notice, however, that we define each $\CO_I$ using a different mirror Lagrangian description of $T_{\{\Delta_i\},\tilde\Pi}$. In any given description, one of the internal edge operators is ``simple,'' being a gauge-invariant product of elementary chiral multiplets. The other operators may appear more complicated, and will in general take the form of monopole operators.

Just as we defined operators for internal edges, we can also define operators $\CO_E$ for any easy external edges (or cusp holonomies) that are positions in $\Pi\subset\tilde\Pi$. In various mirror duality frames, they will appear either as products of chiral fields or monopole operators, and they will be charged under the flavor symmetries of $T_{M,\Pi}$ that correspond to the positions $X_E$ (or $U$, etc.).

Currently, we only have a rigorous construction of operators $\CO_I$ and $\CO_E$ for triangulations with easy edges. Indeed, it appears that if we try to define a theory $T_{M,\Pi}$ using a triangulation of $M$ with hard edges, the theory will be slightly degenerate --- and potentially missing some expected operators. We will see an example of this behavior in Section \ref{sec:gauge41}. Fortunately, it seems that we can always refine a given triangulation of a 3-manifold $M$ so that no hard edges are present, and then use this triangulation to construct $T_{M,\Pi}$. \\

One of our central claims is that the theories $T_{M,\Pi}$ constructed here are topological invariants of a three-manifold $M$ (and a polarization of its boundary), which do not depend on the actual triangulation being used to define them --- or on the choice of refinement, should a given triangulation include hard edges. In particular, we claim that different triangulations lead to different mirror-symmetric descriptions of the same underlying 3d SCFT. To understand this, we now proceed to analyze the simplest and most important example of a triangulated 3-manifold: the bipyramid.

\subsection{The bipyramid and the 2--3 move}
\label{sec:gauge23}

Let's consider the theory of the bipyramid, as constructed from two different triangulations. To keep things simple, we will focus on the ``equatorial'' polarization $\Pi=\Pi_{\rm eq}$ for the bipyramid, as defined in \eqref{bipcoord3} or \eqref{bipcoord2}. In particular, the three equatorial edges of the bipyramid are position coordinates in $\Pi$. We keep the same notation as in Section \ref{sec:geom}, and repeat Figure \ref{fig:Pach} here as a visual reference.

\begin{figure}[htb]
\centering
\includegraphics[width=5in]{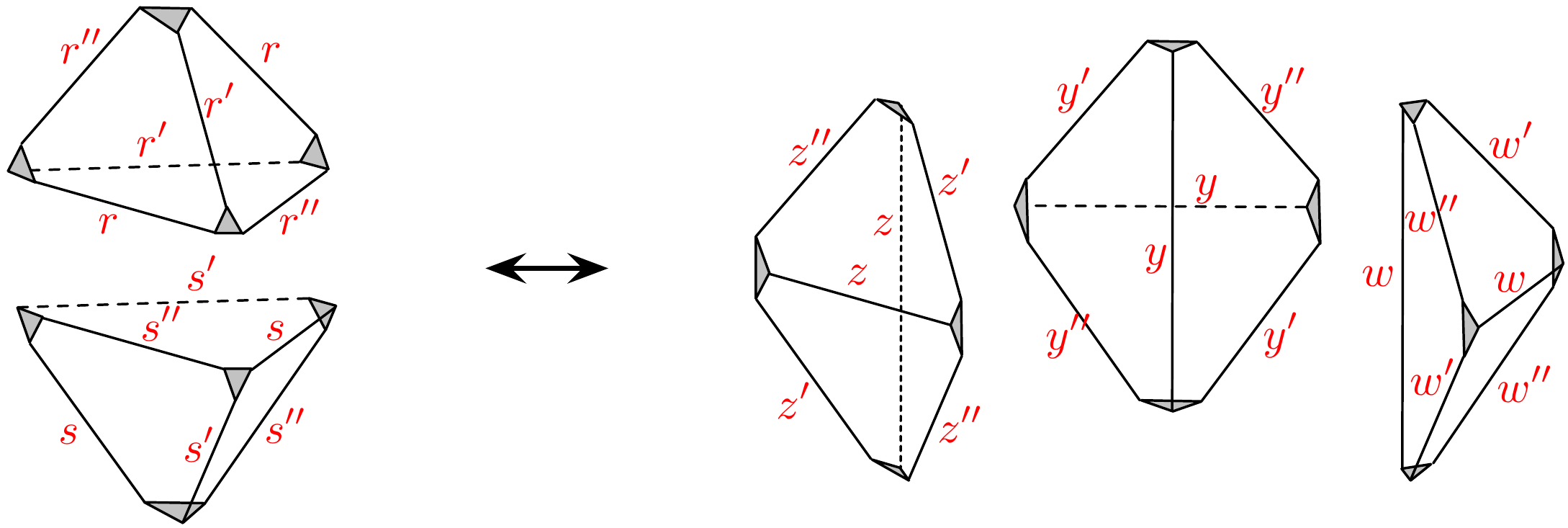}
\caption{Decompositions of the bipyramid, with labelled edge coordinates (Figure \protect\ref{fig:Pach}).}
\label{fig:Pach2}
\end{figure}

If we decompose the bipyramid into three tetrahedra, then according to our rules $T_{ \{\Delta_i\} , \Pi}$ is a theory of three free chiral multiplets, coupled to a background $U(1)^3$,
with some extra CS couplings determined by our choice of momenta in $\Pi$. The operator associated to the unique internal edge is simply the product of the three chiral fields.
Hence $T_{M,\Pi} $ is simply the XYZ model, with appropriate coupling to the unbroken $U(1)^2$ flavor symmetry.
The operators associated to the external edges are the three chiral multiplets themselves.

Being more explicit, we can start with a product polarization $\{\Pi_i\} = \{\Pi_Z,\Pi_W,\Pi_Y\}$, such that $Z,W,Y$ are coordinates and $Z'',W'',Y''$ are momenta. In the equatorial polarization $\Pi_{\rm eq}$, we know that $X_1 = Z$ and $X_2=W$ are positions while $P_1=Z''+Y'$ and $P_2=W''+Y'$ are momenta; we therefore choose a compatible polarization $\tilde\Pi$ on $\CP_{\{\pd\Delta_i\}}$ with positions $X_1,X_2,C$ and momenta $P_1,P_2,\Gamma$, where $C=X+Y+Z$ and $\Gamma=-Y'$. The affine symplectic transformation $g$ from $\{\Pi_i\}$ to $\tilde\Pi$ is encoded as
\be \begin{pmatrix} X_1\\X_2\\C\\P_1\\P_2\\\Gamma\end{pmatrix}= \begin{pmatrix} 1&0&0&0&0&0\\ 0&1&0&0&0&0\\1&1&1&0&0&0 \\0&0&0&1&0&-1\\0&0&0&0&1&-1\\0&0&0&0&0&1\end{pmatrix}\begin{pmatrix} 1&0&0&0&0&0\\0&1&0&0&0&0\\0&0&1&0&0&0\\0&0&0&1&0&0 \\0&0&0&0&1&0\\0&0&1&0&0&1\end{pmatrix} \begin{pmatrix} Z\\W\\Y\\Z''\\W''\\Y''\end{pmatrix}+\begin{pmatrix} 0\\0\\0\\i\pi\\i\pi\\-i\pi\end{pmatrix}\,, \label{Spbip3} \ee
which involves a $T$-type transformation, a $GL$-type transformation, and a shift that will not be visible at the level of Lagrangians. Thus, starting with a Lagrangian description
\begin{align} \CL_{\{\Pi_i\}}[V_Z,V_W,V_Y] &= \frac{1}{4\pi}\int d^4\theta\Big(-\frac12 \Sigma_ZV_Z-\frac12 \Sigma_WV_W-\frac12\Sigma_YV_Y\Big)+ \notag\\ &\hspace{.8in} \int d^4\theta \big( \phi_Z^\dagger e^{V_Z}\phi_Z+\phi_W^\dagger e^{V_W}\phi_W+\phi_Y^\dagger e^{V_Y}\phi_Y\big) \end{align}
for $T_{\{\Delta_i\},\{\Pi_i\}}$, we construct the Lagrangian for $T_{\{\Delta_i\},\tilde\Pi} = g\circ T_{\{\Delta_i\},\{\Pi_i\}}$ simply as
\be \CL_{\tilde\Pi}[V_{X_1},V_{X_2},V_C] = \CL_{\{\Pi_i\}}[V_{X_1},V_{X_2},V_C-V_{X_1}-V_{X_2}]+\frac{1}{4\pi}\int d^4\theta(\Sigma_C-\Sigma_{X_1}-\Sigma_{X_2})\,(V_C-V_{X_1}-V_{X_2})\,, \label{Lbip3eq}\ee
in other words by adding a level 1 Chern-Simons term for $V_Y$, and redefining $V_Z=V_{X_1},\,V_W=V_{X_2}$, and $V_Y=V_C-V_{X_1}-V_{X_2}$. It is trivial to see that the elementary operator
\be \CO_C \equiv \phi_Z\phi_W\phi_Y \label{WXYZ} \ee
exists in $T_{\{\Delta_i\},\tilde\Pi}$, as do the individual operators $\phi_Z,\phi_W,\phi_Y$ associated to the equatorial external edges. The bipyramid theory $T_{M,\Pi_{\rm eq}}$ is then defined by adding the superpotential $\CW = \CO_C$ to \eqref{Lbip3eq}, which forces $V_C=\theta\bar\theta m_C=0$; direct calculation then shows
\begin{align} \CL_{M,\Pi_{\rm eq}}[V_{X_1},V_{X_2}]&=\frac{1}{4\pi}\int d^4\theta\, \Sigma_{X_1}V_{X_2} + \int d^4\theta \big( \phi_Z^\dagger e^{V_{X_1}}\phi_Z+\phi_W^\dagger e^{V_{X_2}}\phi_W+\phi_Y^\dagger e^{-V_{X_1}-V_{X_2}}\phi_Y\big) \notag\\
 &\hspace{.8in} + \int \big(d^2\theta \,\phi_Z\phi_W\phi_Y +c.c.\big)\,.
\label{XYZbip}
\end{align}
This is the promised XYZ model, with slightly redefined $U(1)^2$ symmetries, and a mixed Chern-Simons term.

If we decompose the bipyramid into two tetrahedra instead of three,
we need no superpotential. On the other hand, the transformation $g$
from the polarization $\{\Pi_R,\Pi_{S''}\}$ for the two tetrahedra to $\Pi_{\rm eq}$ is non-trivial:
as the positions are $X_1=R + S''$, $X_2=R'' + S$, and $R'+S'$, it is easy to see that
$g$ involves gauging (with no CS coupling) the $U(1)$ under which the two chiral multiplets have opposite charge.
Hence with this definition $T_{M,\Pi}$ is simply $N_f=1$ SQED, with appropriate coupling to the
$U(1)^2$ flavor symmetry. The operator associated to the edge coordinate $X_1=R+S''$ is simply the meson operator.

Again, one can go through explicit Lagrangian manipulations as above. Starting from a polarization $\{\Pi_i\}=\{\Pi_R,\Pi_{S''}\}\sim (R,S'';R'',S')$ we reach the equatorial polarization $\Pi_{\rm eq}\sim (X_1,X_2;P_1,P_2)$ via a symplectic transformation $g = g_S\,g_T\,g_U$, with
\be \label{gbip2}
g_S = \begin{pmatrix} 1&0&0&0\\0&0&0&-1\\0&0&1&0\\0&1&0&0\end{pmatrix},\quad
g_T = \begin{pmatrix} 1&0&0&0\\0&1&0&0\\0&0&1&0\\0&1&0&1\end{pmatrix},\quad
g_U = \begin{pmatrix} 1&1&0&0\\0&1&0&0\\0&0&1&0\\0&0&-1&1 \end{pmatrix}\,.
\ee
Therefore, we obtain $\CL_{M,\Pi_{\rm eq}}[V_{X_1},V_{X_2}]$ by starting with
\be \CL_{\{\Delta_i\},\{\Pi_i\}}[V_R,V_{S''}]=\frac{1}{4\pi}\int d^4\theta\Big(-\frac12 \Sigma_RV_R-\frac12 \Sigma_{S''}V_{S''}\Big)+ \int d^4\theta \big( \phi_R^\dagger e^{V_R}\phi_R+\phi_{S''}^\dagger e^{V_{S''}}\phi_{S''}\big)\,,\ee
redefining the $U(1)^2$ symmetry, adding a Chern-Simons term, and gauging a $U(1)$. A straightforward calculation produces%
\footnote{\label{foot:half}%
In the last step of the derivation of \eqref{SQEDbip}, we shifted the dynamical gauge multiplet $V\to V+\frac12 V_{X_1}$, thereby adding to the `$X_1$' flavor current a half-integral multiplet of the gauge current. This non-integral shift is not necessary, but can be made sense of because the multiplet $V_{X_1}=\theta\bar\theta m_{X_1}$ is nondynamical. In the form \eqref{SQEDbip} of the Lagrangian, the identification of $\frac12 V_{X_1}$ with an axial flavor multiplet becomes immediate.} %
\begin{align} \CL_{M,\Pi_{\rm eq}}[V_{X_1},V_{X_2}] &= \frac{1}{4\pi}\int d^4\theta\big(\Sigma_{X_1}V_{X_2}+(\Sigma_{X_1}+2\Sigma_{X_2})\,V\big) \notag \\
 &\qquad\qquad +\int d^4\theta\big( \phi_R^\dagger e^{V+\frac12{V_{X_1}}}\phi_R+\phi_{S''}^\dagger e^{-V+\frac12{V_{X_1}}}\phi_{S''}\big)\,, \label{SQEDbip}
\end{align}
with the $U(1)$ gauge multiplet $V$ dynamical. This is precisely $N_f=1$ SQED, with a mixed Chern-Simons coupling, and slightly redefined $U(1)^2$ symmetry. The meson operator $\CO_{X_1}\equiv \phi_R\phi_{S''}$ is obviously charged under $V_{X_1}$. We know that SQED also has two monopole operators $v_+$ and $v_-$, and from the form of the FI term in \eqref{SQEDbip} we see that they must be charged under the combinations $V_{X_2}$ and $-V_{X_1}-V_{X_2}$, respectively. Thus, they correspond to the remaining two equatorial edges.

Thanks to the basic $\CN=2$ mirror symmetry statement \eqref{XYZmir}, our construction gives the same theory
$T_{M,\Pi}$ for the bipyramid, \emph{no matter how we triangulate it}.
By carefully comparing the Lagrangian descriptions \eqref{XYZbip} and \eqref{SQEDbip}, we see that the three equatorial edge operators --- elementary fields in the XYZ model and a meson/monopoles in SQED --- are mapped to each other by mirror symmetry, and their coupling to the background $U(1)$ gauge multiplets $V_{X_1}$ and $V_{X_2}$ coincide perfectly.

One can also repeat the exercise for the longitudinal polarization. The two triangulations give respectively
$\CN=4$ SQED with $N_f=1$ and the theory of a free hypermultiplet, {\it i.e.} the basic $\CN=4$ mirror pair.
This is a useful exercise in order to show that the operators associated to
longitudinal edges by the two polarizations are also mapped into each other by mirror symmetry.

With this result, we are in position to argue that the theories $T_{M,\Pi} $
defined by different triangulations of the same three-manifold $M$ are mirror to each other.
Different triangulations are related by a sequence of $2-3$ moves.%
\footnote{Strictly speaking, we should only consider triangulations that have easy edges, as discussed in Section \ref{sec:defTM}. It is very plausible --- although not mathematically proven --- that to connect two ``easy'' triangulations, one can always find a chain of $2-3$ moves that only pass through other easy triangulations.} %
Two triangulations that differ by a $2-3$ move give two definitions of the theory $T_{M,\Pi}$
that differ only by a basic mirror symmetry relation. The mirror symmetry  acts on the degrees of freedom
associated to the particular bipyramid that is decomposed in two different ways in the course of a $2-3$ move.

\subsection{The flip}
\label{sec:flip}

Just as $2-3$ moves change the internal triangulation of a 3-manifold, the flips described in Section \ref{sec:geom23} can change the triangulation of its (geodesic) boundary. This has a very simple effect on a theory $T_{M,\Pi}$.

For example, suppose that $T_{M,\Pi}$ has an operator $\CO_X$, charged under a global symmetry $U(1)_X$, that corresponds to an external edge with position coordinate $X$. We want to add a tetrahedron $\Delta_Z$ to flip this edge, as in Figure \ref{fig:flip}. Following our gauge theory dictionary, this means that we form the combined theory $T_{M,\Pi}\otimes T_{\Delta_Z,\Pi_Z}$, and add a superpotential coupling
\be \CW = \CO_X\phi_Z\,. \label{Wflip} \ee
The new theory now has a chiral operator $\phi_Z$ that transforms under the anti-diagonal subgroup of $U(1)_X\times U(1)_Z$ that is unbroken by \eqref{Wflip}.

This transformation simply describes the $F$ operation of Section \ref{sec:F}. Just as $F^2$ is a trivial operation on a 3d SCFT, flipping a diagonal twice is a trivial operation on the boundary of a 3-manifold.

\subsection{$T_M$ as a boundary condition}
\label{sec:gauge4d}

In section \ref{sec:ops} we learned some useful facts about the relation between
three dimensional theories and boundary conditions for four-dimensional theories.
We saw that all the 3d theories in an orbit of the $Sp(N,Z)$ action can be
thought of as representing the same boundary condition in different electric-magnetic duality frames
of a four-dimensional abelian gauge theory.
We also saw that the $F$ transformation on three-dimensional theories
can be thought of as relating two mirror description of the
same boundary condition for one hypermultiplet.

We can use these facts to try to liberate $T_{M,\Pi}$ from the dependence on the polarization $\Pi$,
and even on the choice of triangulation of the geodesic boundary $\CC$ of $M$. To remove the polarization dependence, we can
couple $T_{M,\Pi}$ to a four-dimensional gauge theory, whose symplectic  lattice of electric-magnetic charges
is modeled on the lattice generated by the edge coordinates of the triangulation of $\CC$, the geodesic boundary of $M$.
In order to remove the dependence on the triangulation of $\CC$, we need to couple $T_{M,\Pi}$ to a set of hypermultiplets as well,
one for each edge of the triangulation of $\CC$. In order for the flip to coincide with an $F$ move,
each hyper must be coupled by a superpotential to $\CO_E$, and hence have four-dimensional gauge charges equal or opposite to
the charge associated to the edge itself.

Thus we find it natural to couple $T_{M,\Pi}$ to an apparently bizarre four-dimensional theory:
an $\CN=2$ abelian gauge theory coupled to hypermultiplets of several dyonic charges, one for each edge of the triangulation of $\CC$.
This theory is less bizarre than it seems. Indeed, \cite{GMNIII}, the symplectic  lattice generated by a triangulation of $\CC$
coincides naturally with the lattice of IR electric-magnetic charges for the four-dimensional theory obtained from two M5 branes wrapping $\CC$.
Furthermore, in a large patch of the 4d Coulomb branch, the whole spectrum of IR BPS particles can be thought of
as bound states of a basis of hypermutliplet particles, each associated to an edge of the triangulation, and carrying the corresponding charges.

Thus there is a sense in which the abelian gauge theory with the hypermultiplets associated to the edges of the triangulation
is a complete IR description of the four-dimensional theory associated to $\CC$. And thus $T_{M,\Pi}$ can be thought as the description of a boundary condition
for the four-dimensional theory, in a given duality frame. This is a property which we surely expect to be true of $T[M,\mathfrak{su}(2)]$.
In later sections we will reinforce the connection further. For example, the moduli space of vacua of $T_{M, \Pi}$ compactified on a circle
naturally defines a boundary condition for the four-dimensional gauge theory compactified on a circle.

\subsection{The octahedron}
\label{sec:oct}

We include two more brief examples of three-manifold theories. The first, the octahedron, demonstrates how $2-3$ moves can be used in the interior of a manifold, resulting in interesting chains of $\CN=2$ SCFT dualities. The second, the figure-eight knot complement, will illustrate how potential difficulties with ``hard'' edges can be resolved.

\begin{figure}[htb]
\centering
\includegraphics[width=3.5in]{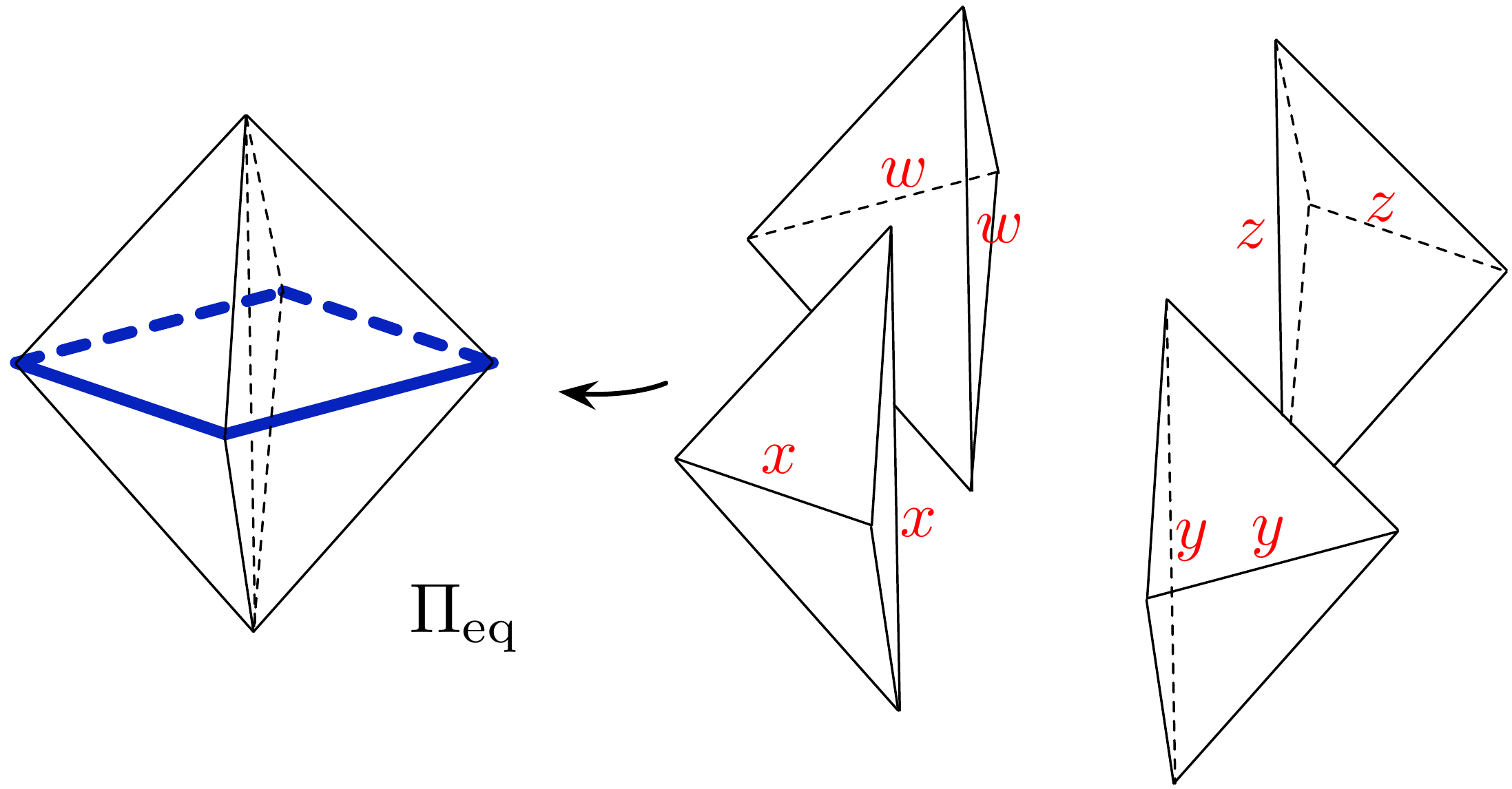}
\caption{The octahedron from four tetrahedra}
\label{fig:oct4}
\end{figure}

The simplest way to construct an octahedron is from four tetrahedra, glued together along a central edge (Figure \ref{fig:oct4}). Suppose we we work in an equatorial polarization $\Pi_{\rm eq}$ as shown, with independent positions $(X,Y,Z,C)$, where the internal edge has parameter $C=X+Y+Z+W$. The resulting theory $T_{{\rm oct},\Pi_{\rm eq}}$ is a simple generalization of the bipyramid theory \eqref{XYZbip}. It starts with four chirals $\phi_X,\phi_Y,\phi_Z,\phi_W$ and four background gauge multiplets $V_X, V_Y, V_Z, V_W$. The multiplet $V_W$ is redefined as $V_W \to V_C-V_X-V_Y-V_Z-V_W$, and then we add a quartic superpotential
\be \CW_{\rm eq} = \phi_X\phi_Y\phi_Z\phi_W \ee
to break the global symmetry $U(1)_C$. We are still left with $U(1)_X\times U(1)_Y\times U(1)_Z$.

To be more specific, we should fix conjugate momenta in $\Pi_{\rm eq}$, taking (say) $(X+W'',Y+W'',Z+W'',-W'')$. This choice of momenta will add some background Chern-Simons couplings to the Lagrangian of $T_{{\rm oct},\Pi_{\rm eq}}$, which we encourage the careful reader to work out.

\begin{figure}[htb]
\centering
\includegraphics[width=5.5in]{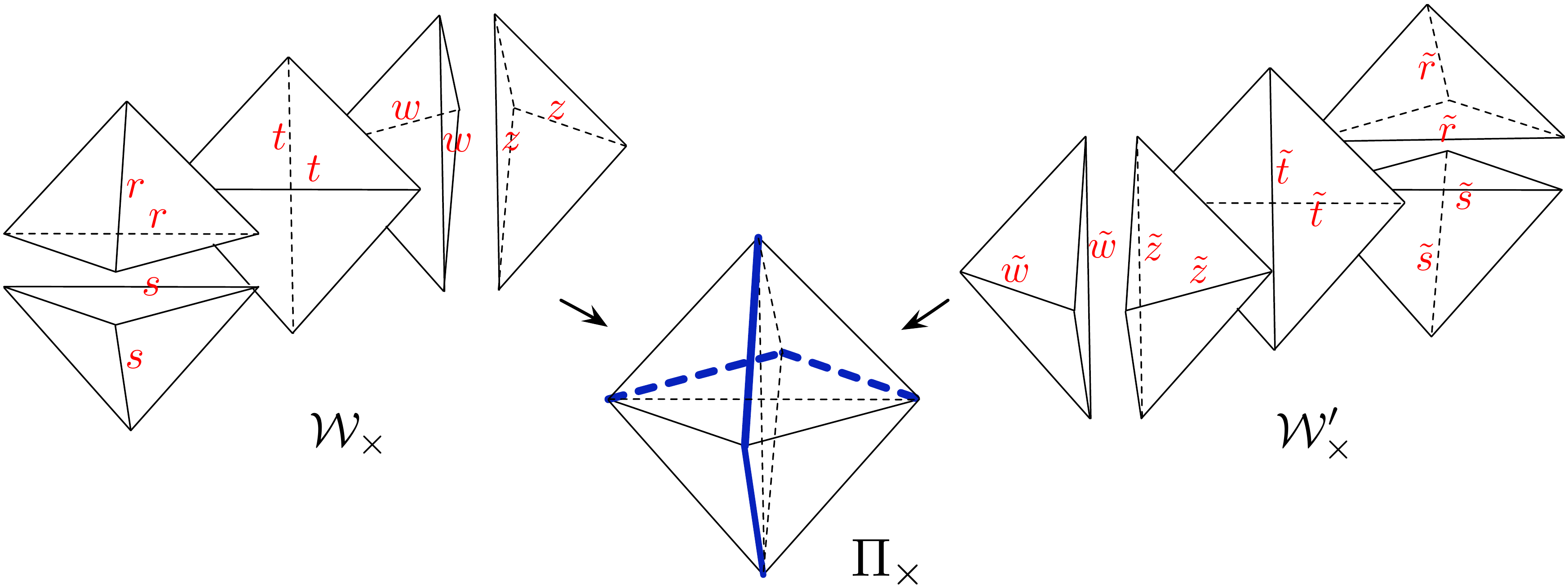}
\caption{The octahedron from five tetrahedra, two ways. Positions of the polarization $\Pi_\times$ are indicated in the middle.}
\label{fig:oct5}
\end{figure}

Now, if we change to a different polarization $\Pi_\times$, as in the center of Figure \ref{fig:oct5}, we must perform an $Sp(6,\Z)$ transformation on the theory $T_{{\rm oct},\Pi_{\rm eq}}$. This transformation, call it $g_\times$, gauges the $U(1)$ symmetry under which $(\phi_X,\phi_Y)$ transform as a hypermultiplet. Thus, we obtain a new theory $T_{{\rm oct},\Pi_\times} = g_\times\circ T_{{\rm oct},\Pi_{\rm eq}}$ which has a subsector that looks like $N_f=1$ SQED. By the basic $\CN=2$ mirror symmetry (acting on this subsector), if must be equivalent to a theory of five chirals, with no dynamical gauge group, and superpotential
\be \CW_\times = \phi_T\phi_Z\phi_W + \phi_T\phi_{ R}\phi_{ S}\,. \label{Woct5} \ee
From the perspective of SQED, $\phi_T \equiv \phi_X\phi_Y$ is a meson, and the new fields $\phi_{R},\,\phi_S$ are monopole operators; the second term in \eqref{Woct5} is just the ``XYZ'' superpotential that we must add during mirror symmetry.

By looking at the left-hand side of Figure \ref{fig:oct5}, we should immediately identify the description of $T_{{\rm oct},\Pi_\times}$ using five chirals as arising from a five-tetrahedron triangulation of the octahedron. The two terms in the superpotential $\CW_\times$ come directly from the two internal edge coordinates $C_1 = T+Z+W$ and $C_2 = T+R+S$ in this triangulation.

To go a bit further, we notice that there another possible triangulation into five tetrahedra, shown on the right side of Figure \ref{fig:oct5}. In a sense, it is maximally incompatible with the polarization $\Pi_\times$. If we try to use triangulation to define $T_{{\rm oct},\Pi_\times}$, we will again start with five chirals $\phi_{\tilde R},\phi_{\tilde S},\phi_{\tilde T},\phi_{\tilde Z},\phi_{\tilde W},$ but will have to gauge the two $U(1)$ symmetries which treat the respective pairs $\phi_{\tilde R},\phi_{\tilde S}$ and $\phi_{\tilde Z},\phi_{\tilde W}$ as hypermultiplets. What results is a mirror description of $T_{{\rm oct},\Pi_\times}$ as a \emph{dynamical} $U(1)^2$ gauge theory with two hypermultiplets and a neutral chiral $\phi_{\tilde T}$, coupled by a superpotential
\be \CW_\times' = \phi_{\tilde T}\phi_{\tilde Z}\phi_{\tilde W} + \phi_{\tilde T}\phi_{\tilde R}\phi_{\tilde S}\,. \label{Woct5g} \ee
It is not too hard to recognize that these two descriptions of $T_{{\rm oct},\Pi_\times}$ correspond to the case $N=2$, $r=0$, $k=1$ of the infinite family of mirror pairs \eqref{mirrorR}.

There are infinitely more splittings of the octahedron, all giving dual descriptions of $T_{{\rm oct},\Pi_\times}$ and its $Sp(6,\Z)$ images. We could similarly analyze triangulations of larger polyhedra or more general 3-manifolds to generate a huge class of 3d $\CN=2$ mirror symmetries. We expect, in particular, that the family of dual theories mentioned in \eqref{mirrorR} is realized as a (small!) subset of these.

\subsection{Figure-eight knot}
\label{sec:gauge41}

As our final example, we consider the theory associated to a manifold with a torus cusp boundary: the complement of the figure-eight knot $M=S^3\bs \mb{4_1}$.

The minimal triangulation of $M$ into two tetrahedra, discussed in Section \ref{sec:cuspbdy}, has two internal edges and both of them are hard:
\be C_1 = 2Z+Z''+2W+W''\,,\qquad C_2 = 2Z'+Z''+2W'+W'' \,.\ee
We could certainly try to write down a gauge theory from this triangulation. Indeed, starting with $T_{\Delta_Z,\Pi_{Z'}}\otimes T_{\Delta_W,\Pi_W}$, we can change the polarization to $\wt\Pi$ with (positions; momenta)$=(U,C_1;v,\Gamma)$, where $U=Z'-W$, $v=Z-Z'$ as in \eqref{Uv41}, and $\Gamma_1=-W$ is the conjugate to $C_1$. The resulting theory $T^{(2)}_{\mb{4_1},\tilde\Pi}$ is a $U(1)$ gauge theory with two chiral matter fields both of charge $+1$, and no dynamical Chern-Simons coupling. The factors in the global symmetry group $U(1)_{\rm vector}\times U(1)_{\rm top}$ correspond to position coordinates $\tfrac12U$ and $-C_1-\tfrac32U$, respectively. Explicitly, we find a Lagrangian
\begin{align}
\CL^{(2)}_{\mb{4_1},\tilde\Pi}[V_U,V_{C_1}] &= \frac{1}{4\pi}\int d^4\theta\,\Big(-\frac32 \Sigma_U V_U-(2\Sigma_{C_1}+3\Sigma_U)V\Big) \notag \\
 &\hspace{1in} +\int d^4\theta \big( \phi_{Z'}^\dagger e^{V+V_U}\phi_{Z'}+ \phi_W^\dagger e^{V} \phi_W\big)\,,
\label{41tet2}
\end{align}
with $V$ dynamical. Unfortunately, we are hard-pressed to find two monopole operators $\CO_{C_1},\,\CO_{C_2}$ in this theory that could be added to a superpotential. Their existence is crucial to break the (essentially topological) $U(1)_{C_1}$ symmetry, to set $V_{C_1}\to 0$ and to complete the gluing procedure.

\begin{figure}[htb]
\centering
\includegraphics[width=5in]{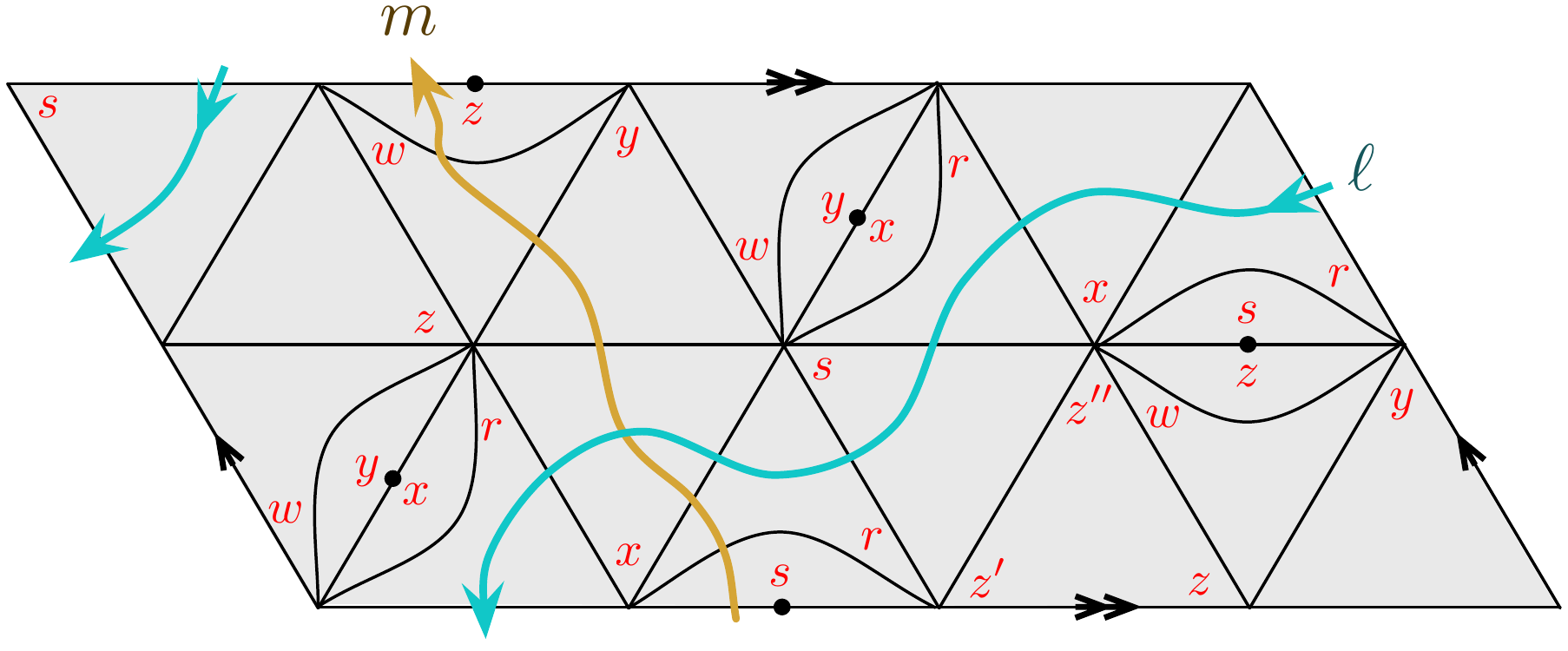}
\caption{The torus cusp for the figure-eight knot complement, triangulated into six tetrahedra. The cyclic order of edge parameters $(z,z',z'')$, etc., is always the same, so we only indicate one parameter per vertex triangle.}
\label{fig:cusp6}
\end{figure}

To resolve this problem, we must resolve the triangulation. For example, we have found a decomposition of the figure-eight knot complement into six tetrahedra, such that all internal edges are easy. We sketch a developing map of the resulting cusp neighborhood in Figure \ref{fig:cusp6}, from which we read off the six internal edge coordinates%
\footnote{We invite the reader to check that this triangulation produces the same A-polynomial as in \eqref{A41}.}
\be
\begin{array}{l@{\qquad}l} C_1 = X+W+2(R'+S'+Z'')\,, &
C_2 = R+Y+2(Z'+W'+S'')\,, \\
C_3 = S+W+2(R''+X''+Y')\,, &
C_4 = R+Z+2(Y''+W''+X')\,, \\
C_5 = X+Y\,, &
C_6 = S+Z\,.
\end{array}
\ee
We also find eigenvalues for the meridian and longitude cycles on the boundary $T^2$,
\be
U = S'+R'-X''+Y''-W'+Z''\,,\qquad v = X+R'-S-R''\,.
\ee
Using the combinatorial data for this gluing, it is straightforward (if tedious) to follow the rules of Section \ref{sec:defTM} to define the actual figure-eight knot theory $T_{\mb{4_1},\Pi}$, where $\Pi$ has position $U$ and momentum $v$. This theory has six operators $\CO_{C_1},...,\CO_{C_6}$ that can be added to the superpotential to break the $U(1)$ symmetries corresponding to the internal edges.

%%%%%%%%%%%%%%%%%%%%%%%%%%%%%%%%%%%%%%%%%%%%%%%%%%%%%%%%%%%%%%%%%%%%%%%%%%
%%%%%%%%%%%%%%%%%%%%%%%%%%%%%%%%%%%%%%%%%%%%%%%%%%%%%%%%%%%%%%%%%%%%%%%%%%

\section{Moduli space on $\R^2\times S^1$}
\label{sec:moduli}

One simple way to test the correspondence $M \longleftrightarrow  T_M$
is to associate a moduli space to each side. In the analogous construction \cite{Gaiotto-dualities}
of the 4d $\CN=2$ superconformal theory from a Riemann surface $\CC$,
there is a similar test of the correspondence $\CC  \longleftrightarrow T(\CC)$
based on comparing the moduli space of complex (equivalently, conformal) structures on $\CC$
with the moduli space of marginal couplings of the theory $T(\CC)$.

In the present case, there is a similar test of the correspondence $M  \longleftrightarrow T_M$
based on comparing moduli spaces of complex flat connections on $M$ and the moduli space
of supersymmetric vacua of the theory $T_M$. To be more precise, the space of complex flat connections on $M$
can be identified with the space of SUSY moduli in the theory $T_M$ on $\R^2 \times S^1$ \cite{DGH}:
\be
\boxed{\phantom{\int}
\CM_{{\rm flat}} (M,SL(2,\C)) \; = \; \CM_{{\rm SUSY}} (T_M)\,.
\phantom{\int}}
\label{MMclaim}
\ee
While the definition of the moduli space $\CM_{{\rm flat}} (M,SL(2,\C))$ is clear (and was reviewed in section \ref{sec:geom})
we need to properly interpret the right-hand side of \eqref{MMclaim}.

Upon compactification on $\R^2\times S^1$, the $\CN=2$ theory $T_M$ becomes effectively two-dimensional. Supersymmetry then requires that the vevs of chiral and twisted chiral fields, whether dynamical or not, are complex valued. For example, 3d real mass parameters associated to a background $U(1)$ gauge multiplet $V$ become complexified by the holonomies of the background photon on $S^1$. Therefore, moduli spaces parametrized by vevs of chiral and twisted chiral fields are always complex manifolds.
Here, we are mostly interested in the moduli space parameterized by vevs of twisted chiral fields --- the descendants of 3d gauge multiplets --- and denote this space $\CM_{\rm SUSY}$.

For example, if $M$ is a closed 3-manifold without boundaries or cusps, the corresponding field theory $T_M$
on $\R^2 \times S^1$ has the moduli space of supersymmetric vacua $\CM_{{\rm SUSY}} (T_M)$
obtained by minimizing the twisted superpotential $\widetilde{\CW}$.
Since the twisted superpotential is a holomorphic function, the variety defined by
the equations $\partial \widetilde{\CW}=0$ is a complex variety, just like the moduli space
of flat $SL(2,\C)$ connections on $M$.

More generally, if $M$ is a 3-manifold with boundary $\CC = \partial M$, it is natural
to project the moduli space $\CM_{{\rm flat}} (M,SL(2,\C))$ onto the moduli space of
flat connections on $\CC$, {\it i.e.} consider those flat connections on $\CC$ which can be extended to all of $M$. In Section \ref{sec:geom}, this projection was cut out by the Lagrangian submanifold
\be \CL_M\;\subset\;\CP_{\pd M} = \CM_{\rm flat}(\CC,SL(2,\C))\,.\ee
Correspondingly, in the $\CN=2$ gauge theory $T_M$, it is natural to ask for which values of
the parameters $v_i$ (= vevs of non-dynamical fields) the theory $T_M$ has SUSY vacua on $\R^2 \times S^1$.
In order to answer this question, we introduce the effective twisted superpotential $\widetilde{\CW}_{{\rm eff}}$
obtained by minimizing $\widetilde{\CW}$ with respect to all {\it dynamical} fields, and then define \cite{DG-Sdual}:
\be
\CM_{\rm SUSY}^{\text{(param)}}(T_M)\,:\quad u_i = \frac{\partial \widetilde{\CW}_{{\rm eff}}}{\partial v_i} \,.
\label{uidef}
\ee
In the the case where $v_i$ is the twisted mass in a background $U(1)$ gauge field, the coordinate $u_i$ should be thought of as the background FI parameter for this field; then it is clear that \eqref{uidef} is the condition for unbroken supersymmetry.
As we illustrate in a number of examples below, $\widetilde{\CW}_{{\rm eff}}$
is a transcendental function, generically a sum of dilogarithm functions.
However, after taking the derivatives in \eqref{uidef} and introducing the new coordinates
\be
\ell_i = e^{v_i}\,, \qquad m_i = e^{u_i}
\label{lmviauv}
\ee
(which are natural, because the complexified vevs $u_i$ and $v_i$ are periodic),
one finds a nice algebraic variety that is identical to $\CL_M$.

Geometrically, it should be clear that the Lagrangian submanifold $\CL_M$ cannot depend on the coordinates and polarization used to describe the phase space $\CP_{\pd M}$ when $M$ has a boundary. Changing coordinates will simply re-parametrize $\CL_M$. Similarly, the space $\CM_{\rm SUSY}^{(\rm param)}(T_{M,\Pi})$ should not depend on the polarization $\Pi$ (or boundary triangulation, etc.) used in previous sections to define a theory $T_{M,\Pi}$. One way to see this is to interpret $T_{M,\Pi}$ on $\R^2\times S^1$ as describing a boundary condition $\CB[T_{M,\Pi}]$ for a 4d $\CN=2$ theory $T[\CC]$ ($\CC=\pd M$) compactified on $\R^3\times S^1$, as in Section \ref{sec:gauge4d}. 
With a little bit of work, one can show that the coordinates $\ell_i,\, m_i$ become boundary values of natural coordinates (\eg\ $\CX_E$ of \cite{GMN, GMNII}) on the moduli space of the compactified 4d theory.
From this point of view, $\CM_{\rm SUSY}^{(\rm param)}(T_M)=\CL_M$ becomes a complex Lagrangian submanifold in the four-dimensional moduli space%
\footnote{As discussed (\eg) in \cite{GMN, GMNII}, this 4d moduli space actually has the structure of a hyperkahler manifold. The space $\CM_{\rm SUSY}^{(\rm param)}(T_M)$ is then embedded into $\CM_{\rm SUSY}(T[\CC])$ as a brane of type $(A,B,A)$.}
$\CM_{\rm SUSY}(T[\CC])\simeq \CP_{\pd M}$. This Lagrangian characterizes the boundary condition itself, rather than any specific realization of it via a 3d SCFT. In particular, changing the polarization $\Pi$ merely shifts the duality frame of the combined 4d-3d system, and must map $\CM_{\rm SUSY}^{(\rm param)}(T_M)$ to an isomorphic space. \\

The present discussion of supersymmetric vacua, particularly as given by equations \eqref{uidef} with $\wt W$ a sum of dilogarithm functions, is highly reminiscent of recent work relating effective 2d field theories to  quantum integrable systems \cite{NS-I, NShatashvili, NRS}. For example, 3d $\CN=2$ theories much like $T_{M}$ compactified on a circle are related to the XXZ spin chain. A precise connection between our present constructions and integrable systems would be very interesting, but has yet to be established.

\subsection{The tetrahedron}

Now, let us illustrate this in a few concrete examples, starting with the theory $T_{\Delta,\Pi_Z}$ that we associate
to a single tetrahedron. The theory $T_{\Delta,\Pi_Z}$ is a single chiral multiplet $\phi_Z$ coupled to a $U(1)$ background gauge field
that also has a (supersymmetric) Chern-Simons interaction at level~$-\tfrac{1}{2}$.
On a circle of finite radius $\beta$, this theory has the effective twisted superpotential (\cf\ \cite{Nek-5d, LN, NS-I, DG-Sdual})
\begin{equation}
T_{\Delta,\Pi_Z}\;:\quad \widetilde{\CW}_{{\rm eff}} (Z) \; = \; \Li_2(e^{-Z}) = \Li_2(z^{-1})\,, \label{CWT1}
\end{equation}
where
\be Z := \beta\,\tilde m_Z \ee
is proportional to the twisted mass in the 2d background gauge multiplet (which contains the real mass $m_Z=\Re(\tilde m_Z)$ of the 3d chiral $\phi_Z$). Note that the superpotential \eqref{CWT1} includes an infinite tower of Kaluza-Klein modes on the circle $S^1$, which have been re-summed.

According to \eqref{uidef} the effective complexified FI parameter in the IR is given by
\begin{equation}
Z'' = \frac{\partial \widetilde{\CW}_{{\rm eff}}}{\partial Z'} = \log(1-e^{-Z})
\end{equation}
The relation between $Z$ and $Z''$ can be conveniently written as
\begin{equation} \label{Aone}
\CM_{\rm SUSY}^{\rm (param)} ~:~\quad
e^{Z''} + e^{-Z} -1=z''+z^{-1}-1=0\,,
\end{equation}
and, as promised, describes a nice algebraic curve in the variables \eqref{lmviauv}.
This is precisely the curve \eqref{LDelta} that describes the space of $SL(2,\C)$ structures on a tetrahedron.
Hence, we just verified \eqref{MMclaim} in a basic example of a tetrahedron and its gauge theory counterpart $T_{\Delta,\Pi_Z}$:
\be
\CL_{\Delta} \; = \; \CM_{{\rm SUSY}}^{\rm (param)} (T_{\Delta,\Pi_Z})\,.
\ee

Equation \eqref{Aone} appears to allow any value of the twisted mass $Z$ (given appropriate FI parameter $Z''$) except $Z=0$. At $Z=0$, we hit a singular point, where it looks like the FI parameter must run off to infinity to preserve supersymmetry. 
This can be understood directly in the gauge theory: at $Z=0$ the chiral field $\phi_Z$ is massless, and hence we were not supposed to integrate it out. The effective description of a gauge theory theory with massive vacua breaks down there.

Had we chosen any other polarization for the tetrahedron theory, say $\Pi'=g\circ \Pi_Z$ with position $X$ and momentum $P$ such that
\be \begin{pmatrix} X\\P \end{pmatrix} = \begin{pmatrix}a & b\\ c & d\end{pmatrix} \begin{pmatrix}Z\\Z''\end{pmatrix}\,, \ee
the Lagrangian \eqref{Aone} would be mapped to the isomorphic curve
\be p^ax^{-c}+p^b x^{-d}-1 = 0\,.\ee
As a beautiful example of this behavior, we can consider the particular transformation
\be \sigma\;:\quad \begin{pmatrix}Z\\Z''\end{pmatrix}\mapsto
\begin{pmatrix}Z'\\Z\end{pmatrix}= \begin{pmatrix}-1 & -1\\ 1 & 0\end{pmatrix} \begin{pmatrix}Z\\Z''\end{pmatrix} + \begin{pmatrix} i\pi \\ 0 \end{pmatrix}\,, \label{sST} \ee
which is an affine extension of $ST = \big(\begin{smallmatrix} -1 & -1\\ 1 & 0\end{smallmatrix}\big)\in SL(2,\Z)$ that generates the triality symmetry \eqref{STonT1}. (Note that, just like $ST$ itself, $\sigma$ satisfies $\sigma^3 = id$.)

From the general $Sp(2N,\Z)$ action on theories $T_{M,\Pi}$ \eqref{Sp2T}--\eqref{Sp2GL}, it is easy to see how the twisted superpotentials $\wt\CW_{\rm eff}$ on $\R^2\times S^1$ should transform. For example, the element $T$ adds a level 1 Chern-Simons term $\frac1{4\pi}\int d^4\theta \,\Sigma_ZV_Z$ to the Lagrangian of $T_{\Delta,\Pi_Z}$, which descends (with proper normalization) to
\be T\;:\quad \wt\CW_{\rm eff}(Z) \;\mapsto\; \wt\CW_{\rm eff}'(Z)=\wt\CW_{\rm eff}(Z) + \frac12 Z^2\,.\ee
Similarly, $S$ adds a mixed Chern-Simons term $\frac1{2\pi}\int d^4\theta\,\Sigma_{Z'}V_Z$ and makes $V_Z$ dynamical. Since we now should extremize with respect to $Z$, this must act as a Legendre transform,
\be S\;:\quad \wt\CW_{\rm eff}(Z) \;\mapsto\; \wt\CW_{\rm eff}'(Z') = \left[\wt\CW_{\rm eff}(Z)+Z'Z \right]_{\frac{\pd}{\pd Z}=0}\,. \ee
Finally, we have affine shifts. While these were unimportant for defining Lagrangians on $\R^3$, the do show up in the theory on $\R^2\times S^1$. Namely, shifts by $i\pi$ in ``position'' and ``momentum'' coordinates appear as half-integral shifts in Wilson loops and theta angles, respectively. Thus, for the tetrahedron theory on $\R^2\times S^1$, it is the affine $\sigma$ in \eqref{sST} that implements mirror symmetry,
\be \sigma \circ T_{\Delta,\Pi_Z} \,\simeq\, T_{\Delta,\Pi_Z}\,, \ee
rather than simply $ST$

Putting together the above ingredients, we find that
\be \sigma\;:\quad \wt\CW_{\rm eff}(Z)\;\mapsto\; \wt\CW_{\rm eff}'(Z') \equiv \left[\wt\CW_{\rm eff}(Z)+\frac12 Z^2+(Z'-i\pi)Z\right]_{\frac{\pd}{\pd Z}=0}\,.\ee
Setting $Z=\pd \wt\CW_{\rm eff}(Z') /\pd Z'$ and exponentiating, we obtain
\be \CM_{\rm SUSY}^{\rm (param)}(T_{\Delta,\Pi_{Z'}})\;:\quad z+z'{}^{-1}-1=0\,.\ee
As expected, this transformation leaves the moduli space invariant.

\subsection{The bipyramid}

To find the moduli space for the bipyramid theory, let us work in the equatorial polarization $\Pi_{\rm eq}$, as discussed in Section \ref{sec:geom} and Section \ref{sec:gauge23}. We closely follow the notation in those sections. We can start with the decomposition into two tetrahedra, and use the Lagrangian description \eqref{SQEDbip} of $T_{M,\Pi_{\rm eq}}$ as $N_f=1$ SQED, with a shift $V\to V-V_{X_1}/2$, to obtain a twisted superpotential
\be \wt\CW(X_1,X_2;\sigma) = \Li_2(e^{\sigma})+\Li_2(e^{-\sigma+X_1})+\frac12\sigma^2+(X_2-i\pi)\sigma\,.\ee
Here we have extended the symplectic transformation \eqref{gbip2} with an affine shift by $-i\pi$ for the twisted mass $X_2$. By requiring $\pd\wt\CW/\pd \sigma =0$ (because $\sigma$ is the vev of a dynamical field), and setting $P_1=\pd\wt\CW/\pd X_1$ and $\pd\wt\CW/\pd X_2$, it is straightforward to derive the moduli space
\be \hspace{-.2in}\CM_{\rm SUSY}^{(\rm param)}(T_{M,\Pi_{\rm eq}})\;: \quad\;\; p_1+\frac{p_2}{x_1}-1=0\,,\qquad p_2+\frac{p_1}{x_2}-1 = 0\,.
\label{SUSYbipeq}
\ee
This is the same as the Lagrangian $\CL_M$ appearing in \eqref{Lbipeq}. An easier way to derive \eqref{SUSYbipeq} would be to begin with the product of moduli spaces for two tetrahedra
\be r''+r^{-1}-1=0\,,\qquad s'+s''^{-1}-1=0\,,\ee
and simply apply the affine $Sp(4,\Z)$ transformation $r\to\frac{x_1}{p_2},\, r''\to p_1,\, s_2\to p_2,\, s_1\to-\frac{p_1}{x_2p_2}$\,.

Equivalently, we can take the decomposition of the bipyramid into three tetrahedra, and the corresponding XYZ model. The twisted superpotential corresponding to the Lagrangian \eqref{XYZbip} is
\be \wt\CW_{\rm eff}(X_1,X_2,C) = \Li_2(e^{-X_1})+\Li_2(e^{-X_2})+\Li_2(e^{C-X_1-X_2})+i\pi(X_1+X_2-C)\,. \ee
Note that, according to the shifts in the symplectic transformation \eqref{Spbip3}, we have turned on a half-integral theta angle for the combination $\Sigma_{X_1}+\Sigma_{X_2}-\Sigma_C$. Setting $P_1=\pd\wt\CW/\pd X_1,\,P_2=\pd\wt\CW/\pd X_2,\, \Gamma=\pd\wt\CW/\pd C$ and exponentiating, we find equations
\be \gamma p_1+\frac1{x_1}-1=0\,,\qquad \gamma p_2+\frac{1}{x_2}-1=0\,,\qquad -\frac{\gamma x_1x_2}{c}+\frac{x_1x_2}{c}-1=0\,. \label{preL3} \ee
Now, however, the (ordinary) cubic superpotential of the XYZ model tells us that we must set the twisted mass $C=0$ (modulo $2\pi i$), or $c=e^C=1$. By appending this to equations \eqref{preL3} and eliminating $\gamma$, we then obtain
\be (x_1-1)\Big(p_1+\frac{p_2}{x_1}-1\Big)=0\,,\qquad (x_2-1)\Big(p_2+\frac{p_1}{x_2}-1\Big)=0\,. \label{SUSYbipeq3}\ee
These are equivalent to \eqref{SUSYbipeq} as long as $x_1\neq 1$ and $x_2\neq 1$. We recall, however, that $x_{1,2}=1$ (or $X_{1,2}=0$) are precisely the analogues of the singular points in moduli space discussed below \eqref{Aone}. There, either supersymmetry is broken or new Higgs branches of dynamical vacua open up. Away from this singular locus, equations \eqref{SUSYbipeq3} reduce to \eqref{SUSYbipeq}.

%%%%%%%%%%%%%%%%%%%%%%%%%%%%%%%%%%%%%%%%%%%%%%%%%%%%%%%%%%%%%%%%%%%%%%%%%%%%%%%%%%%%%%%%%%%

\section{$S^3_b$ partition functions}
\label{sec:S3b}

In the previous section, the correspondence $(M,\Pi) \; \leftrightarrow \; T_{M,\Pi}$
was tested by comparing moduli spaces attached to each side of the correspondence.
A more refined test could be obtained by associating certain functions to each side.
For example, on the gauge theory side one can associate either an equivariant partition
function or an index (an analog of the elliptic genus) to the 3d $\CN=2$ theory $T_M$,
by analogy with what was done in \cite{AGT} or \cite{Rastelli-2dQFT, Rastelli-qYM} in the context of 4d $\CN=2$ gauge theory.
Then, these functions are expected to match the corresponding topological invariants of $M$.

In this section, we discuss one such test based on comparing the
partition function of the 3d $\CN=2$ theory $T_{M,\Pi}$ on a squashed three-sphere (or ``ellipsoid'') $S^3_b$
with the $SL(2)$ Chern-Simons partition function of the 3-manifold $M$:
\be
\boxed{\phantom{\int}
Z^{SL(2)}_{\rm CS} (M) \; = \; Z_{S^3_b} (T_{M,\Pi})\,,
\phantom{\int}}
\label{ZZclaim}
\ee
where the squashing parameter $b$ is related to the Chern-Simons coupling coupling strength $\hbar$ as
\be \hbar = 2\pi ib^2\,. \ee
This relation is a direct generalization of the AGT correspondence \cite{AGT} to three dimensions.
In fact, it is fully consistent with the AGT correspondence, which corresponds to taking $M = \R \times \CC$
to be a product of the ``time'' direction and a Riemann surface $\CC$ (possibly with punctures),
through a somewhat lengthy chain of correspondences \cite{DGG-defects}, \cite{DG-Sdual}, reviewed e.g. in \cite{Yamazaki-3d}

Various aspects of partition functions in $SL(2)$ Chern-Simons theory are discussed in \cite{gukov-2003, DGLZ, Wit-anal, Dimofte-QRS}. Given a 3-manifold $M$ with boundary phase space $\CP_{\pd M}$, as defined here in Section \ref{sec:geom}, Chern-Simons theory should promote $\CP_{\pd M}$ to a Hilbert space
\be \CP_{\pd M}\;\leadsto\; \CH_{\pd M}\,, \label{HdM} \ee
and the partition function $Z^{SL(2)}_{\rm CS}(M)$ can be thought of as a distinguished wavefunction in $\CH_{\pd M}$. In particular, $Z^{SL(2)}_{\rm CS}(M;X_1,X_2,...)$ is a function of half the coordinates on $\CP_{\pd M}$, the ``positions'' in a given polarization $\Pi$. An affine $Sp(2N,\Z)$ change of polarization acts on $Z^{SL(2)}_{\rm CS}(M;X_1,X_2,...)$ in the standard Weil representation \cite{Shale-rep, Weil-rep}; for example, $S$-type elements act as Fourier transform, and $T$-type elements act as multiplication by quadratic exponentials $\sim \exp \frac{X_i^2}{2\hbar}$.

Similarly, the $S^3_b$ partition function of $T_{M,\Pi}$ depends on the twisted masses $m_\CO$ of various chiral operators that transform under $U(1)$ flavor symmetries. These real masses are naturally complexified by the R-charge, due to the background curvature of the ellipsoid \cite{Kapustin-3dloc, HHL}. Indeed, if we describe $S^3_b$ geometrically as
\be b^2|z_1|^2+b^{-2}|z_2|^2 = 1\,,\qquad z_1,z_2\in\CC\,, \label{defS3b}\ee
then $Z_{S^3_b}(T_{M,\Pi})$ depends holomorphically on the combinations $\tilde m_\CO \equiv m_\CO+\frac{iQ}{2}R_\CO$, with $Q=b+b^{-1}$. These complexified masses become identified with the ``positions'' in $\CP_{\pd M}$ or $\CH_{\pd M}$, as%
\footnote{Throughout this section, we work in units such that the ``average'' radius of the ellipsoid is $\rho=1$. Otherwise, it would appear on the right-hand side of \eqref{defS3b}, and would multiply $m_{\CO_X}$ in \eqref{XmO}.}
\be \label{XmO}
X = 2\pi b\,
 \tilde m_{\CO_X}= 2\pi b\,m_{\CO_X}+\Big(i\pi+\frac\hbar2\Big)R_{\CO_X}\,,
\ee
where $\CO_X$ is (say) the operator we associated to a boundary position $X$ in Section \ref{sec:defTM}. We will see that the ellipsoid partition function $Z_{S^3_b}(T_{M,\Pi};\tilde m_{X_1},\tilde m_{X_2},...)$ transforms as a wavefunction under changes of the polarization $\Pi$, in exactly the same way as $Z^{SL(2)}_{\rm CS}(M;X_1,X_2,...)$.

Both sides of \eqref{ZZclaim} are eminently computable. In fact, \cite{Dimofte-QRS} developed a general state integral model for $SL(2)$ Chern-Simons theory that directly quantizes the semi-classical construction of flat connections from ideal tetrahedra, as described in Section \ref{sec:geom}. Similarly, \cite{HHL} derived a prescription for ellipsoid partition functions of Chern-Simons-matter theories, using equivariant localization. It is not hard to see that the two constructions become equivalent when applied to our theories $T_{M,\Pi}$. We proceed to study a few aspects of this equivalence, starting with basic $T_\Delta$ building blocks and then forming more general theories/manifolds.

\subsection{Chirals and tetrahedra}
\label{sec:S3tet}

Consider a free chiral multiplet $\phi_Z$ with twisted mass $m_Z$ for a $U(1)$ flavor symmetry, and R-charge $R_Z$. This R-charge assignment enters in a fundamental way when putting the chiral on an ellipsoid. We set $\tilde m_Z = m_Z+\frac{iQ}{2}R_Z$, and find a partition function%
\footnote{Here and in the following, we will ignore overall numerical constants in front of the partition function.} %
\cite{HHL}
\be Z_{S^3_b}(\text{chiral multiplet}) = s_b\big(\tfrac{iQ}{2}-\tilde m_Z\big)\,, \label{Zchiral} \ee
where
\be s_b(x) = \prod_{m,n \in \Z_{\ge 0}} \frac{mb + nb^{-1} + \tfrac{Q}{2}-ix}{mb + nb^{-1} + \tfrac{Q}{2}+ix} = e^{-\frac{i\pi}{2}x^2}\prod_{r=1}^\infty\frac{1+e^{2\pi bx+2\pi i b^2(r-\tfrac12)}}
 {1+e^{2\pi b^{-1}x+2\pi ib^{-2}(\tfrac12-r)}}
\label{sbdef}
\ee
is a variant of the noncompact quantum dilogarithm function \cite{Barnes-QDL, Fad-modular} commonly used in Liouville theory.

Two of the properties enjoyed by the function $s_b(x)$ are
\begin{subequations}
\be
s_b(x) s_b(-x) \; = \; 1\,,
\label{sbsb1}
\ee
\be
s_b(x) \sim
\begin{cases}
e^{i \pi x^2/2} & \mbox{as } x \to + \infty \\
e^{-i \pi x^2/2} & \mbox{as } x \to - \infty\,,
\end{cases}
\label{sbasympt}
\ee
\end{subequations}
which have a nice interpretation in 3d $\CN=2$ gauge theory.
According to \eqref{Zchiral}, the first property \eqref{sbsb1}
implies that the partition function of two chiral fields $\phi$, $\phi'$ of opposite flavor charge and R-charge adding to 2 is trivial.
Indeed, this R-charge assignment allows one to add a marginal superpotential
\begin{equation}
\CW \; = \; M \phi \phi'
\end{equation}
which makes both fields arbitrarily massive and decouples them.
The second property \eqref{sbasympt} agrees with an important fact:
a Chern-Simons action of level $k$ for the background gauge field gives a contribution
\begin{equation}
e^{- i \pi k \tilde m^2}
\label{z3dCSk}
\end{equation}
to the partition function. Therefore, we see that at large positive $\sigma$ the chiral multiplet
contributes as a Chern-Simons coupling of level $+\tfrac{1}{2}$, while at large negative $\sigma$
as a Chern-Simons coupling of level $-\tfrac{1}{2}$, as expected \cite{AHISS} (\cf\ our discussion of such couplings in Section \ref{sec:MS}).
In a similar way, many beautiful identities obeyed by the special function \eqref{sbdef} --- in turn related to the combinatorics of 3-manifolds triangulations ---
find physical interpretation as dualities among 3d $\CN=2$ gauge theories.

The actual theory $T_{\Delta,\Pi_Z}$ associated to a tetrahedron has an extra level $-\tfrac12$ Chern-Simons coupling for the background gauge field, leading to a partition function
\be \boxed{ Z_{S^3_b}(T_{\Delta,\Pi_Z};\tilde m_Z) = e_b\big(\tfrac{iQ}{2}-\tilde m_Z\big) \equiv e^{\frac{i\pi}{2}\big(\frac{iQ}2-\tilde m_Z\big)^2}s_b\big(\tfrac{iQ}{2}-\tilde m_Z\big) \,.}
\label{ZDelta}\ee
With the identification \eqref{XmO}, this is equivalent to the Chern-Simons partition function of a single tetrahedron, found in \cite{Dimofte-QRS}.

In order to consider other polarizations for $T_{\Delta}$, we should analyze how the $SL(2,\Z)$ action on gauge theories affects partition functions. It is already clear from \eqref{z3dCSk} that the $T$-move sends
\be T\;:\quad Z_{S^3_b}(\tilde m)\;\mapsto\; Z_{S^3_b}'(\tilde m)= e^{-i\pi \tilde m^2}Z_{S^3_b}(\tilde m)\,. \label{Z3T} \ee
Similarly, the $S$-move adds a factor $e^{-2\pi i \tilde m\tilde m'}$ to the partition function, and dictates that we integrate over the vev $m$, since its gauge multiplet has become dynamical. In other words, $S$ acts as a Fourier transform:
\be S\;:\quad Z_{S^3_b}(\tilde m)\;\mapsto\; Z_{S^3_b}'(\tilde m')=\int d\tilde m\, e^{-2\pi i \tilde m\tilde m'}\, Z_{S^3_b}(\tilde m)\,. \label{Z3S} \ee
Note that this an integral along the real line, which could be deformed to a contour in the complex plane. In addition to $S$ and $T$, affine shifts in polarization also act nontrivially on the ellipsoid, by redefining the R-charge used to couple a theory to background curvature.
For example, a classical shift by $\pm i\pi$ in a position coordinate $Z$ corresponds to sending $R_Z\mapsto R_Z\pm 1$, or $\tilde m_Z\mapsto \tilde m_Z\pm \tfrac{iQ}{2}$.

The above action of the affine symplectic group shows that the ellipsoid partition function transforms as a wavefunction under changes of polarization, precisely as claimed. In particular, the above transformations are identical to those that appear in $SL(2)$ Chern-Simons theory. As a simple example, we can consider the affine $ST$ action that sends the polarization $\Pi_Z$ to $\Pi_{Z'}$ for the tetrahedron theory. This affine action was called $\sigma$ in \eqref{sST}. We find
\begin{align} Z_{S^3_b}(T_{\Delta,\Pi_{Z'}};\tilde m_{Z'}) &=\sigma \circ Z_{S^3_b}(T_{\Delta,\Pi_Z}) \notag \\ &= \int\, dm_Z\,e^{-i\pi \tilde m_Z(\tilde m_Z+2\tilde m_{Z'}-iQ)}\,
e_b\big(\tfrac{iQ}{2}-\tilde m_Z\big)  \notag \\
&= e_b\big(\tfrac{iQ}{2}-\tilde m_{Z'}\big)\,, \label{STZDelta}
\end{align}
up to a constant factor.
The last equality follows from a standard functional identity for $e_b(x)$ \cite{FKV}, and verifies the prediction from mirror symmetry that the transformation $\sigma$ leaves the tetrahedron theory invariant.

\subsection{Gluing and bipyramids}

In Section \ref{sec:MS}, we derived $ST$-invariance of the tetrahedron theory $\CT_1\simeq \CT_{\Delta,\Pi_Z}$ by starting with $\CN=4$ mirror symmetry, translating to $\CN=2$ mirror symmetry for the XYZ model and SQED with $N_f=1$, and and then reducing further to the theories $\CT_1$ and $ST\circ \CT_1$ via a mass deformation. It is somewhat instructive to now do the same at the level of partition functions. In the process, we will see how gluing of partition functions should work.

Let's begin with the partition function of a hypermultiplet,
with (complex) vector twisted mass denoted by $x$ and axial twisted mass by $y$:
\begin{equation}
Z_{S^3_b} (\, \text{hypermultiplet} \, ) \; = \; s_b \big( \tfrac{iQ}{2} -x -y \big) s_b \big( \tfrac{iQ}{2} +x -y \big)\,.
\label{Zhyper}
\end{equation}
The ${\cal N}=2$ R-charge and axial charge are a linear combination of the Cartan generators of the $SU(2)_H \times SU(2)_C$ R-charges of the 
${\cal N}=4$ theory. We are using a convention where in the ${\cal N}=2$ language the R-symmetry of chiral multiplets in the standard hypermultiplet is 
absorbed in their axial twisted mass $y$. Then the scalar field in the vectormultiplet has ``bare'' R-charge $2$, and axial charge $-2$, i.e. complex twisted mass $i Q -2y$.
This is also required for the basic superpotential coupling required by an ${\cal N}=4$ gauging. 

Hence if we add a full $\CN=4$ gauge multiplet to gauge the flavor symmetry,
the chiral multiplet in it contributes a $s_b(2 y - i Q/2)$. 
The partition function is
\begin{equation}
s_b(2 y - i Q/2) \int s_b(i Q/2-x -y) s_b(i Q/2+x -y) e^{- 2 i \pi z x } dx\,.
\end{equation}

The basic $\CN=4$ mirror symmetry should match this to the partition function of a twisted hypermultiplet,
{\it i.e.} a hypermultiplet with the opposite axial charge \cite{IS}.
The chiral fields in a twisted hypermutliplet have ``bare'' R-charge $1$ and axial charge $-1$, i.e. complex twisted mass $\tfrac{iQ}{2} -y$.
Hence we should replace $y$ with $\tfrac{iQ}{2}-y$ in \eqref{Zhyper}
and write the basic $\CN=4$ mirror symmetry relation as
\begin{equation} \label{twothree}
s_b(2 y - i Q/2) \int s_b(i Q/2-x -y) s_b(i Q/2+x -y) e^{- 2 i \pi z x } dx = s_b(y-z) s_b(y+z)
\end{equation}
As a check, we are supposed to obtain either the partition functions of $\CN=2$ SQED with $N_f=1$ flavor
or the partition function of the XYZ model by acting with $S$ or with $F$ on the above relation.
If we act with $S$, {\it i.e.} with the Fourier transform, we get
\begin{equation}
s_b(2 y - i Q/2)  s_b(i Q/2-x -y) s_b(i Q/2+x -y) \; = \; \int s_b(y-z) s_b(y+z) e^{- 2 i \pi z x }  dx \,. \label{ZXYZ}
\end{equation}
The left-hand side is the partition function of the XYZ model. The real masses of the three chiral fields
add to zero, and the R-charges to $2$, as it should be to allow the superpotential interaction $\CW = \mu u\tilde u$, \cf\ Section \ref{sec:MS}. Equation \eqref{ZXYZ} happens to be another well known identity for quantum dilogarithm functions \cite{FKV, PonsotTeschner}.

Now, if we redefine $x\to x-y$, $z\to z+y-\tfrac{iQ}{2}$, and take $y$ to be large and positive in \eqref{twothree}, we replicate the mass deformation that reduces us to the theory $\CT_1 \simeq \CT_{\Delta,\Pi_Z}$. Expression \eqref{twothree} becomes
\be \int dx\, e^{-i\pi x\big(x+2(z-\frac{iQ}{2})\big)}\,e_b\big(\tfrac{iQ}{2}-x\big) = e_b\big(\tfrac{iQ}{2}-z\big)\,,\ee
which is precisely \eqref{STZDelta}, expressing the mirror symmetry $\sigma\circ \CT_1 \simeq  \CT_1$.

We could also add Chern-Simons terms on both sides of \eqref{ZXYZ} in order to reproduce the exact partition function of the bipyramid theory, as discussed in Section \ref{sec:gauge23}. Namely, we find an identity
\begin{align} & e^{i\pi (i Q)\tilde m_3}e_b\big(\tfrac{iQ}{2}-\tilde m_1\big)e_b\big(\tfrac{iQ}{2}-\tilde m_2\big)e_b\big(\tfrac{iQ}{2}-\tilde m_3\big)\Big|_{\tilde m_3=iQ-\tilde m_1-\tilde m_2} \\
 &\hspace{1in} =\int d\sigma\, e^{-i\pi \sigma^2-2\pi i\sigma\big(\tilde m_2-\frac{iQ}{2}\big)}e_b\big(\tfrac{iQ}{2}+\sigma\big)e_b\big(\tfrac{iQ}{2}-\sigma+\tilde m_1\big)
\notag \end{align}
The two sides correspond to the theories of three and two tetrahedra, respectively, both in the equatorial polarization $\Pi_{\rm eq}$, with external edge positions $X_1 = 2\pi b\, \tilde m_1$ and $X_2=2\pi b\,\tilde m_2$. For the left-hand side, the superpotential $\CW = \CO_C = \phi_Z\phi_W\phi_Y$ \eqref{WXYZ} implements the constraint $\tilde m_1+\tilde m_2+\tilde m_3=iQ$.

More generally, the rules for constructing theories $T_{M,\Pi}$ in Section \ref{sec:glue} lead to the following rules for calculating the corresponding ellipsoid partition functions:
\begin{itemize}
\item[1)] Multiply together partition functions $Z_{S^3_b}(T_{\Delta_i,\Pi_i};\tilde m_{Z_i}) = e_b\big(\tfrac{iQ}{2}-\tilde m_{Z_i}\big)$, one for each tetrahedron in the triangulation of $M$.
\item[2)] Act with $Sp(2N,\Z)$ in the Weil representation (\ie\ by generalizing the quadratic exponentials and Fourier transforms of \eqref{Z3T}--\eqref{Z3S}), to transform to the polarization $\tilde\Pi$ in which all internal edges are ``positions.''
\item[3)] Set the complex masses $\tilde m_I$ now associated to internal edges equal to $iQ$.
\end{itemize}
We note that the specialization in Step 3 is the only consequence of adding a superpotential $\CW = \sum_I \CO_I$ to the theory $T_{M,\Pi}$. Indeed, such a superpotential sets the real masses of the $\CO_I$ to zero and the R-charges equal to $2$. Otherwise, the ellipsoid partition function is completely independent of superpotential terms, and cares only about gauge and matter content.

These rules for constructing $Z_{S^3_b}(T_{M,\Pi})$ are identical to the rules presented in \cite{Dimofte-QRS} for building the $SL(2)$ Chern-Simons partition function of $M$. One can see even subtle quantum effects matching in the two descriptions. For example, in quantum Chern-Simons theory, the classical internal edge constraints $C_I = 2\pi i$ become corrected to $C_I = 2\pi i+\hbar$, and this follows immediately from the dictionary \eqref{XmO} between edge parameters and complexified masses $\tilde m_I$.

\subsection{Figure-eight knot}

We should be able to reproduce the well known Chern-Simons wavefunction for the figure-eight knot complement from the theory $T_{\mb{4_1},\Pi}$ described in Section \ref{sec:gauge41}. The definition of the actual theory, including internal edge operators, required a decomposition of the knot complement into six tetrahedra. However, since ellipsoid partition functions do not depend in a crucial way on superpotential terms, we might hope to get away with the simpler decomposition into two tetrahedra, also discussed in Section \ref{sec:gauge41}. Indeed, this turns out to work.

From the Lagrangian \eqref{41tet2}, we can immediately write down a partition function
\be Z_{S^3_b}(T_{\mb{4_1},\Pi};\tilde m_U) = \int d\sigma\, e^{i\pi\big(\tilde m_U^2+(2\tilde m_C-iQ+2\tilde m_U-\sigma)\sigma\big)}\,e_b\big(\tfrac{iQ}{2}-\sigma-\tilde m_U\big)e_b\big(\tfrac{iQ}{2}-\sigma\big)\,. \ee
Now, there are no operators in the theory to force $\tilde m_{C_1}=iQ$, but we can put this in by hand. Up to a factor of   due to a small change of  polarization, the result is then identical to the figure-eight wavefunctions described in \cite{hikami-2006, DGLZ, Dimofte-QRS} (see also \cite{Yamazaki-layered, SpirVar-knots}).

\subsection{Relation to moduli spaces on $\R^2\times S^1$}

Finally, we point out that our tests of the proposed duality $(M,\Pi) \; \leftrightarrow \; T_{M,\Pi}$
here and in section \ref{sec:moduli} are not entirely unrelated.
Indeed, in the semi-classical limit $\hbar = 2\pi i b^2 \to 0$, the partition function of
the theory $T_M$ behaves exactly in the same way as the partition function of Chern-Simons theory on $M$,
\be
Z_{S^3_b} (T_M) \,\overset{\hbar\to0}{\sim}\, \exp\Big(\frac1\hbar \widetilde{\CW}_{{\rm eff}} + \CO(\log\hbar)\Big)\,,
\label{zscw}
\ee
where $\widetilde{\CW}_{{\rm eff}}$ is the effective twisted superpotential of the theory $T_M$ on $\R^2 \times S^1$.
Hence, if $\widetilde{\CW}_{{\rm eff}}$ matches the classical $SL(2)$ Chern-Simons action on $M$,
\be
\widetilde{\CW}_{{\rm eff}} (T_M) \; = \; S_0 (M) \,,
\ee
then the relation between moduli spaces \eqref{MMclaim} follows automatically.
Indeed, the moduli space $\CM_{{\rm flat}} (M,SL(2,\C))$ is a graph of $d S_0$
and, similarly, the moduli space $\CM_{{\rm SUSY}} (T_M)$ is a graph of $d \widetilde{\CW}_{{\rm eff}}$.
In terms of gauge theory, the reason for \eqref{zscw} is that, in the limit $b \to 0$,
the squashed 3-sphere $S^3_b$ degenerates into $\R^2 \times S^1$,
\be
S^3_b \quad \leadsto \quad \R^2 \times S^1 \,.
\ee

The relation between moduli spaces $\CM_{\rm flat}(M,SL(2,\C)) = \CM_{\rm SUSY}(T_M)$ of Section \ref{sec:moduli} has a ``quantum'' analog
that does not require taking the limit $\hbar \to 0$.
Indeed, the full quantum partition functions discussed here obey a set of $q$-difference equations:
\be
\boxed{\phantom{\int}
\widehat{A}_i \; Z \; = \; 0
\phantom{\int}}
\label{AZclaim}
\ee
for some operators $\widehat{A}_i$ that in the classical limit become defining polynomials of our moduli spaces.
In Chern-Simons theory, \eqref{AZclaim} is known as the generalized / quantum volume conjecture \cite{gukov-2003}
(sometimes also called the AJ-conjecture \cite{Gar-Le,garoufalidis-2004} in the math literature),
whereas in $\CN=2$ gauge theory it expresses Ward identities for line operators. We consider these line operators next.

%%%%%%%%%%%%%%%%%%%%%%%%%%%%%%%%%%%%%%%%%%%%%%%%%%%%%%%%%%%%%%%%%%%%%%%%%%%%%%%%%%%%%%%%%%%%%%%%

\section{Line operators and $q$--difference equations}
\label{sec:lines}

In order to understand the meaning of operator identities \eqref{AZclaim}
in 3d $\CN=2$ theory, we need to incorporate line operators in our correspondence \eqref{MvsTM}.

Given a triangulated 3-manifold $M$ with nonempty boundary $\pd M$, each equation in \eqref{AZclaim}
is written in terms of quantum holonomy operators%
\footnote{For example, in the context of knot complements, these operators are often denoted as $\hat m=e^{\hat u}$ and $\hat \ell=-e^{\hat v}$.} %
that, from the viewpoint of Chern-Simons theory on $M$, are obtained by quantizing the space of flat $SL(2,\C)$ connections $\CP_{\pd M}$ on the boundary. These operators act on the Hilbert space \eqref{HdM}.
We illustrate this with a simple example that plays a key role in this paper, namely
with the $\CN=2$ theory $T_{\Delta,\Pi_Z}$ that we associate with a single tetrahedron.

In particular, in the previous section we identified the $S^3_b$ partition function
of this theory \eqref{ZDelta} with the wave function of the $SL(2)$ Chern-Simons theory on a tetrahedron. From the explicit form of the partition function \eqref{ZDelta}, it is easy to
see that it satisfies the functional equation%
\footnote{We simply abbreviate $Z_{S^3_b}(T_{\Delta,\Pi_Z},\tilde m_Z)$ as $\CZ(\tilde m_Z)$.}
\be
\CZ(\tilde m_Z + ib \big) = \left( 1 - e^{-2\pi b \tilde m_Z} \right) \CZ(\tilde m_Z ) \,.
\ee
Using $\hat Z''= ib\partial_{\tilde m_Z}$ and $\hat Z=2\pi b\tilde{m}_Z$, we can write this equation in
a more convenient form:
\be
\left( e^{\hat Z''} + e^{-\hat Z} - 1 \right) \CZ(\tilde m_Z) \; = \; 0 \,,
\label{psiAhat}
\ee
which is clearly reminiscent of the familiar equation \eqref{Aone}
that describes the space of SUSY moduli in the theory $T_{\Delta, \Pi_{Z'}}$.
Indeed, for reasons that we reviewed at the end of section \ref{sec:S3b},
in the semi-classical limit $\hbar \sim b^2 \to 0$ the equation \eqref{psiAhat}
gives precisely \eqref{Aone}:
\be
\CM_{{\rm SUSY}}
~:~ \quad
e^Z + e^{- Z'} - 1 \; = \; 0 \,.
\label{apoltetr}
\ee
In terms of geometry, we know from Section \ref{sec:geom} that $Z$ and $Z'$ are the complexified ``shear coordinates'' or edge parameters on the boundary $\pd \Delta$ of the tetrahedron; and indeed \eqref{psiAhat} is just the quantization of the tetrahedron's classical Lagrangian \eqref{LDelta} \cite{Dimofte-QRS}. More generally, if a 3-manifold $M$ has a triangulated geodesic boundary, it is the quantization of external edge coordinates $\exp(\hat X_E)$ on the boundary that appears in the operator equations \eqref{AZclaim}.

From a different perspective, the classical external edge coordinates $x_E=\exp(X_E)$ on a triangulated geodesic boundary $\CC=\pd M$ also correspond to vevs of line operators in the \emph{four-dimensional} $\CN=2$ theory $T[\CC,\mathfrak{su}(2)]$; and the quantized $\hat x_E=\exp(\hat X_E)$ correspond to the quantum line operators themselves \cite{DMO, AGGTV, DGOT, GMNIII}. To be more precise, it was shown in \cite{GMNII, GMNIII} that every edge $E$ of $\CC$ determines an IR line operator $\exp(\hat X_E)$ in the abelian $\CN=2$ theory on the Coulomb branch of $T[\CC,\mathfrak{su}(2)]$. This operator carries the electric and magnetic charges associated to the edge $E$, exactly as described in Section \ref{sec:gauge4d}. Using this relation, we propose to interpret operator equations \eqref{AZclaim} as Ward identities
for line operators in a 4d theory coupled to the 3d boundary theory $T_M$.

\begin{figure}[htb]
\centering
\includegraphics[width=2.5in]{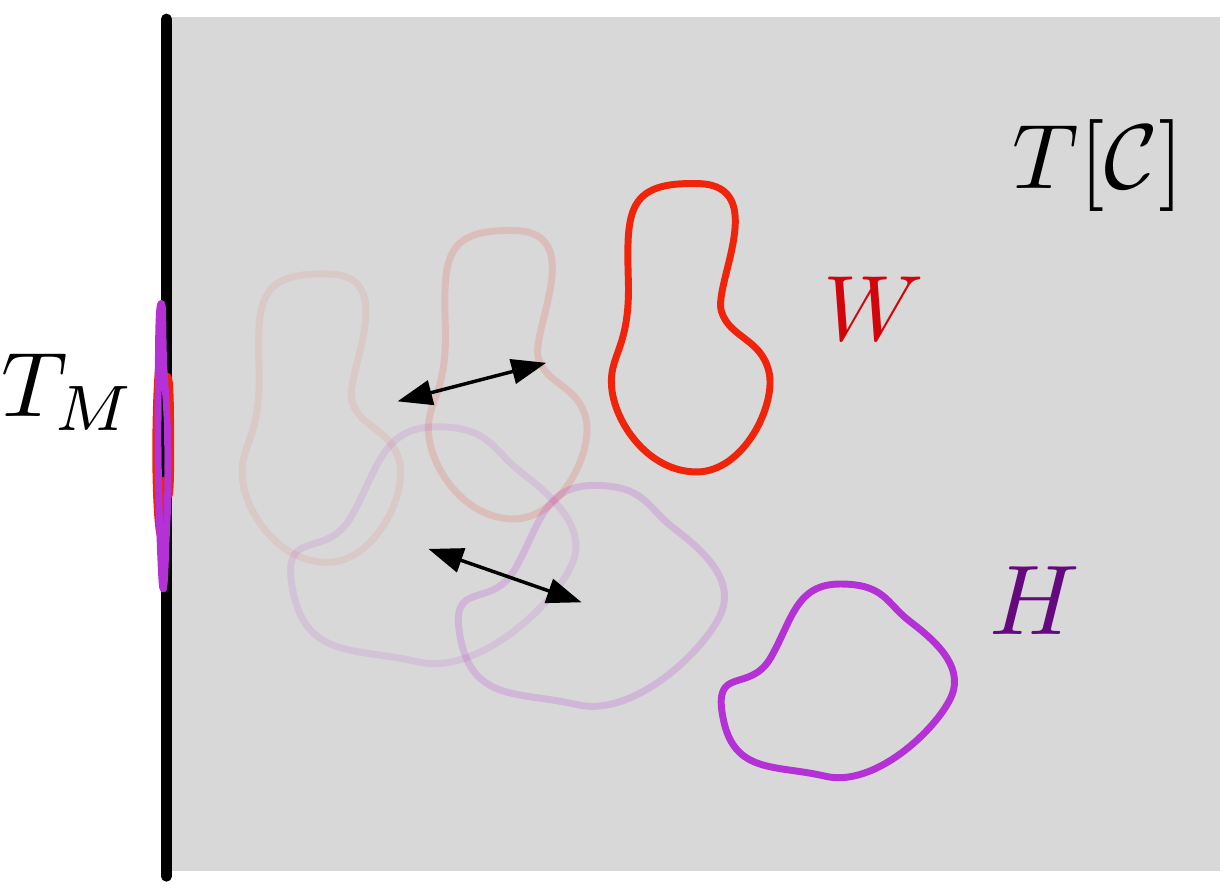}
\caption{Line operators in 4d becoming identified in the boundary theory $T_M$.}
\label{fig:bdyop}
\end{figure}

In the presence of boundary conditions, not all line operators of the bulk $\CN=2$ gauge theory in
four dimensions are independent. Indeed, one can start with a line operator $L$ (or, more generally, a collection of line operators $L_i$)
in the 4d $\CN=2$ gauge theory and then bring it to the three-dimensional boundary where the theory $T_M$ lives (Figure \ref{fig:bdyop}).
Due to the boundary conditions (which {\it e.g.} may identify some of the 4d fields),
vevs of line operators that were independent in the bulk become related on the boundary.
This can be summarized in the form of Ward identities
\be
\sum c_i L_i \; = \; 0 \,.
\label{cLi}
\ee
For example, in our favorite example of the theory $T_{\Delta}$ the equation \eqref{psiAhat} can be written in
the form \eqref{cLi} as
\be
W + H^{-1} - 1 \; \simeq \; 0
\label{WHinT1}
\ee
where we used the identification of $\hat Z$, $\hat Z'$, and $\hat Z''$ with the corresponding abelian
Wilson / 't Hooft line operators:
\be
\begin{array}{l@{\qquad}l}
\underline{\text{edge}} & \underline{\text{line operator}} \\[.1cm]
\hat z = e^{\hat Z} & W=\text{Wilson} \\[.1cm]
\hat z' = e^{\hat Z'} & \text{Wilson-'t Hooft} \\[.1cm]
\hat z'' = e^{\hat Z''} & H=\text{'t Hooft}
\end{array}
\label{whhlines}
\ee
(Thus, $\hat z^{-1}=H^{-1}$ denotes an 't Hooft operator of magnetic charge $-1$. Similarly, $W^0=H^0=1$ denotes a trivial line operator.)
The above dictionary \eqref{whhlines} corresponds to the polarization $\Pi_{Z}$ for $T_\Delta$.
The triality symmetry of $T_\Delta$ \eqref{STonT1}, generated by the $ST$ element of the 4d electric-magnetic duality group $SL(2,\Z)$, permutes Wilson, 't Hooft, and Wilson-'t Hooft operators.

To explain the origin of Ward identities like \eqref{WHinT1}, it is instructive to simplify the
theory $T_{\Delta}$ (which consists of a chiral multiplet and Chern-Simons coupling) even further
and consider only the Chern-Simons part of the theory. As we discussed in section \ref{sec:S3b},
a supersymmetric Chern-Simons interaction at level $k$ for the background gauge field contributes
to the partition function a factor \eqref{z3dCSk}:
\begin{equation}
\CZ_{CS_k} = e^{- i \pi k \tilde m^2} \,.
\end{equation}
Much like the partition function of the theory $T_{\Delta}$, it obeys the following $q$-difference equation:
\be
\left( \hat z'' - q^{\frac k2} \, \hat z^k \right) \CZ_{CS_k} =
\left( e^{ib \partial_{\tilde m}} - e^{i\pi b^2 k+2\pi bk\tilde m} \right) \CZ_{CS_k} = 0 \,.
\ee
According to \eqref{whhlines},
this identity should be interpreted as a statement that at a 3d boundary with Chern-Simons term at level $k$
a 't Hooft operator with one unit of a magnetic flux is equivalent to a Wilson operator of electric charge $k$,
\be
H - e^{i\pi b^2 k} \, W^{k} \; \simeq \; 0 \,.
\ee
This is indeed correct, as one can easily verify by doing a direct path integral manipulation.
Notice, it is important here that supersymmetric Chern-Simons theory
lives on the boundary of the 4d space-time where Wilson and 't Hooft operators belong.

\begin{figure}[htb]
\centering
\vspace{.2in}
\includegraphics[width=5.2in]{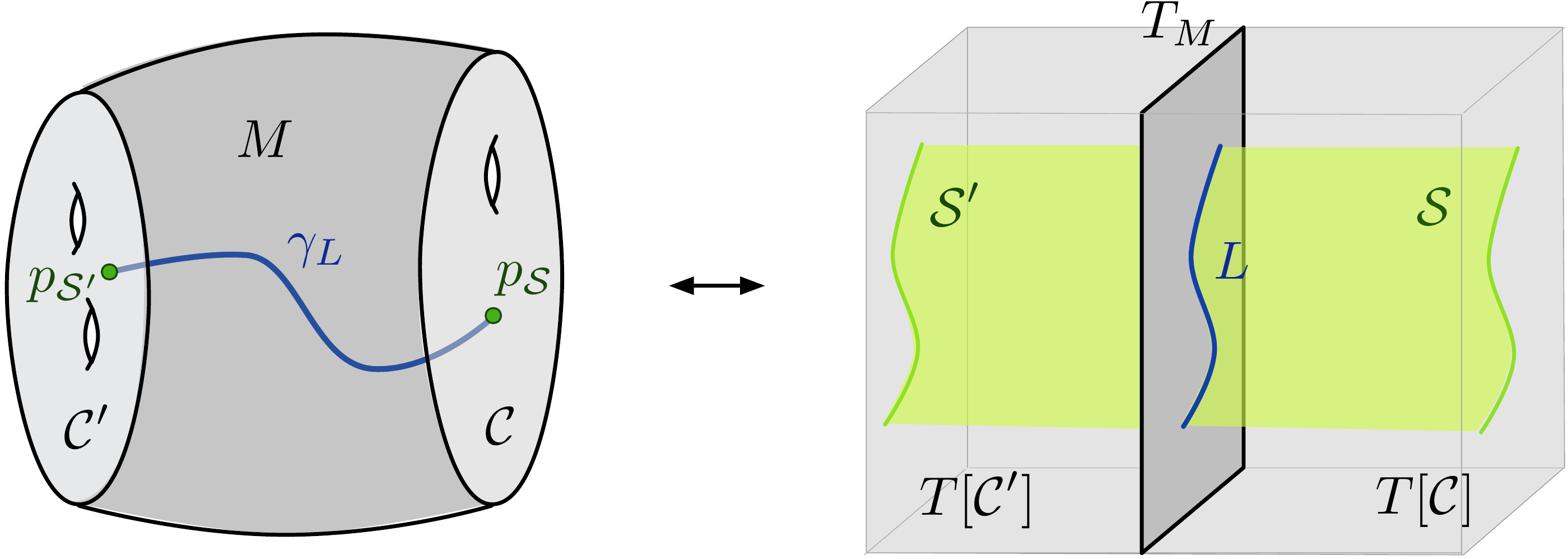}
\caption{Line operators in both $M$ and $T_M$.}
\label{fig:MTMop}
\end{figure}

Most of our discussion in this section was based on interpreting $T_M$ as a boundary theory
in the 4d $\CN=2$ theory on the Coulomb branch of $T[\CC,\mathfrak{su}(2)]$,
where $\CC = \partial M$ is the geodesic boundary of $M$.
This interpretation can be easily extended to 3-manifolds with ``small'' boundaries ({\it a.k.a.} cusps)
and also to 3-manifolds with several boundary components.
For example, in the latter case, each boundary component is a 2-dimensional Riemann surface $\CC$
to which we associate either IR or UV limit of the 4d $\CN = 2$ gauge theory $T[\CC]$
depending on whether the boundary $\CC$ is ``big'' or ``small.''

Within this framework, we could also look at a different class of line operators, corresponding to curves in a 3-manifold $M$ itself.
In general, a 1-dimensional curve $\gamma_L$ inside a cobordism $M$ may have end-points
on various boundary components of $M$, as shown in Figure \ref{fig:MTMop}.
In order to find its interpretation in 3d $\CN=2$ theory $T_M$, we recall that
a point $p \in \CC$ defines a surface operator in 4d $\CN=2$ theory $T[\CC]$,
whereas the cobordism itself defines a domain wall between two different $\CN=2$ theories
in four dimensions ({\it cf.} Figure \ref{fig:cobordism1}).
In four-dimensional space-time, a surface operator meets the domain wall over
a 1-dimensional curve, which is precisely the line operator $L$ associated to $\gamma_L \subset M$, see Figure \ref{fig:MTMop}.
In this description of $T_M$ as a theory on a duality wall,
the line operator $L$ arises as an interface between two different surface operators.

The interplay between line operators on $M$ and line operators in 3d $\CN=2$ theory $T_M$ can be
easily motivated by thinking about $T_M$ as the effective theory
$T[M,\mathfrak{su}(2)]$ obtained by reduction of the six-dimensional $(2,0)$ theory on a 3-manifold $M$.
This is very similar to the correspondence between line operators in Liouville theory on $\CC$
and line operators in 4d $\CN=2$ theory $T[\CC,\mathfrak{su}(2)]$,
where 6d theory again turns out to be very useful \cite{DMO,AGGTV,DGOT}.
Indeed, six-dimensional $(2,0)$ theory contains two-dimensional surface operators.
Upon compactification on a $d$-dimensional manifold $M_d$, the support of a surface operator
can have the form $\gamma_L \times L$, where $\gamma_L \subset M_d$ is a 1-dimensional curve on $M_d$
and $L \subset \R^{6-d}$ is a line in the $(6-d)$ dimensional space-time where the theory $T[M_d,\mathfrak{su}(2)]$ lives.
Surface operators of this form give rise to a large class of line operators in $T[M_d,\mathfrak{su}(2)]$
labeled by curves $\gamma_L$ on $M_d$.

%%%%%%%%%%%%%%%%%%%%%%%%%%%%%%%%%%%%%%%%%%%%%%%%%%%%%%%%%%%%%%%%%%%%%%%%%%%%%%%%%%%%%%%%%%%%%%%%

\acknowledgments{We wish to thank A. Kapustin, N. Seiberg, C. Vafa, R. van der Veen, and E. Witten for many helpful and enlightening discussions.
The work of TD is supported in part by NSF Grant PHY-0969448.
The work of DG is supported in part by NSF grant PHY-0503584
and in part by the Roger Dashen membership in the Institute for Advanced Study.
The work of SG is supported in part by DOE Grant DE-FG03-92-ER40701 and in part by NSF Grant PHY-0757647. TD and SG thank the Kavli Institute for Theoretical Physics (research supported by DARPA under Grant No.
HR0011-09-1-0015 and by the National Science Foundation under Grant
No. PHY05-51164) and the Simons Center for Geometry and Physics for their hospitality in the summer of 2011. TD also acknowledges the Max Planck Institut f\"ur Mathematik for its hospitality and support during June, 2011.
Opinions and conclusions expressed here are those of the authors and do not necessarily reflect the views of funding agencies.}

\bibliographystyle{JHEP_TD}
\bibliography{toolbox}

\end{document}